\documentclass[a4paper,11pt]{article}
\pdfoutput=1
\usepackage[utf8]{inputenc}
\usepackage[T1]{fontenc}
\usepackage{lmodern}
\usepackage[british]{babel}
\usepackage{amsmath,slashed}
\usepackage{amssymb}
\usepackage{amsfonts}
\usepackage{float,physics}
\usepackage{stmaryrd}
\usepackage{url}
\usepackage{graphicx,psfrag}
\usepackage{placeins,bbm}
\usepackage{subcaption}

\usepackage{jheppub}
\usepackage{todonotes}

\DeclareOldFontCommand{\bf}{\normalfont\bfseries}{\mathbf}

\newcommand{\Ls}{\ensuremath{L_s}}
\newcommand{\Lr}{\ensuremath{L_r}}
\newcommand{\Nc}{\ensuremath{N}}
\newcommand{\SU}[1]{\ensuremath{\text{SU}(#1)}}

\newcommand{\be}{\beta}

\newcommand{\Figref}[1]{Fig.~\ref{#1}}

\def\be{\begin{equation}}
\def\ee{\end{equation}}
\DeclareOldFontCommand{\bf}{\normalfont\bfseries}{\mathbf}

	\title{A $T_2 \times R^2$ roadmap to Confinement in SU(2)
Yang-Mills theory}

    \author[a]{Georg Bergner}
    \author[b,c]{Antonio Gonz\'alez-Arroyo }
    \author[d]{Ivan Soler}

    \affiliation[a]{University of Jena, Institute for Theoretical Physics, \\
     Max-Wien-Platz 1, D-07743 Jena, Germany}
     
    \affiliation[b]{ Instituto de F\'{\i}sica Te\'orica UAM/CSIC,  Nicol\'as
      Cabrera 13-15, \\
      Universidad Aut\'onoma de Madrid, E-28049 Madrid, Spain}
      
    \affiliation[c]{
     Departamento de F\'{\i}sica Te\'orica,  M\'odulo 15,
     \\
      Universidad Aut\'onoma de Madrid, Cantoblanco, E-28049 Madrid, Spain}

     \affiliation[d]{Dipartimento di Fisica dell’Universit`a di Pisa and INFN, \\
     Largo Pontecorvo 3, I-56127 Pisa, Italy}

    \emailAdd{georg.bergner@uni-jena.de}
    \emailAdd{A.Gonzalez-Arroyo@pm.me}
    \emailAdd{ivan.soler@df.unipi.it}

    \abstract{We study the behaviour of \SU{2} Yang-Mills fields on a $T_2\times R^2$ geometry where the two-torus is equipped with twisted boundary conditions. We monitor the evolution of the dynamics of the system as a function of the torus size $l_s$.  For small sizes the behaviour of the system is well understood in terms of semiclassical predictions. In our case, the long distance structure is that of a  two-dimensional gas of
    vortex-like fractional instantons with size and density growing with $l_s$.  Our lattice Monte Carlo simulations  confirm the semiclassical predictions and allow the determination of the relevant scale signalling the transition to the non-dilute situation.
    At low densities the string tension takes the standard value of a 2D center-vortex gas, growing with the density and approaching the value measured at infinite volume.  
    Our work includes preliminary studies of the extension to \SU{N} and to the region of  large sizes in which boundary conditions are irrelevant, and all physical scales  are determined by the $\Lambda$ parameter.}
     
\begin{document}
	\maketitle
\section{Introduction}
In this paper, we study the behaviour of \SU{\Nc} Yang-Mills fields living
on $T_2\times R^2$, a four dimensional space in which two dimensions are compactified as a
flat square  torus equipped with twisted boundary conditions
(TBC)~\cite{tHooft:1979rtg, tHooft:1980kjq, tHooft:1981sps}.
Our purpose is to monitor the evolution of the system as a function of the linear size
of the 2-torus measured in physical units. Since the theory is
asymptotically free, we expect perturbative and semiclassical methods
to be a good approximation to the system when the size is small, and by following the track towards large volumes we expect to gain insight into the complete understanding
of the fascinating non-perturbative properties of the theory. Our numerical analysis considers mainly the gauge group \SU{2}, while we also comment on generalizations to larger \Nc.

This strategy has a long history.  In the 90's, a similar pathway was
followed with an equivalent goal in mind.  It was first explored by
Lüscher~\cite{Luscher:1982uv} in the study of two-dimensional (2D) asymptotically free non-gauge field theories,  and later extended to the case of a Yang-Mills system in four dimensions (4D)~\cite{Luscher:1983gm}. All these studies were done within a Hamiltonian type
of setting, in which the euclidean time direction is non-compact while 
the spatial directions are compactified on a 3-torus whose size-dependence is being monitored ($T_3\times R$). One advantage is that one does not
expect a phase transition to take place at a particular size separating
the weak coupling region from the large volume one, and thus spoiling 
the connection one is trying to establish. Studies were done both with 
and without twisted boundary conditions using  analytic~\cite{Luscher:1982ma,
Luscher:1983gm,vanBaal:1986cw,
Koller:1987yk, Koller:1987fq, GonzalezArroyo:1987ycm,Daniel:1989kj,
Daniel:1990iz} and numerical~\cite{Berg:1988cp,Stephenson:1989pu} methods. It feels natural
to consider TBC for various reasons. First of all, perturbative studies 
with periodic boundary conditions have to address the peculiar
orbifold structure of the zero-energy  configurations. These zero
energy states (called {\em torons} in Ref.~\cite{Gonzalez-Arroyo1983,Coste:1985mn})  are
indeed the particular case of flat connections on the torus. Performing 
perturbative/semiclassical studies in this setting is a tour-de-force that some
researchers, such as Pierre van Baal, had to address to pursue the
program~\cite{vanBaal:1986cw,
Koller:1987yk, Koller:1987fq}. These problems are absent if suitable TBC are employed, and perturbative studies of the low energy states can be  performed quite straightforwardly~\cite{GonzalezArroyo:1987ycm,Daniel:1989kj,
Daniel:1990iz}. There is another very
puzzling phenomenon related the volume dependence of the system  that suggests the use of TBC. That is the fact that this volume
dependence disappears  at large \Nc. Periodic boundary conditions fail
to preserve the necessary center-symmetry for this phenomenon to take
place~\cite{Eguchi:1982nm, Bhanot:1982sh}. However, using TBC,  one  can enforce enough symmetry to  attain volume independence~\cite{GonzalezArroyo:1982hz}. How exactly this fact matches with the semiclassical program is still to be elucidated. In any case it is obvious that even at finite \Nc, TBC helps enforcing 
center-symmetry and hence smoothens the path towards large volumes.
Indeed, the studies performed on the 90's focused on the glueball spectrum,  showing that as the size is varied, level crossings occur in the low-lying  spectrum for  periodic boundary conditions that are absent for TBC~\cite{Koller:1987yk}.

However, even for the case of TBC, it soon became clear that one had to go beyond perturbation theory to connect small to large
volumes: semiclassical methods are necessary. This is somehow related
to the recovery of the spatial $(Z(\Nc))^3$ center-symmetry on $T_3\times R$. The case of \SU{2} is particularly simple, since in that case one finds that there are two gauge inequivalent zero-energy states, which are invariant under a $Z(2)\times Z(2)$ subgroup. The remaining $Z(2)$ group is broken spontaneously by
choosing any of the two gauge inequivalent vacua (zero-energy states). However, there is no real breaking of this symmetry since it costs a finite amount of action to tunnel between both zero-energy states. The configuration with minimum action
that tunnels between these vacua is indeed a fractional instanton with
topological charge $Q=1/2$~\cite{GarciaPerez:1989gt, GarciaPerez:1992fj}. The situation
resembles pretty much the problem of a 1-dimensional scalar field with
a double well-potential. The finite action of the kink solution, which
tunnels among the two minima, restores the symmetry, generating a
mass-gap proportional to the exponential of this kink-action.  
For the Yang-Mills system the equivalent of the field is played by a
particular Polyakov loop aligned along the direction of the `t Hooft
magnetic flux, and the kink is replaced by the fractional instanton. 
Notice also that the role of the mass-gap is now the energy  
with the correlator of two of the particular Polyakov lines separated
over time (an electric flux energy in `t Hooft terminology). However, 
the minimum energy associated with the exponential decay of the
correlator of two Polyakov lines is what we associate with the string
tension. For large volumes, the energy grows linearly with the size of
the torus with a slope given by the string tension. This is what
connects the semiclassical study to the recovery of the Confinement
regime. 

Indeed, a semiclassical study using similar formulas that we will use
here was already performed in the 90's within that spatial volume
dependence program on $T_3\times R$~\cite{RTN:1993ilw,GarciaPerez:1993jw}. 
At small torus sizes, the system gives rise to a 1-dimensional gas of kink-like fractional instantons with a Poissonian distribution and whose density, controlled by the semiclassical formulas, grows with
the spatial size. Remarkably, the analytic expressions matched with the
numerical results obtained by a Monte Carlo analysis of the system using lattice methods. However, the lattice methodology allows us to explore the system all the way towards large volumes. Interestingly
enough, for physical sizes that are not too far from the region where
the dilute gas approximation breaks down, the
electric flux energy dependence on the  torus size flattens up to a
linear one, giving values of the string tension close to those that other researchers had measured for the large volume
string tension~\cite{GarciaPerez:1993jw,
Gonzalez-Arroyo:1995ynx}. In trying to 
understand this  smooth transition, one of the present authors and his students proposed a
mechanism to explain it~\cite{Gonzalez-Arroyo:1995ynx, GonzalezArroyo:1995ex}. The main ideas are the following.  At small volumes
the size of the fractional instantons is fixed by the torus size. 
Semiclassical formulas include  quantum corrections which  imply that bigger fractional instantons have a higher probability (as is the case for instantons). However, this does not occur all the way to infinite size. At large enough sizes  the most probable FI size and separation is  a dynamical scale given in terms of the  Lambda parameter of the theory. Hence, once the torus becomes larger than this size, 
fractional instantons 
do not grow and tend to fill in the full four dimensional space. This
then explains how the system recovers rotational invariance and the 
twisted boundary conditions on the 3-torus become irrelevant.
Effectively, the resulting picture of the vacuum is described by the 4D
fractional instanton liquid model (FILM). Such a liquid would induce and relate quantities, like the string tension, the topological susceptibility and the condensate, which appear unrelated in other scenarios. We point to a recent
review~\cite{Gonzalez-Arroyo:2023kqv} of the proposal including additional references. Some recent lattice gauge theory results has been argued to provide additional evidence 
in favor of the FILM model~\cite{Nair:2022yqi} (see also 
Ref.~\cite{Unsal:2020yeh}).

More recently, a new set of papers has explored a different roadmap
with a similar strategy in mind. This originated in the rather
complementary study of an $S_1\times R^3$ geometry and using the size 
of the short circle as a monitoring parameter to connect the region of
weak coupling towards the strongly-coupled fully infinite volume  one. 
It is important that this evolution is free of phase transitions. The
connection is then smooth: a process that has been labelled {\em
adiabatic continuity}~\cite{Poppitz:2012nz, Poppitz:2012sw}. One possible way to prevent phase 
transitions is to focus on gauge theories with  adjoint fermions: 
Adjoint QCD. It is known that the gluonic interactions induce the
Polyakov  loop to concentrate on elements of the center, effectively
breaking the center symmetry and inducing a phase transition. Adjoint 
fermions act in the opposite direction and can neutralize the effect. This
is clear for the case of $\mathcal{N}=1$ Supersymmetric Yang-Mills, at
least when the boundary conditions are chosen periodic rather than
antiperiodic to preserve supersymmetry~\cite{Anber:2014lba}. The claim is that 
even for massive fermions this is indeed the case~\cite{Bergner:2018unx}. 
Another possibility explored is to deform the Yang-Mills interaction by
adding double trace terms that give additional weight to the
center-symmetric situations~\cite{Unsal:2008ch,Shifman:2008ja}. This $S_1\times R^3$
roadmap has been recently reviewed in Ref.~\cite{Poppitz:2021cxe}, focusing 
specially on the SU(2) gauge group like in our case. We address
the reader to this review for an updated list of references. 
However, explicit mention is necessary for some recent  papers that 
have considered and advocated a  
$T_2 \times R^2$ strategy   like the one addressed in this
paper~\cite{Tanizaki:2022ngt,Hayashi:2024qkm,Hayashi:2024yjc}. This might turn out to be a good way 
to put together the two different  semiclassical strategies  described in this introduction.

The paper is organized as follows. In the next section (Sec.~\ref{seq:semiclassical}) we start by making  a short review of  the semiclassical approach. This is then applied to our specific $T_2\times R^2$ geometry and boundary conditions. This shows the relevance of some particular fractional instanton solutions, some of whose properties are analysed.  In Sections~\ref{subseq:other_solutions} to \ref{subseq:twist_Q1} we explain how for small torus sizes a picture of a dilute  gas emerges, whose building blocks are the afore-mentioned fractional instantons. 
Qualitative and quantitative predictions are derived from this picture. In particular, in Section~\ref{subseq:Continuum_limit}, we specify the semiclassical predictions on the density of the gas both on the lattice and in the continuum.  

The next sections are devoted to presenting our results based on Monte-Carlo simulations of the SU(2) Yang-Mills theory, 
in order to test the predictions of the previous one. A partial collection of results was presented in a recent publication~\cite{Soler:2025vwc}. In Section~\ref{subseq:MC_Sim} we describe the procedure followed  in the generation and analysis of our numerical data. We explain in detail our  simulation method, the smoothing/filtering methods applied, and the identification of structures and fitting of parameters. In this study, we have mainly focused on SU(2) Yang-Mills theory, but some small tests and comments are included about the extension to SU(N).  Only some qualitative pictures considering general distributions of fluctuations and peaks are shown for the discussions of the methods in this Section.

Our main numerical results and analysis are presented in Section~\ref{seq:results}. The primary goal is an investigation for small torus sizes since it provides a clean and controlled setup for the semiclassical approach.
It is shown how the numerical data matches with the predictions from the semiclassical analysis of an effective two dimensional gas of fractional instantons and anti-instantons,  measuring some of the unknown parameters. All observables are analysed, with particular attention being paid to the string tension and its relation with the density of the gas.

Although our main goal has been the study of the small torus sizes, for which the semiclassical predictions are supposed to hold better, in Section~\ref{seq:larger_torus} we show some results concerning larger torus sizes. Our analysis allows us to determine the necessary modifications of our techniques in order to investigate the approach towards the limit of large torus sizes, in which we recover  the full isotropic four-dimensional Yang-Mills theory.

In the last Section~\ref{sec:conclusions} we give a brief review of our results, and conclude with final discussions of the general picture and possible extensions of our work. We collect in appendices some interesting aspects that complement our work.

\section{The semiclassical regime on $T_2\times R^2$}
\label{seq:semiclassical}

\subsection{Generalities about the semiclassical approximation}
\label{subseq:generalities}
In this section we will focus on the semiclassical regime for which
analytic predictions can be obtained and tested against our numerical
results. This occurs when the size of the small torus is small enough
so that the effective coupling is also small. Perturbation theory
becomes a good approximation for short distance observables. However,
certain properties of the theory are zero to all orders of
perturbation theory and that is where semiclassical methods play a role.
That is a well-known technique which has been essential to understand the
properties of many field theories. Classical quantum field theory
text-books like~\cite{Zinn-Justin:2002ecy} dedicate a good deal of space to it. The method is a generalization of the well-known saddle point
approximation for integrals over the complex plane. Recently, the field has been gaining focus and stronger versions,
under the label
{\em resurgence}, have been studied in certain examples of quantum systems and field theories (see Refs.
~\cite{Dunne:2014bca, Dunne:2016nmc, Aniceto:2018bis} for a  review).

A simple description of the idea implies that the path-integral is split into a sum of terms labelled by extrema of the action (classical
configurations) $A_\mu=A_\mu^\mathrm{clas}$ which include small quantum
fluctuations around them $A_\mu=A_\mu^\mathrm{clas}+
A_\mu^\mathrm{quant}$:
\begin{align}
Z\sim \sum_{A_\mu^\mathrm{clas}} \int d A_\mu^\mathrm{quant}
e^{-\frac{1}{g^2} S(A_\mu^\mathrm{clas}+
A_\mu^\mathrm{quant})},
\end{align}
This is particularly relevant when computing observables for which the
splitting is not terribly sensitive to the quantum part provided it is
small:
\begin{align}
O(A_\mu^\mathrm{clas}+A_\mu^\mathrm{quant})=O(A_\mu^\mathrm{clas})+
\mathcal{O}(g^2), 
\end{align}
On the other hand the action can also be expanded for small quantum
fluctuations
\begin{align}
\frac{1}{g^2} S(A_\mu^\mathrm{clas}+
A_\mu^\mathrm{quant})= \frac{1}{g^2} S(A_\mu^\mathrm{clas})+
\frac{1}{g^2} S^{(2)}(A_\mu^\mathrm{clas}, A_\mu^\mathrm{quant}) +
\ldots
\end{align}
The leading correction $S^{(2)}$ is quadratic in the quantum fluctuations since
the classical part is an extremum of the action. This quadratic piece
can be dealt with by a rescaling the quantum fields
$A_\mu^\mathrm{quant}=g Q_\mu$ and performing a Gaussian  integration over
$Q_\mu$. The Jacobian of the transformation induces a factor $g$ for
each degree of freedom, and the Gaussian integration over $Q$ a factor which we write as 
\begin{align}
e^{-S_Q(A_\mu^\mathrm{clas})},
\end{align}
Altogether we have
\begin{align}
\label{eq:semiclas_sum}
Z\sim  \sum_{A_\mu^\mathrm{clas}} e^{-\Gamma(A_\mu^\mathrm{clas})}
g^\mathrm{nqdf} + \mathrm{higher\ orders\ in \ g},
\end{align}
where the free energy of each configuration is 
\begin{align}
\label{eq:free_energy}
\Gamma(A_\mu^\mathrm{clas})
= \frac{1}{g^2} S(A_\mu^\mathrm{clas}) + S_Q(A_\mu^\mathrm{clas})
\end{align}
The factor $g^\mathrm{nqdf}$ is irrelevant if the number of quantum
degrees of freedom ($\mathrm{nqdf}$) is the same for all classical
fields. However, there are certain directions for which the second
order term $S^{(2)}$ vanishes. These are the zero-modes, and in
practice imply a weight  $g^{-\mathrm{nzm}}$ in terms of the number of
zero-modes ($\mathrm{nzm}$) of each classical field. 
As a matter of fact, the $S^{(2)}$  is given by a quadratic form involving 
the Dirac operator in the adjoint representation. Hence,  the number of 
zero modes can be obtained  from  the index theorem. 

Our presentation has been quite schematic. For example, the fact that  there are 
infinitely many classical fields turns our symbolic sums into
integrals. For certain families  of classical solutions these integrals 
range  over collective coordinates labelling the ensemble. Other
important questions that have been ignored are gauge fixing, ghosts,
and the presence of ultraviolet divergences. 
We point the  reader  to  the beautiful paper by `t
Hooft~\cite{tHooft:1976snw} where all the technical issues are
addressed in computing the quantum fluctuations around the BPST
instanton solutions.

The semiclassical approach  also applies on the lattice. Even more so,
since in this case the path integral reduces to an integral  over real 
variables  and many of the divergences are absent. On a finite volume also
the number of degrees of freedom  becomes finite. 
However, even in this case unexpected difficulties might appear as
seen in papers as ~\cite{Gonzalez-Arroyo1983,Coste:1985mn} for which the space of
classical solutions (torons) becomes an orbifold and the collective coordinate
formalism breaks down. One of the advantages of using twisted boundary conditions 
is that it eliminates  some of these
difficulties\cite{Coste:1986cb,Luscher:1985wf}.

\subsection{Classical solutions}
\label{subseq:classical_solutions}
After this rather general presentation, let us apply it to the case at
hand. We have to look for  solutions of the euclidean classical 
equations of motion. It is in this context that the BPST instanton
solution was found~\cite{Belavin:1975fg}. However, contrary to what many people
think, this is not the only basic solution  and not even the most
relevant for our study. We are looking at solutions on the torus of
the Yang-Mills euclidean equations of motion. 
As mentioned earlier, `t Hooft put forward that the
topology is not only given by the instanton number $Q$, but also by 
certain $Z_{\Nc}$ fluxes traversing the planes of the torus. If these
fluxes are present, we say that we have twisted boundary conditions.
In our study we are considering a small 2-torus with non-zero flux ($-1$). Although we are considering the range of the remaining 2 directions as infinite, we might as well consider that they are compactified in a very large torus. This is a practical option which is typically adopted in lattice gauge theory studies. The physics of the theory is not affected by the choice of boundary conditions in this big 2-torus. However, it affects the classical solutions and this can be used as a tool as we will see.  
If we put also a non-zero flux in this large  torus,
then the topological charge becomes fractional $Q=n+1/2$. Bogomolny
bound implies that the minimal action configuration would be
non-trivial and self-dual, we call such a configuration a 
fractional instanton. This problem was studied in
Ref.~\cite{GonzalezArroyo:1998ez}, where it was found that the basic $Q=1/2$ solution has an action density which peaks at a certain point (its center) and tends to zero action density as we move away from its center in the large torus. Furthermore, if one scales all coordinates by the torus size $l_s$ the solution converges to a fixed limit as the size of the big torus goes to infinity. It is  
this limiting solution what we call a $T_2 \times R^2$ fractional
instanton.  Because of its properties it was also referred as
a vortex-like fractional instanton. The properties of the solution
(as described in Ref.~\cite{GonzalezArroyo:1998ez}) are the  following:
\begin{enumerate}
\item To the level of precision of the action minimization method (at
the permille level) the configuration is self-dual.
\item The action density and all gauge invariant observables 
are functions of the coordinates $x_\mu/l_s$,
 where $l_s$ is the size of the small torus.
\item The action density has a peak at a given location (its center) and has circular invariance under rotations in the big plane around this point.
\item Furthermore, the action density decreases  as one 
moves away from this point in the radial direction $r/l_s$, and in an exponential manner for large enough distances.
\item The value of the Wilson loop in the large plane and centered at
the instanton position tends to $-1$ as the size of the loop grows.
This implies that the instanton behaves like a center-vortex
perpendicular to the big plane.
\item From self-duality and the Bogomolny bound~\cite{Bogomolny:1975de,Prasad:1975kr} one concludes that the action is half that of
an instanton ($4 \pi^2/g^2$). 
\item The solution is certainly non-unique 
because its center breaks translational invariance. However, these are the only continuous degrees of freedom. This follows from the index theorem implying that the space of solutions depends on only 4 parameters ($4Q\Nc$ for our $\Nc=2$ $Q=1/2$ case).
\item Computing the Polyakov loops in the two small torus directions  at the center of the solution we find $P_1=\pm 1$ and $P_2=\pm 1$. $(Z(2))^2$ center symmetry maps the 4 different solutions onto each other.
\end{enumerate}

Given the previous properties, one can compute the weight associated with every fractional instanton (See Eqs.~\eqref{eq:semiclas_sum}\eqref{eq:free_energy}) . 
Given the number of zero modes and the value of the action we conclude that  when
going from the instanton to the fractional instanton contribution one gets
\begin{align}
 \frac{1}{g^8} e^{-8\pi^2/g^2-S_{QI}} \longrightarrow \frac{1}{g^4}
 e^{-4\pi^2/g^2-S_{QF}}\, .
\end{align}
The quantum contribution to the free energy for the instanton $S_{QI}$  has been computed by `t Hooft~\cite{tHooft:1976snw}. Unfortunately,  the corresponding quantity   for the fractional instanton 
($S_{QF}$) is not known, but our work provides a numerical estimate of its value.

Obtaining a precise numerical approximation to the $Q=1/2$ fractional instanton 
can be done as in Ref.~\cite{GonzalezArroyo:1998ez}.
A more recent and complete analysis of the solution will be presented elsewhere~\cite{AGAtoappear}, but for the purpose of this paper we have used various  numerical approximations using $L_t^2\times \Ls^2$ lattices  of 
various sizes.
The lattice size in the short torus directions acts as a lattice spacing error $a=1/\Ls$. Typical
errors in the approximation are given in powers of $a$. The size of the errors depends on the lattice quantity that is being used.
For example, the value of the Wilson action $S_W$ can be measured after minimizing its value, we get 
\begin{align}
  \frac{4}{\beta}  S_W(\Ls)= 4 \pi^2 (0.9999(7) - 0.44(5)/\Ls^2) .
\end{align}
Notice that on the lattice this is the quantity that enters the semiclassical weight factor.
Computing now the  clover defined lattice action $S_C$, we see that  it can be fitted to the 
formula:
\begin{align}
g^2 S_C(\Ls)=4 \pi^2 (1.00027(21)-2.417(22)/\Ls^2) \; .
\end{align}
The lattice approximation to the total topological charge $Q_L$ gives
\begin{align}\label{ex:maxqsemiclass}
Q_L= \frac{1}{2} (1.00006(3)-2.441(3)/\Ls^2) \; .
\end{align}
We see that both formulas extrapolate to the expected self-dual values
with less than permille errors. No significant dependence is seen on the
large torus  lattice size. 
The maximum topological charge density for a single fractional instanton is obtained as\footnote{When comparing to the lattice results, the value has to be divided by $\Ls^4$.}
\begin{align}\label{eq:q0fit}
Q_{\text{max}}=2.274(7)-7.0(8)/\Ls^2 \; .
\end{align}

\begin{figure}
\begin{subfigure}[t]{0.5\textwidth}
\includegraphics[height=6cm]{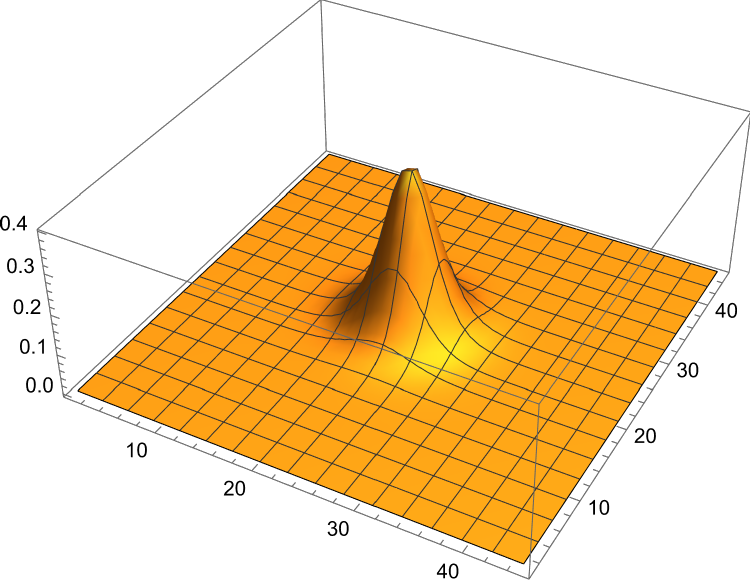} 
\caption{\label{fig:SingleFractionalFitA}}
\end{subfigure}
\quad\quad
\begin{subfigure}[t]{0.5\textwidth}
\includegraphics[height=6cm]{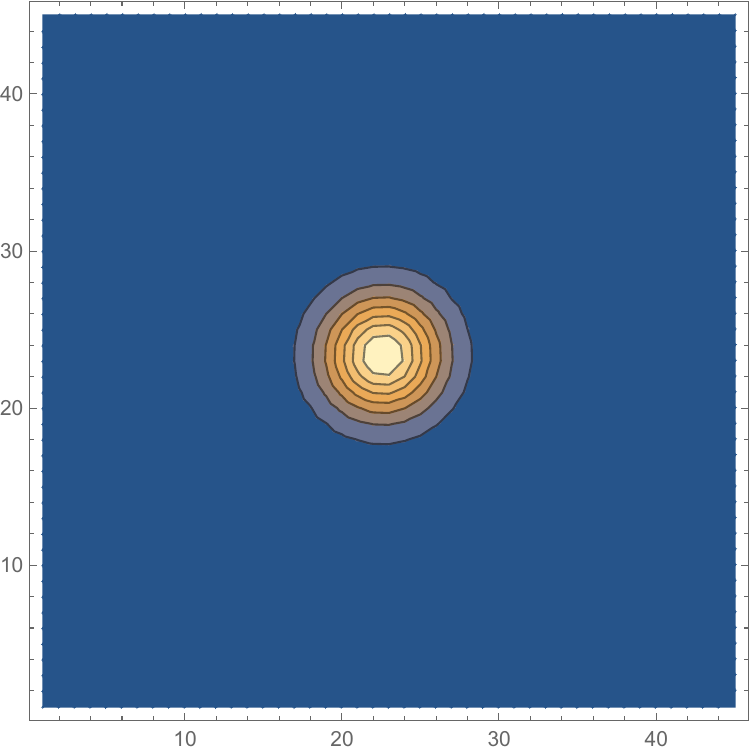}
\caption{\label{fig:SingleFractionalFitB}}
\end{subfigure}
\begin{subfigure}[t]{0.5\textwidth}
\includegraphics[height=6cm]{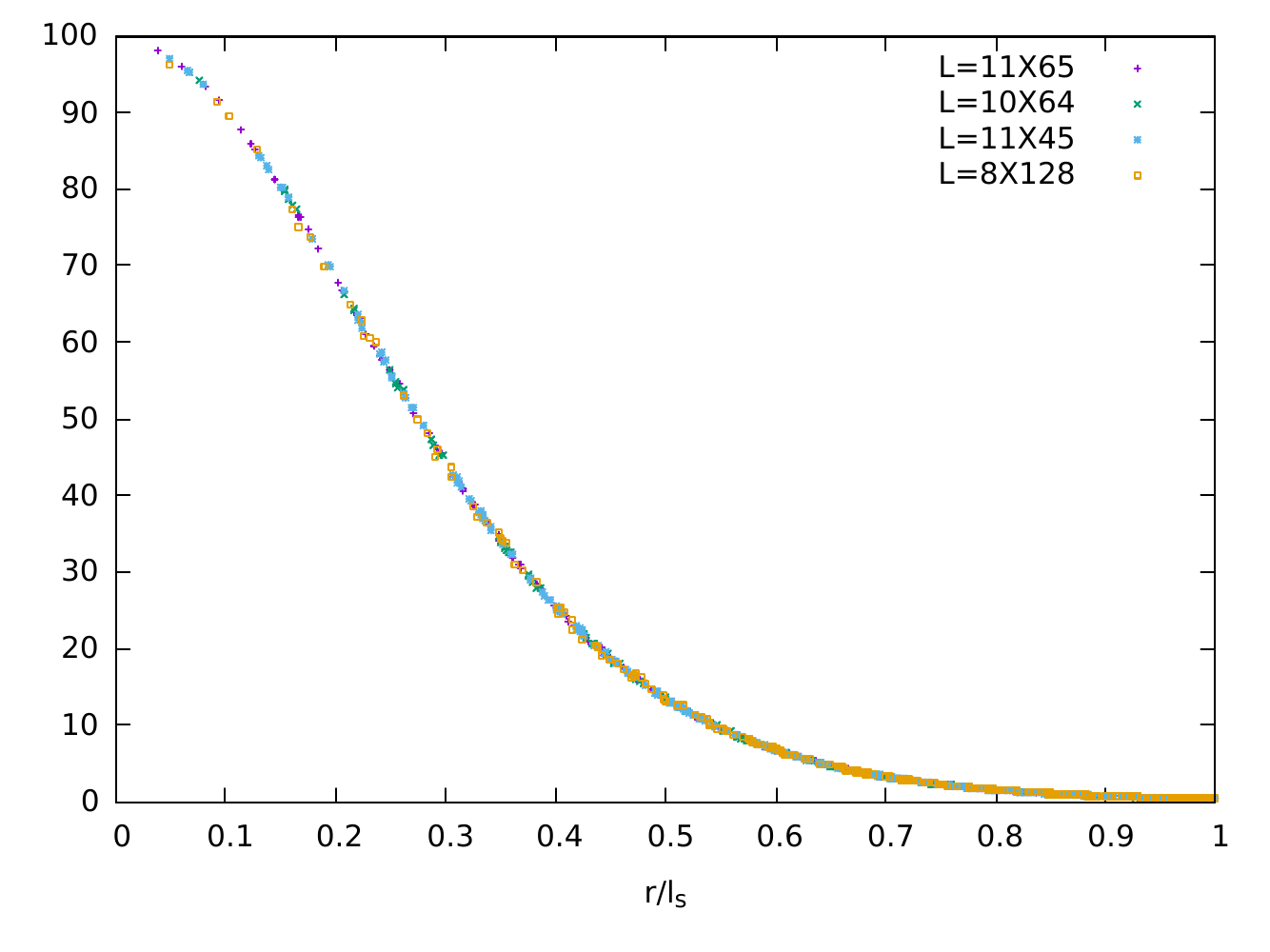}
\caption{\label{fig:SingleFractionalFitC}}
\end{subfigure}
\begin{subfigure}[t]{0.5\textwidth}
\includegraphics[height=6cm]{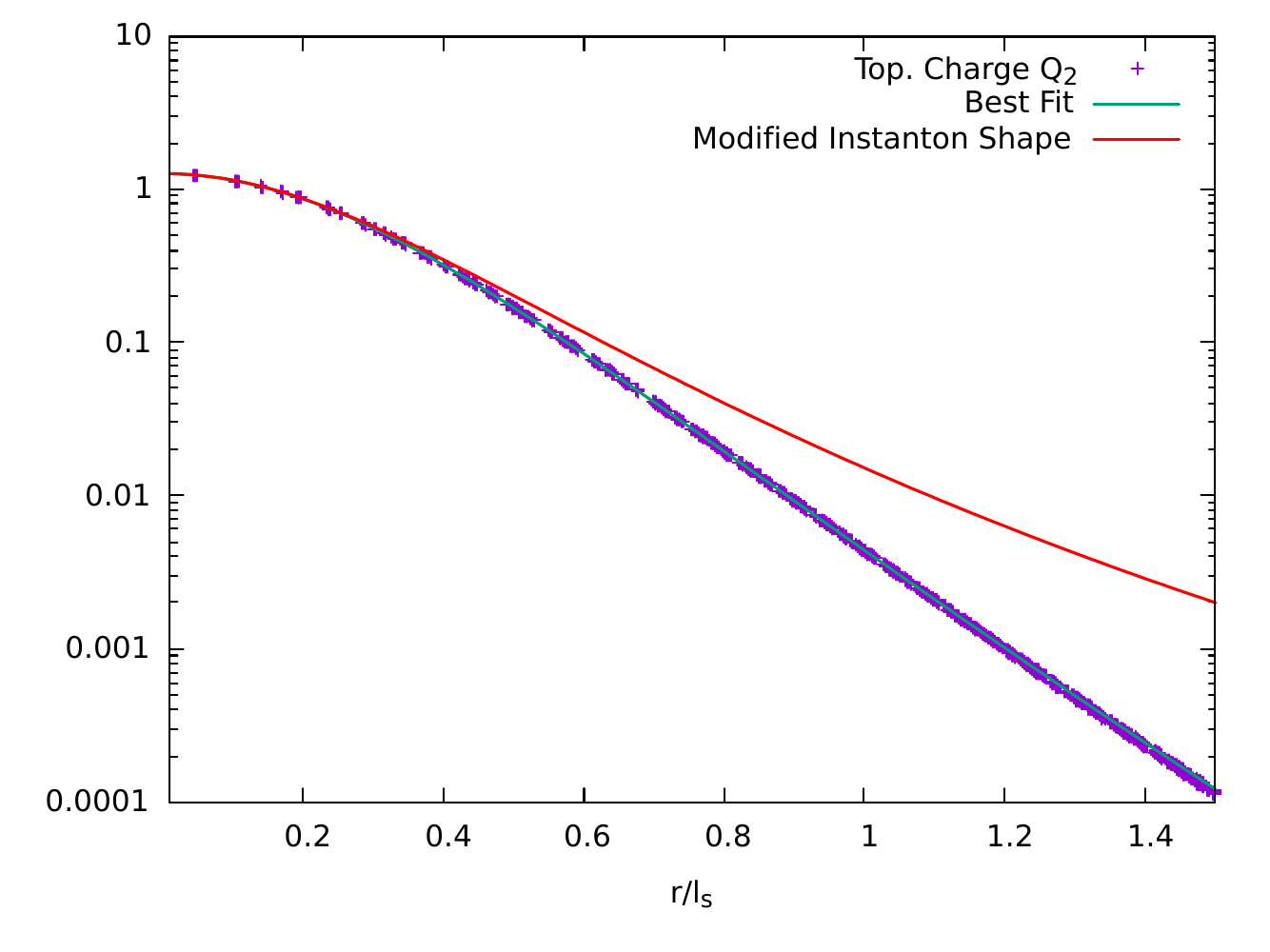}
\caption{\label{fig:SingleFractionalFitD}}
\end{subfigure}
\caption{This figure shows several projections of the SU(2)  vortex-like fractional instanton. a) Integrated  energy density profile $S_2(n_1,n_2)$ in lattice units for a  $11^2\times 45^2$ approximation; b) the corresponding contour plot; c) The integrated profile $S_2(r/l_s)$ as a function of distance to center $r$ for various numerical approximations; d) Topological charge density in logarithmic scale compared with fitting functions (see text). }\label{fig:SingleFractionalFit}
\end{figure}

For the purposes of this paper we will now focus on the 2-dimensional action density obtained after integrating over the two short directions $S_2(x,y)$. On the lattice one measures the sum of the corresponding plaquettes which is given by $S_2(x,y)/\Ls^2$.
This observable will be useful later to
identify the fractional instantons present in the Monte Carlo configurations. In Fig.~\ref{fig:SingleFractionalFit} we show the distribution of this observable in
4 different projections. As an example we take the $11^2\times 45^2$ configuration. Fig.~\ref{fig:SingleFractionalFitA}  displays  $S_2(n_1,n_2)/11^2$  as a function of the lattice coordinates of the 
large plane for this configuration as a 3D plot. The corresponding contour plot is  given in Fig.~\ref{fig:SingleFractionalFitB}, in which  one notices  the  approximate circular symmetry of the solution around its central peak.  This symmetry and the universality of the solution is   more precisely appreciated  in Fig.~\ref{fig:SingleFractionalFitC}. This shows the radial profile of the
observable $S_2(r\cos(\theta), r\sin(\theta))$ obtained from different lattice approximations. Any lattice
correction as well as possible violations of circular symmetry (i.e. $\theta$ dependence) show up
as small fluctuations around a universal curve. In logarithmic scale
one easily sees that for large sizes the curve decays exponentially 
(Fig.~\ref{fig:SingleFractionalFitD}).

An interesting observable is the value $S_2$ at the maximum.\footnote{We use in this case a simple quadratic interpolation of the form $A-B_1(x-x0)^2 -B_2 (y-y0)^2$ to get the approximate peak height and position.} Our
lattice approximations  can be fitted to the formula
\begin{align}\label{eq:S2fit}
S_{2,\text{max}}=100.27(20) -283(20)/\Ls^2\; .
\end{align}
The maximum of the integrated topological charge density follows just from self-duality ($1.270(3))$.
Notice that the extrapolated peak value is not far from $32 \pi$,
which is the same value obtained for a 2D projection of an instanton 
of size $\rho=l_s/\sqrt{2}$. However, the shape of the function does not agree 
with that of an instanton since it decays exponentially with $r$ and the
total integral is $4 \pi^2$ instead of $8 \pi^2$. Even in the neighbourhood of the center, the shape of the topological charge density differs from the corresponding one of  BPST instantons. This is interesting since it can be used to distinguish fractional instantons. 
A simple function that fits the topological charge density 2D profiles $q(x)$ for $r\le 0.2$ is given by  $q(x)=q_{max}(1+r^2/\rho^2)^{-3}$. Extrapolating to the continuum limit our numerical results
we obtain $q_{max}=1.276(5)$ and $\rho^2=0.297(10)$. Notice that the combination $Q_I=\pi \rho^2 q_{max}=0.595(5)$,  which differs considerably from the value $1$ valid for BPST instantons. 

The previous fit only matches the profile for $r\le 0.2$.
This is particularly clear when comparing the logarithm of the topological charge density with the modified instanton shape $q(x)$ displayed in Fig.~\ref{fig:SingleFractionalFitD}.
A good fit to the action density shape covering the full range can be  obtained by approximating the logarithm of
the function by a rational function of the form:
\begin{align}
\label{eq:S2shape}
\log(S_2(r)/S_2(0))=-\frac{B r^2+F\cdot D r^3}{1.0+C r+Dr^2}
\end{align}
For our $11^2\times 45^2$ approximation  the best fit parameters are $B=9.73$, $C=0.63$, $D=1.30$ and
$F=5.2$. The comparison is shown in Fig.~\ref{fig:SingleFractionalFitD}, where the integrated topological density is plotted together with the best global fit described above. We see that the two functions describe well the data for $r/l_s\le 0.2$ but differ considerably at large distances, given the strong
exponential fall-off (obvious given the large value of $F$).  As a matter of fact, using the same function for the full range introduces distortions in the parameter values. A better way to describe the function is to use different fits in different regions. Concerning the exponential tail we found that for $r>1$ the $S_2(r)$ can be described at the 5\% level by $170(8)\ e^{-2 \pi r}/r$.

\subsection{Description  of the semiclassical components}
\label{subseq:other_solutions}
Our results so far have considered only isolated fractional instanton (FI) solutions corresponding to the minimum action with twisted boundary condition in the big and small torus. The solution can be translated in space, which implies an entropy factor that scales with the area of the large torus. This overcomes the price, in terms of probability, of adding more  FIs. Thus, one has to consider also multi-fractional instantons in our semiclassical description. Consequently, as expected from thermodynamic principles, the total action diverges in the infinite volume limit. However, as we will see in the next section, the probability per unit area of containing one FI  is small when the small torus size $l_s$ is small. On the lattice this probability depends both on  $\Ls$ and $\beta$. At these small densities the separation is much larger than the size of the objects and the most probable configurations amount to a dilute gas of FI. 
In this situation, the total free energy for one of these configurations becomes very approximately the sum of the free energies of each FI. Indeed, even at closer distances the action of such a  multi-FI configurations is exactly additive as it follows from the index theorem and the Bogomolny bound.

However, the same arguments given above imply that one should also include  fractional anti-instantons (AFI). At 
these large separations the interaction energies are very small, and the free energy of a typical configuration having both FI and AFI would be very approximately the sum of all its components.
Notice, however, that  these configurations containing both FI and AFI will no longer be classical solutions. There is no topological protection,  and hence it is expected that a FI-AFI pair can be continuously connected to the configuration with both absent, having smaller action. However, the directions of instability are probably very few and the slopes are small for large separations, so they are  amply compensated by quantum fluctuations. These expectations are verified by the study of  the behaviour under  cooling and gradient flow (see for example in a previous paper~\cite{Bergner:2024njc}):  the FI-AFI pair is eventually annihilated by the flow, but this happens at large  times (gradient flow time, steps of cooling) when the  separation  is large compared to the size of the  objects. 

In summary, in the  semiclassical description, the ''classical'' configurations are given by a gas composed of both FI and AFI at random locations. At low densities, expected for  sufficiently small $l_s$, the gas is dilute and the free energy is the sum of that of its constituents. We recall that the objects occupy the full small 2-torus, and most of its properties depend very slightly upon the location of the center within it. 
Thus, effectively one can say that one is dealing with a two-dimensional gas.

\subsection{Properties of the 2D gas of FI and AFI}
\label{subseq:Properties_gas}

In the dilute situation, the probability  of having one  object per unit
area  is a small number,  which should be proportional to the
path-integral  weight of such an object. The same value applies to FI
and to AFI. This together with the randomness of 2D location tells us
that the distribution of the gas is Poissonian. To do calculations it
is sometimes  better to take the distribution to be multinomial, which
converges to Poisson in the dilute situation. 
The multinomial can be thought as a way to  take into account the finite size of the objects. 
We will explain this below.

Let us compute the probability for a given configuration having $n_+$
FI and $n_-$ AFI as follows. Let us take a portion
of the plane of area $A$ and  divide it into a large number $M$
of small portions of the same size  with area $A_0$ (thus $A=M A_0$). The
portions are large enough to contain a FI fully but small enough so
that the probability of having two objects in each portion is
negligible. In each portion of size $A_0$ the probability of having
one FI is $p_0$, the probability of having one AFI is also $p_0$ and
the probability of having nothing is $1-2p_0$. Notice that we can write 
$p_0=\bar{p}A_0$, where $\bar{p}$ is the probability per unit area. 
Then, the probability
that the big area contains $n_+$ FI and $n_-$ AFI is given by the
multinomial distribution:
\begin{align}
P(n_+,n_-)= \frac{M!}{n_+! n_-! (M-n_+-n_-)!} p_0^{(n_++n_-)}
(1-2p_0)^{M-n_+-n_-} .
\end{align}
Notice that the sum over all possible values of $n_\pm$ is 1. The Poisson distribution is obtained when taking the limit $M$ large, $p_0$
small with the product $p_0 M$ fixed:
\begin{align}
P(n_+,n_-) \longrightarrow P'(n_+,n_-)= \frac{(p_0M)^{n_++n_-}}{n_+! n_-!}e^{-2p_0M} .
\end{align}

From these  distributions one can easily  calculate all the moments. 
For that purpose we can replace the
probabilities by variables $x_\pm$ for one FI/AFI and $1-2x_0$ for no object. 
Summing up one gets a generating function
\begin{align}
Z(x_+, x_-,x_0) = (1-2x_0+x_++x_-)^M\; ,
\end{align}
from which all moments can be obtained by differentiation at $x_+=x_-=x_0=p_0$. For example, the mean number of fractional instantons is given by
\begin{align}
\langle n_+\rangle = \langle n_-\rangle = p_0 M =\bar{p} A.
\end{align}
Notice that this mean number is precisely the quantity appearing in the Poisson distribution.
From the previous expression one sees that  the 2D density of objects becomes
\begin{align}
\rho_{2D}= \frac{1}{A} \langle (n_++n_-)\rangle= 2 \bar{p}.
\end{align}
If each object has topological charge $Q=\pm 1/2$ one can also 
compute the distribution of topological charge. The mean value is zero
but the dispersion (i.e. the 2D topological susceptibility) becomes
\begin{align}
\hat{\chi}_{2D}= \frac{1}{4A} \langle (n_+-n_-)^2\rangle = \frac{2 M p_0}{4 A}= 
\bar{p}/2,
\end{align}
which is equal to the density divided by 4.
Higher order moments can be easily computed. Later on we will give the full distribution.

\begin{figure}
    \centering
    {
\input{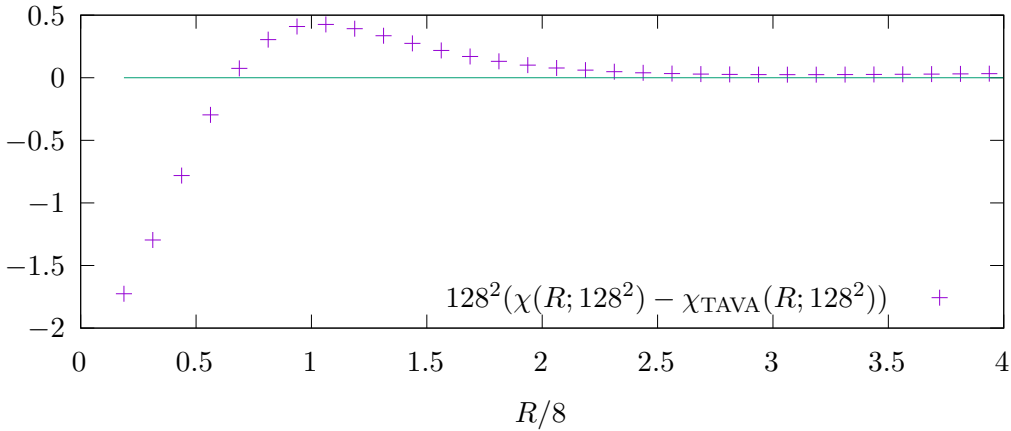}
}
    \caption{Square Creutz ratios $\chi(R)$ averaged over the plane obtained for a numerical approximation to the SU(2) fractional instanton on a $8^2\times 128^2$ lattice after subtracting the same quantity for the thin abelian vortex approximation (TAVA) computed on a torus of size $A=128^2$ and multiplying by the lattice area of the big torus  ($128^2$). } 
    \label{fig:CreutzCompare}
\end{figure}

Finally, we notice that  since the fractional instantons behave as
$Z_2$ vortices, one can evaluate the expectation value of a Wilson
loop $W$ covering the full area $A$ by simply summing up the contribution
of each object:
\begin{align}
\langle W \rangle = \langle (-1)^{(n_++n_-)}\rangle = (1-4p_0)^M  .
\end{align}
This shows that the expectation value for these large loops decays
with the area and the string tension is given by 
\begin{align}
\sigma= -\log(1-4p_0)/A_0\sim 4 p_0/A_0= 4 \bar{p} = 2 \rho_{2D}.
\end{align}
The last result is precisely the result obtained for the Poisson distribution, and can be labelled {\em thin abelian vortex approximation} (TAVA).
The non-abelian character of the FI and their finite size can correct this slightly, but this should be very approximately correct in the dilute
situation. Indeed, it is possible to test this with our numerical FI configurations.  
For practical purposes  it is useful to consider square Creutz  $\chi(R)$ ratios as a good way to estimate the string tension since they get rid of perimeter contributions. Given  $W(R,T)$  Wilson loops of 
$R\times T$ rectangular shape,  the ratios are defined as 
\begin{align}
    \chi(R)\equiv -\log\left(\frac{\langle W(R,R)\rangle \langle W(R-1,R-1)\rangle}{\langle W(R,R-1)\rangle\langle W(R-1,R)\rangle}\right)
\end{align}
Now suppose we have a big torus of area $A$ containing 
a single fractional instanton. Using our  previous analytic computation of Wilson loop expectation values, one can compute the corresponding square Creutz ratios $\chi(R;A)$ (averaged). This is just the exact value for the 2-dimensional $Z_2$
gauge theory. For large area $A$, the result tends to $2/A$ which is
equal to twice the density. However, there are sizeable differences
when $R^2$ becomes comparable with $A$. We compared this result with
the computation of square Creutz ratios for an $8^2\times 128^2$ lattice
containing a single fractional instanton. The area in this case is
$A=128^2$ and the density $2/A$. In Fig.~\ref{fig:CreutzCompare} we show  the 
difference between our measured Creutz ratios minus the result of the
thin abelian vortex approximation. Notice that the results agrees for 
large values of $R$. However, our vortex has a size of order $\Ls=8$
and Creutz ratios for $R/\Ls$ smaller than 1 show a strong size
dependence. Curiously at $R=\Ls$ the Creutz ratio divided by the
density peaks and becomes 2.4 instead of 2. It is important to keep
this in mind when interpreting our Monte Carlo data. 

In summary, as it is well known,  our FI-AFI vortex gas is expected to confine 
with  string tension which is approximately 2 times the density. Furthermore, With our formulas, higher order moments of the density or topological
charge distribution can be computed easily for both the multinomial and the Poisson distribution.

\subsection{Role of twist and of ordinary $Q=1$ instantons}
\label{subseq:twist_Q1}
In this section we will comment on certain slight modifications to
what we have just described. We argued, that to have a fractional
topological charge object, one must have a non-zero twist both in the
small torus as well as in the big torus. The twist in the small torus
is crucial in our study as long as its size is small in terms of the
inverse Lambda parameter of the theory. However, at large sizes,
boundary conditions should not matter and the results must coincide
with those that are obtained using periodic boundary conditions. We
have actually verified this by computing the string tension at
sufficiently large lattices with and without twist. The results agreed
within errors. 

The effect of putting a zero or non-zero twist in the big plane can be understood if one   realizes that twist is just a flux modulo $N=2$. Thus, even if we do not have twist in the large plane, one will still produce fractional instantons, but
always an even number of them. This certainly affects the distributions and
calculations that we did in the previous section. If we put a twist in
the big plane, the total number of objects $n_++n_-$ has to be odd, 
and if we don't, it has to be even. Most of our simulations are done in
this second case. Computing expectation values in the modified
distribution is fairly easy by simply excluding the odd numbers:
\begin{align}
 \langle O  \rangle_{\mathrm{mod}} = \frac{ \langle O
 (1+(-1)^{(n_++n_-)}) \rangle }{\langle
  (1+(-1)^{(n_++n_-)}) \rangle},
\end{align}
where the right-hand side contains expectation values with respect to
the unconstrained distribution of the previous  subsection. Thus, one gets for example 
\begin{align}
 \langle n_+  \rangle_{\mathrm{mod}}  = M p_0 \frac{ 1 -
 (1-4p_0)^{M-1}}{1+(1-4p_0)^M}= \langle n_+ \rangle \,  \frac{ (1 - e^{-4
 \langle n_+  \rangle})}{(1 + e^{-4
  \langle n_+  \rangle})}.
\end{align}
Clearly the correction is negligible when the mean number of objects is
not very small. 

For future purposes, let us evaluate explicitly the distribution of topological charge. That is equivalent to the distribution of the quantity $(n_+-n_-)/2=Q$. 
This distribution can be easily obtained for the constrained Poisson probability distribution using $n_+=n_-+2Q$ 
\begin{align}
P(Q)=\frac{1}{\cosh(2p)}\ \sum_{n_-}\frac{p^{(2Q+2n_-)}}{n_-! (2Q+n_-)!} = \frac{I_{2Q}(2p)}{\cosh(2p)},
\end{align}
where $I_n(x)$ is the modified Bessel function and $p=p_0 M= \bar{p} A$.

Another distribution which turns out to be useful in interpreting our data is the distribution of minimum distance from an object to the closest one. Given a configuration with area $A$ having $n_+$ fractional instantons and choosing one of them, the remaining $n_+-1$ have a probability $1-\pi d^2/A $ of being at a larger distance than $d$ from our selected FI. Then the probability that all of them are outside a circle of radius $d$ from our FI 
is  
\begin {align}
(1-\pi d^2/A)^{n_+-1}.
\end{align}
Convoluting this probability with the one of having $n_+$ fractionals we get 
\begin{align}
P(d)\propto   e^{-p}\sum_{n_+}\frac{(p(1-\pi d^2/A)^{n_+}}{n_+!}=e^{-\pi p d^2/A}.
\end{align}
where $p=\langle n_+ \rangle$.
This combines with a prefactor $d$ (coming from changing measure from area to length) to give a Maxwellian  type of curve. Our derivation assumes that the fractional instanton have zero width and can be put together with other at arbitrarily short distances. Thus, we expect that at long distances the shape is correct but might get corrections for small values of $d$. Notice that from the previous distribution one can deduce 
that the distribution has a peak at $d_\mathrm{peak}=1/\sqrt{\pi \rho_{2D}}=\sqrt{A}/\sqrt{2\pi \langle n_+ \rangle}$  and a mean value $\bar{d}=\sqrt{\pi/2}\, d_\mathrm{peak}$.

In relation to this, an aspect to be addressed is whether ordinary instantons and
anti-instantons of $Q=1$ are also present. The answer is certainly yes. However, the probability to create an instanton goes like the square
of that of a fractional instanton and if this probability is small one
expects many less instantons. Notice, however, that instantons do not
contribute to the string tension but contribute to the topological
susceptibility. To generate a non-zero string tension it is crucial
that the fractional instantons are independent. If all FI appear in
fixed size molecules made of pairs of them, the string tension would
also vanish.

\subsection{Scaling of the densities and the continuum limit}
\label{subseq:Continuum_limit}
As explained in the  previous section, the semiclassical approximation 
can be studied both in the continuum and on the lattice. In this
section we will see how the two are connected. Let us consider 4d
lattice of size $\Ls^2\times \Lr^2$ where $\Lr\gg \Ls$ and SU(2)
lattice Yang-Mills theory is generated with  Wilson action and coupling $\beta$. Now consider  
a dimensionless quantity which we call diluteness $D$ as follows:
\begin{align}
\label{dilutenessdef}
D= \Ls^2 \rho^L_{2D} = \frac{\Ls^2}{\Lr^2}\langle (n_++n_-) \rangle ,
\end{align}
where $\rho^L_{2D}$ is the 2D density computed by dividing the number
of objects by the area of the large torus. The big area is large, so
the diluteness can only depend on $\Ls$ and $\beta$. The
diluteness must be proportional to the probability to create a
fractional instanton per unit area so according to our discussion in Sec.~\ref{subseq:classical_solutions} we expect 
\begin{align}
D= A(\Ls) \beta^2 e^{-\pi^2 \beta}.
\end{align}
An equivalent calculation in the continuum gives a diluteness that depends on 
$l_s$ and the coupling $g$. But as it is well-known, the coupling has to
be defined in a scheme and that introduces a dependence on an energy
scale $\mu$. Because of dimensional reasons, the diluteness can only
depend on $\mu l_s$ and $g(\mu)$. However, $\mu$ is an arbitrary scale 
and the dependence on $\mu$ and $g(\mu)$ can be eliminated in terms of
a new scale: the Lambda parameter $\Lambda$. Thus, at the end of the
day, the diluteness would be a function of $l_s \Lambda$, which is the
only dimensionless combination. A similar
thing happens on the lattice where the scale is the lattice spacing
$a$ and combined with the lattice coupling $\beta$ they should be
traded by the Lambda parameter. This can be done using the beta
function of the theory. We can write 
\begin{align}
a\Lambda_L = \exp\{-\frac{3}{11}\pi^2\beta\}\ \left( \frac{6 \pi^2}{11}
\beta \right)^{51/121}
\end{align}
Using this formula we can rewrite the expression of the diluteness as
follows
\begin{align}
D=A(l_s/a) (a \Lambda_L)^{11/3} \left(\frac{11}{6 \pi^2}\right)^{11/17} \beta^{5/11}
\end{align}
In the continuum limit this should only depend on $l_s \Lambda_L$ and
this implies that, up to logarithmic corrections, the prefactor behaves 
as
\begin{align}
A(\Ls) \sim A_0 \Ls^{11/3}
\end{align}
This implies that in the continuum the diluteness goes as
$(l_s \Lambda)^{11/3}$ and the 2D density as $(l_s \Lambda)^{5/3}$.
This is the result that one obtains in a continuum calculation by
replacing $g(\mu)$ in the naive formula by $g(l_s)$.

\section{Ensemble generation and numerical analysis}
\label{seq:Data_generation}
The ultimate aim of this work is to identify the relevance  of different semiclassical contributions in the Yang-Mills vacuum and their role in inducing   the fundamental mechanisms of the theory,  like confinement.  This can be achieved by the analysis of the  configurations generated by Monte Carlo simulations. The semiclassical approximation is expected  to work  well for small values of $l_s$. As explained in previous sections, in this regime the system can be described 
as a  two-dimensional dilute gas of fractional instantons. Our first 
goal will be to verify the theoretical predictions by the analysis of our configurations and develop the most appropriate methods. We will then monitor the evolution of the system as we gradually increase the size of the two torus. Extending the analysis, we aim at  investigating the evolution of the topological structures as we  move from the semiclassical regime towards the infinite volume limit. This means we want to study the Yang-Mills vacuum as a function of $l_s$ and identify semiclassical contributions for a large set of ensembles. In our simulations we consider mainly \SU{2} Yang-Mills theory; only some outlook on \SU{3} and \SU{4} is presented in the Appendix. Some details of the simulation methods are explained in Sec.~\ref{subseq:MC_Sim}.

Identification of semiclassical structures has been applied in several previous studies. Most  of them  were focused on  instantons, but also  related to fractional instantons as they appear in a $T_3\times R$ setting. 
The procedure  consists of three steps: the first is a filtering of the ultraviolet noise from the configurations, as explained in Sec.~\ref{subseq:MC_Filtering}. The second step is the fitting of structures, see Sec.~\ref{subseq:identification}. In a third step, as detailed in Sec.~\ref{subseq:identifiation_algorithm}, the fit parameters are related to certain semiclassical objects like instantons based in the behavior of the distributions discussed in Sec.~\ref{subseq:Distributions}.

\subsection{Monte Carlo simulations}
\label{subseq:MC_Sim}
The Monte Carlo simulations are performed on a lattice of size $\Ls^2\times \Lr^2$ with $\Lr\gg \Ls$. Periodic boundary conditions are applied in the large $R^2$ directions (each with $\Lr$ lattice points), while twisted boundary conditions are applied in the short $T_2$ directions, having $\Ls$ lattice points. The physical size of the $T_2$ directions is $l_s=a(\beta) \Ls$, where $a(\beta)$ is the lattice spacing at a given coupling. All ensembles presented in this work are generated with a plain Wilson gauge action. The scaling of the lattice spacing has already been determined in several previous studies.\footnote{We have chosen \cite{Allton:2008ty} as a reference. The values are usually expressed in units of the inverse of the square root of the
string tension and converted using a reference scale of $\sigma=5\  \mathrm{fm}^{-2}$ for the string tension.} The full list of lattice sizes, $\beta$ values and the corresponding physical sizes $l_s$ are given in Tab.~\ref{tab:confs}  and \ref{tab:confsfull}. Details about the implementation of twisted boundary conditions have already been explained in several publications \cite{Groeneveld:1980zx,Groeneveld:1980tt,GonzalezArroyo:1982hz}. We have used a standard hybrid Monte-Carlo and a heat-bath algorithm to generate the configurations.

We have carefully monitored fluctuations of the topological charge in our simulations. This is essential to ensure thermalization and proper sampling. We have tested that both hot (random) and cold (zero-action) initial boundary conditions lead to compatible results and measured the autocorrelation time of the topological charge. This limits our simulations to the parameter range of $\beta\leq2.7$ since substantial topological freezing is observed at larger $\beta$.

\begin{table}
    \centering
    \begin{tabular}{|c|c|c|c|}
        \hline
        $\beta$ & $a$ (fermi) & $t_{gf}$\\\hline
        2.4 & 0.11630& 3.52\\
        2.45 & 0.985 & 5.34 \\
        2.5 & 0.08194 & 7.34 \\
        2.55 & 0.07089 &  10.87\\
        2.6 & 0.05938 & 15 \\
        2.65 & 0.05072 & 20.39 \\
        2.7 & 0.04337 & 27.37 \\
        2.75 & 0.03745 & 36.37 \\
        2.8 & 0.03238 & 47.84\\
        \hline
    \end{tabular}
    \begin{tabular}{ |c|c|c| c | } 
        \hline
        \Ls & \Lr & $\tau=\sqrt{8t_{gf}}\;a$ & twist \\
        \hline
        4-14 & 64, 104 & 0.650 fermi & (2,3), $\vec{k}=0$, $\vec{m}=(1,0,0)$  \\
        \hline
    \end{tabular} 
    \caption{LEFT: List of lattice couplings used, together with the corresponding lattice spacing values and lattice gradient flow times applied (further details in the Appendix Tab.\ref{tab:confsfull}) RIGHT: List of Lattice sizes simulated, twist employed and approximate smearing radius in physical units. (Fermi unit$\equiv \sqrt{5}/\sqrt{\sigma}$) }
    \label{tab:confs}
\end{table}
\begin{figure}
    \centering
    \includegraphics[width=0.8\linewidth]{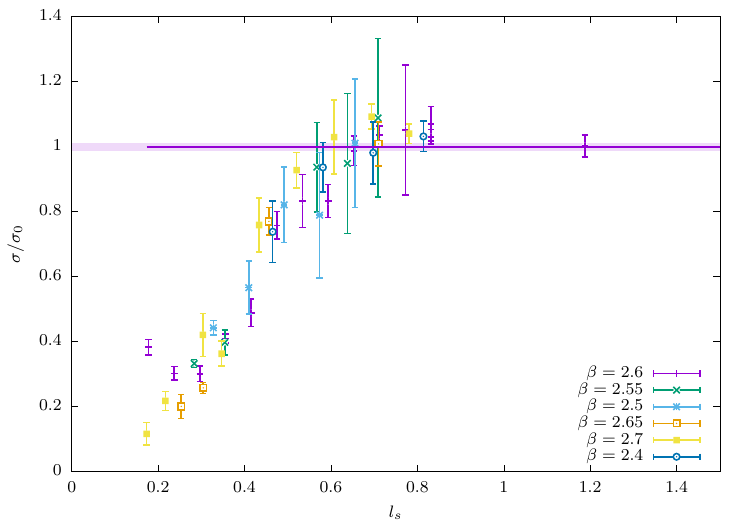}
    \caption{String tension  $\sigma$ as a function of $l_s$ relative to the largest volume result  $\sigma_0$ in \SU{2} Yang-Mills theory. The data are obtained from a two-step fit. In a first step the static quark anti-quark potential is obtained from the Wilson loops in $R^2$ plane. The second step is a fit of the potential to $V(r)=A/r+B+\sigma r$. This result is obtained with APE smearing in directions orthogonal to $R^2$, but without gradient flow. }
    \label{fig:fit_stringtension}
\end{figure}

The generated ensembles allow for a direct measurement of observables such as the expectation values of Wilson loops, Polyakov loops, topological charge, etc.
In Fig.~\ref{fig:fit_stringtension} we show the string tension to illustrate the basic effect of the finite size $l_s$. As expected, the string tension becomes smaller towards the semiclassical regime. However, the interpolation to infinite volume seems smooth and continuous.
Our main analysis requires in addition a filtering of short-distance (ultraviolet) noise from the gauge configurations in order to reveal semiclassical structures. We have used gradient flow for this purpose. The details of the filtering procedure are discussed in the next section.

\subsection{Gauge noise filtering}
\label{subseq:MC_Filtering}
It is well known that the lattice discretization of Yang-Mills theory does not provide a well-defined notion of topology in the presence of short-range fluctuations introducing lattice artefacts. Smoothing methods are required to remove these artefacts. Methods like cooling, gradient flow, or smearing correspond to a modification of the gauge configurations, while other methods like the adjoint zero-mode method~\cite{Gonzalez-Arroyo:2005fzm,GarciaPerez:2009mg,Bergner:2024njc} rely on modified observables that coincide with the standard ones in smooth backgrounds and decrease the effect of high-frequency noise. However, the methods based on zero modes are computationally more demanding. In this work, we have used the gradient flow~\cite{Luscher:2010iy} as a filtering method since we want to scan a large parameter range efficiently. It is an evolution of the gauge fields in a fictitious time $t_{gf}$ towards saddle points of a gauge action. The effects are comparable to cooling \cite{Alexandrou:2015yba}.  We have applied standard integration techniques explained in these references. The choice of the lattice gauge action in the gradient flow, which is independent of the gauge action used in the simulations, leads to different flow types. The leading order contribution of these flow types is the same, but subleading parts introduce relevant differences distorting some topological structures. The gradient flow evolution leads to smearing over a length-scale $\tau=\sqrt{8t_{gf}}$, which can be called the smearing radius~\cite{Narayanan:2006rf}.  

The gradient flow has several effects on the gauge ensemble and introduces an additional dependence on the parameter $\tau$. Different approaches can be used to resolve this dependence on the flow time. One can fix the flow time in lattice units 
and extrapolate to the continuum limit. This would correspond to a modification of the lattice discretization at the ultraviolet scale, which should leave the universal continuum limit unaffected. Alternatively one could try to identify an intermediate range of $\tau$, where lattice artefacts and ultraviolet noise are already removed, but the leading order in the flow time expansion around $\tau=0$ can still be determined. This is in particular useful for ultraviolet finite observables, which can be extrapolated to zero flow time. The third possibility is to keep $\tau$ fixed in physical units and extrapolate the continuum limit. The dependence on the reference scale introduced in this way, can be monitored afterwards independent of the lattice regularization. For finite quantities, the scale can be removed in a second extrapolation step afterwards.

In our present work we consider first a fixed $\tau$ in physical units in order to provide a clear picture of the topological structures and the semiclassical limit. Afterwards, we estimate effects of the flow in an investigation of the dependence on $\tau$. The relation between the density of topological structures and observables should hold for any flow time, while the density itself is slightly changed during the flow.

We also present and discuss some main effects of the gradient flow in the following sections. Some of these  effects can be summarized as follows:
\begin{enumerate}
    \item The fractional instantons in the semiclassical regime are very stable under the flow, even up to very large flow times.
    \item A pair of fractional instanton and fractional anti-instanton with a sufficiently small distance between the two objects is annihilated during the gradient flow. This can be considered  a physical effect related to the smoothing procedure and hence appears for any type of gradient flow. Similar effects occur for a pair of (BPST) instanton and anti-instanton.
    \item Depending on the type of flow, instantons get deformed. 
     This effect has been discussed in the context of cooling in~\cite{Teper:1985rb}, later also related to gradient flow in~\cite{Bornyakov:2015xao}, with some recent verification in~\cite{Tanizaki:2024zsu}. 
    This phenomenon was  studied in detail and explained in  Ref.~\cite{GarciaPerez:1993lic}. 
    In the absence of twist this is due to the fact that self-dual solutions of $Q=1$ on the torus do not exist~\cite{Braam:1988qk}. The action decreases with the torus size and only saturates the Bogomolny limit in the singular case of $\rho=0$. On top of this, lattice artefacts also induce  a dependence of the lattice action on the instanton size. Thus, Wilson flow also produces a decrease of the instanton size until it disappears. Fortunately, changing the lattice action the effect can be reversed  and the instanton size can be made to increase even compensating the continuum phenomenon mentioned before. For this reason, this is called "overimproved" flow.   
\end{enumerate}
Typically, lattice artefact corrections are particularly relevant along the directions of the moduli space of the continuum solutions. However, for the 
$Q=1/\Nc$ fractional instantons the moduli coincides with translations, a discrete version of which is also a symmetry on the lattice. This explains their stability under  gradient flow.\footnote{The stability of such an object is obvious with twist in both, $T_2$ and $R^2$, planes since the topological charge can not decrease below $Q=1/\Nc$, but it also holds in general.}

\subsection{Fitting topological structures}
\label{subseq:identification}

\begin{figure}
    \begin{subfigure}[t]{0.5\textwidth}
    \includegraphics[width=\textwidth]{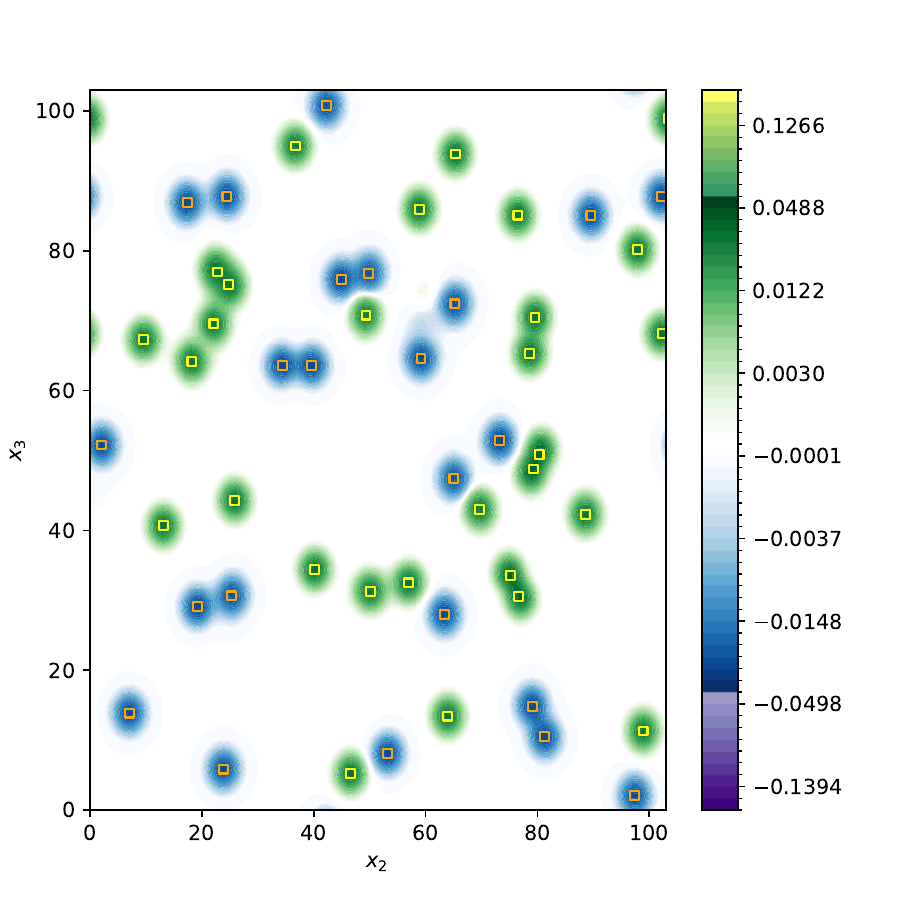
    }
    \caption{\label{fig:examplens6}$\Ls=6$}
    \end{subfigure}
        \begin{subfigure}[t]{0.5\textwidth}
    \includegraphics[width=\textwidth]{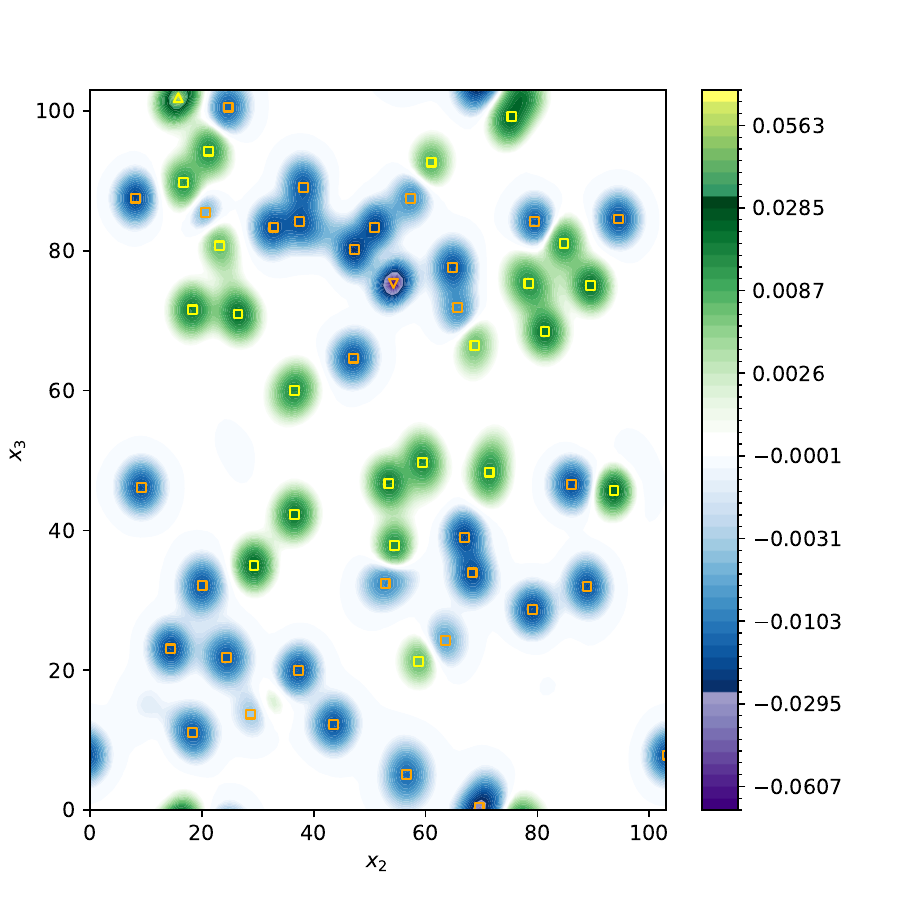}
    \caption{\label{fig:examplens8}$\Ls=8$}
    \end{subfigure}
    \begin{subfigure}[t]{0.5\textwidth}
        \includegraphics[width=\textwidth]{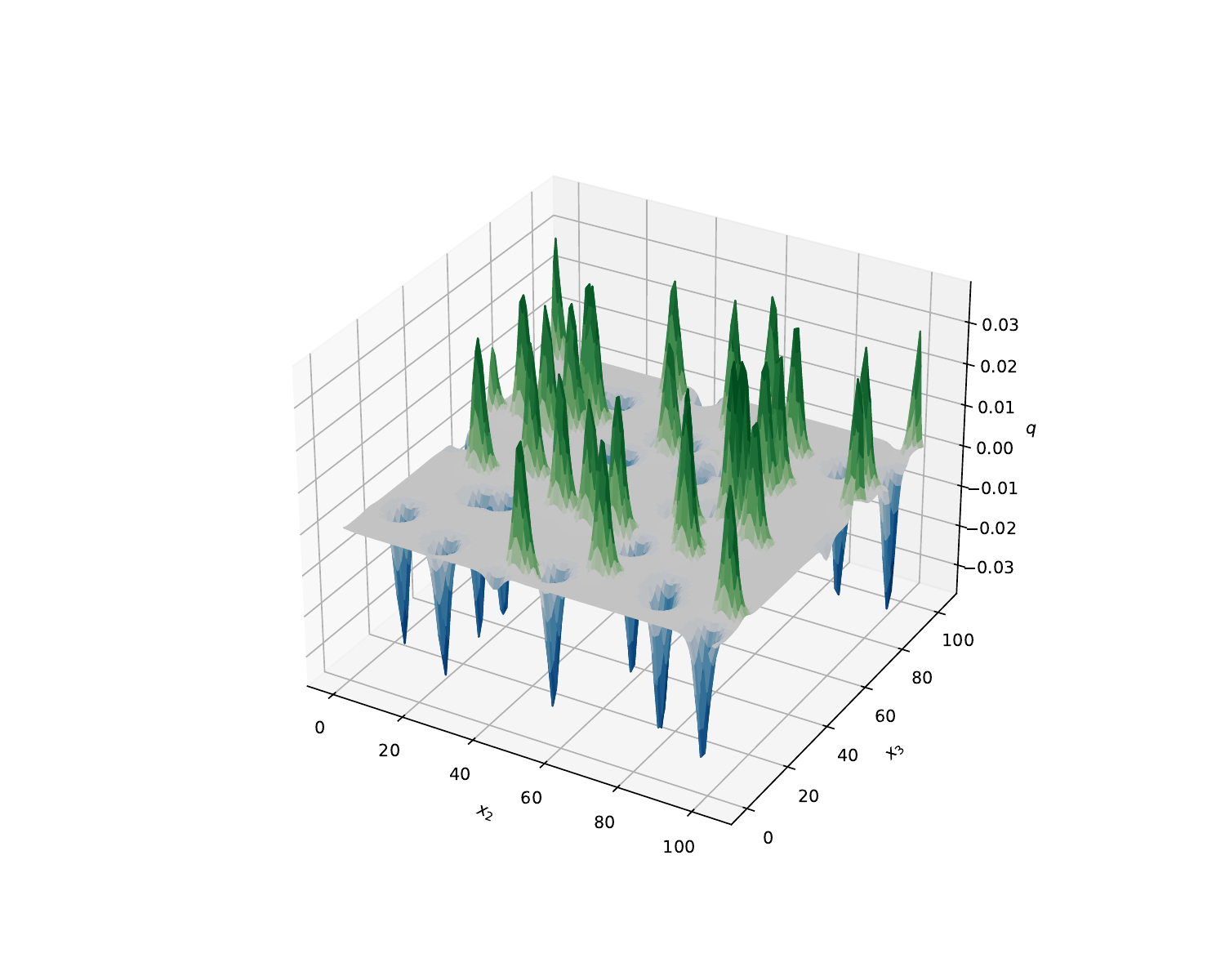
        }
        \caption{\label{fig:examplens6_3d}$\Ls=6$}
        \end{subfigure}
            \begin{subfigure}[t]{0.5\textwidth}
        \includegraphics[width=\textwidth]{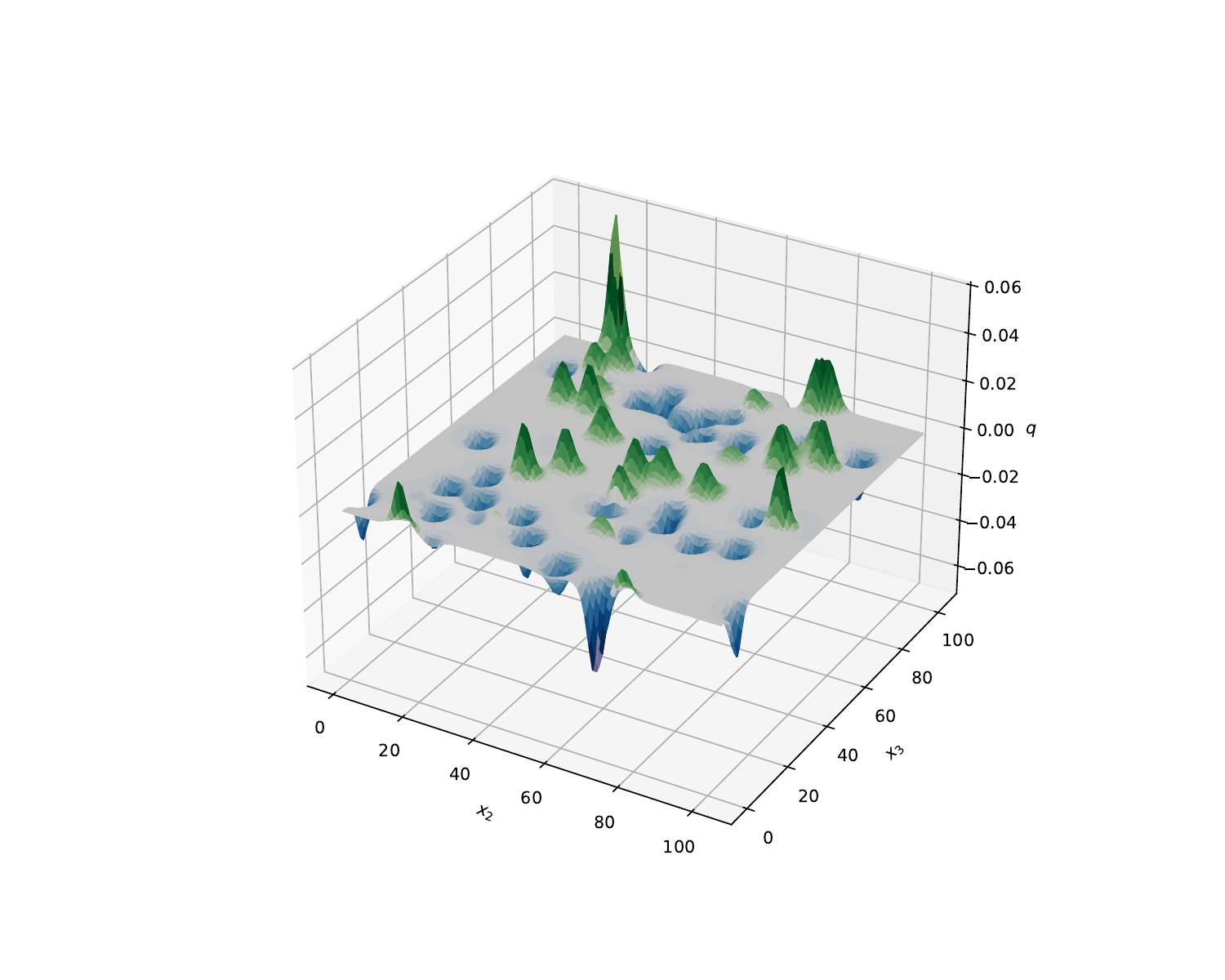}
        \caption{\label{fig:examplens8_3d}$\Ls=8$}
        \end{subfigure}
    \caption{Contour plot of topological charge density $q$ for a single configuration at $\beta=2.6$, $t_{gf}=15$, $\Lr=104$, and two different $\Ls$. The result of the identification algorithm is shown in terms of additional markers: boxes indicate fractional instantons (yellow fractional instantons, orange fractional anti-instantons); triangles indicate instantons (yellow instantons, orange anti-instantons).}
    \label{fig:exampledensitiesns6ns8}
\end{figure}
After a certain number of gradient flow steps, we determine the topological charge density distribution $q(x)$ on each configuration. In some cases, we also consider the action density. We take either the full four dimensional distribution or sum over both small $T_2$ directions. Example of the integrated topological charge density for a single configuration is shown in Fig.~\ref{fig:exampledensitiesns6ns8}. The distribution clearly shows distinct structures, which should be characterized by relevant parameters and identified with semiclassical objects. As a first step, we identify local maxima in the distribution. These are points, which are larger than all other points in the surrounding area defined by going one lattice spacing in each direction. A more precise location of the maximum can be obtained from a quadratic interpolation. A height distribution can be already obtained form these data, but a more detailed analysis is obtained with fit approach.

Therefore, as a next step, we fit the lattice data around each local maximum. The fit function predicts a maximum (peak) value, its coordinates (not necessarily a lattice point), and a width of the distribution. The expected topological charge profile is known in the case of instantons. Since the sizes of the two planes ($T_2$ and $R^2$) are very different one might expect anisotropic modifications of the solution. A modified version, which accounts for anisotropies, has been derived in \cite{deForcrand:1997esx},
\newcommand{\xmax}{x_\text{max}}
\newcommand{\xmaxmu}{(x_\text{max})_\mu}
\newcommand{\qmax}{q_\text{max}}
\newcommand{\rhop}{\tilde{\rho}}
\newcommand{\Qi}{Q_I}
\newcommand{\qmaxp}{\tilde{q}_\text{max}}
\begin{align}\label{eq:fit4d}
    q_{4}(x)=\qmaxp\left(1+ \sum^4_{\mu=1}\left(\frac{x_\mu-\xmaxmu}{\rho_\mu}\right)^2\right)^{-4}\;.
\end{align}
Integration over the small $T^2$ directions  yields
\begin{align}\label{eq:fit2d}
    q_{2}(x)=\qmax\left(1+ \sum_{\mu=2,3}\left(\frac{x_\mu-\xmaxmu}{\rho_\mu}\right)^2\right)^{-3}\;.
\end{align}
We define 
\begin{align}\label{eq:QiBPST}
\Qi=\frac{\qmaxp\pi^2\rhop^4}{6}=\frac{\qmax\pi\rho^2}{2}
\end{align}
with $\rhop^4=\prod^4_{\mu=1}\rho_\mu$ and $\rho^2=\prod_{\mu=2,3}\rho_\mu$. 
The BPST (anti-)instanton solution corresponds to $\Qi=1$ ($\Qi=-1$) and the same $\rho=\rho_\mu$ in all directions. The deviation from this solution is characterized by the deviation of $\Qi$ from one and a non-zero excentricity \cite{deForcrand:1997esx}, defined as
\begin{align}
\epsilon=\left[\sum_\mu (\frac{\rho_\mu}{\rho}-1)^2\right]^{\frac12}\, .
\end{align}
In case of the vortex-like fractional instantons, we lack an analytical expression but can be obtained numerically as explained in  Sec.~\ref{subseq:classical_solutions} and in Ref.~\cite{GonzalezArroyo:1998ez}. The shape is notably different to the BPST instanton profile.  Its integrated shape is more localized than the instanton one, and falls off exponentially at large distances from the center. In any case our fits only affect the behaviour in the vicinity of the center, because they are obtained only from the lattice peak position and its neighbours contained in a local hypercube of side length of two lattice spacings. Thus, using the modified instanton formula we have enough freedom to describe the behaviour of the fractional instanton solution in the neighbourhood of its center as well which can be seen from Fig.~\ref{fig:SingleFractionalFit}. Indeed we  have checked that using a Gaussian fitting profile, basically the same results emerge after  appropriate rescaling.
Thus, for the sake of this paper we have used the modified instanton shape to adjust all the topological structures present in our data.\footnote{We have tested further methods and improvements to account for the different shapes and plan present them in a future publication.} 
We denote objects with parameters according to \eqref{eq:QiBPST} as instanton-like if $|\Qi|=1$. For vortex-like fractional instantons the fit to this formula has been studied in Section 2.  One obtains a value  $|\Qi|\sim 0.6$ close to the value $1/2$ of its topological charge.
In the following we will discuss general properties of the obtained distributions of fit parameters. Some less physical relevant discussions on the gradient flow parameters can be found in Appendix \ref{sec:gfdependence}. 

\subsection{Characterizing the distributions}
\label{subseq:Distributions}
\begin{figure}
    \centering
    \includegraphics[width=\textwidth]{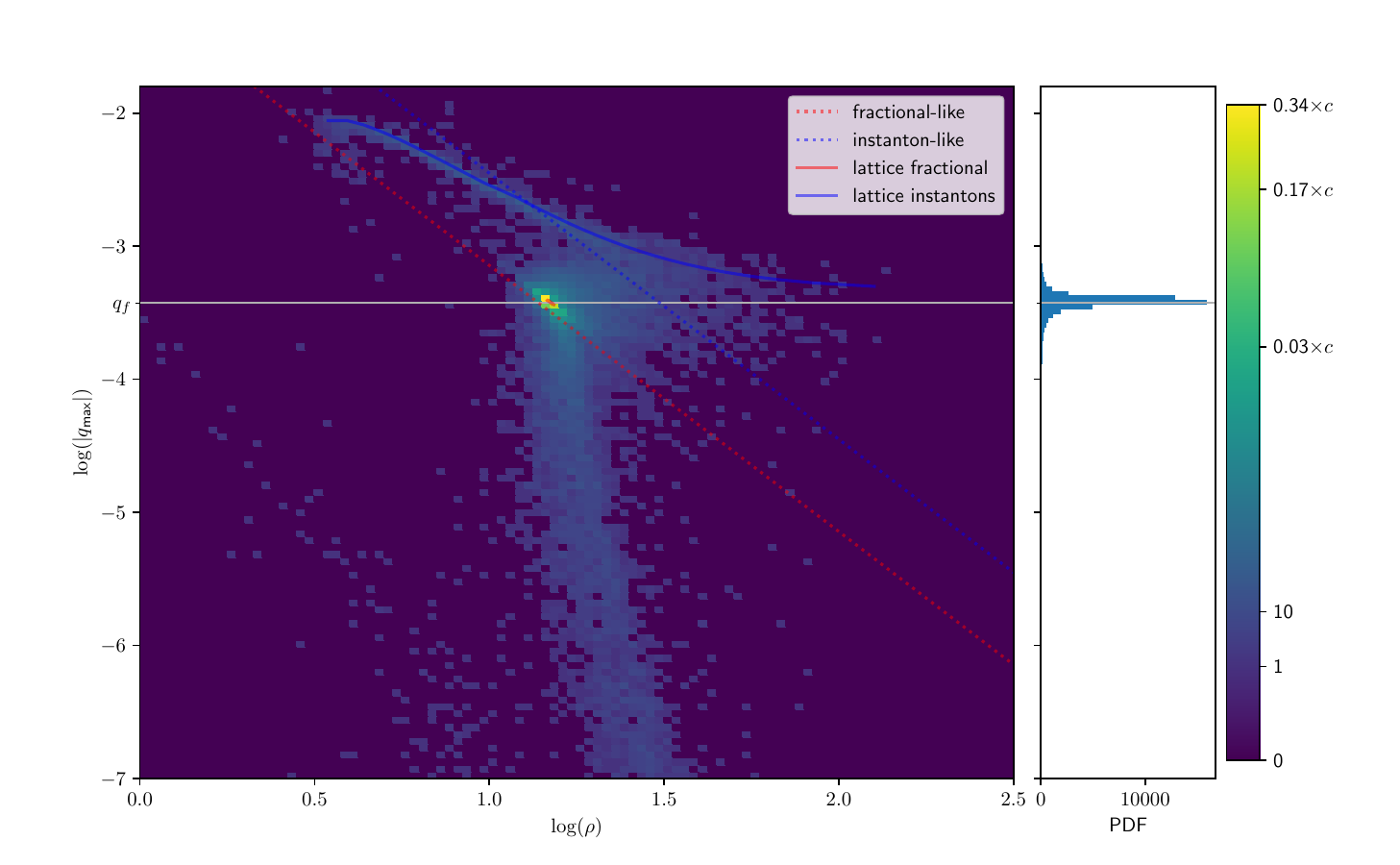}
    \caption{Density plot (2D histogram) of the fit parameters obtained from configurations at $\Ls=6$, $\Lr=104$, $\beta=2.6$, and $t_{gf}=15$ of Wilson gradient flow. The lattice instanton line is obtained from a single instanton configuration using Wilson and overimproved gradient flow. The upper end of the lattice instanton line corresponds to the point where the instanton dissolves under the Wilson flow and the lower end to the point where the excentricity increases rapidly until two peaks emerge under the overimproved flow. The lattice fractional points result from a fit of a set of configurations with a single fractional instanton generated numerically at $\Ls=6$ and the vertical line represents the value obtained from \eqref{eq:S2fit}. An exponential map has been applied to the normalization of the 2d histogram as indicated by the colorbar to visualize small densities ($c$ is the total number of counts). A histogram of the $\qmax$ values is shown on the right.}
    \label{fig:fit_param_ns6tau15}
\end{figure}

As a first step, we are now considering the raw distribution of parameters obtained from the fit as shown in Fig.~\ref{fig:fit_param_ns6tau15} neglecting the excentricity.

 BPST instanton contributions are characterized by $\Qi=\pm 1$, i.~e.\ the line $\log( |\qmax|)=\log\left(\frac{2}{\pi}\right)-2\log(\rho)$ in double log representation of the fit results. Similarly one can introduce further lines for contributions similar to topological charge $\Qi$ using $\log(|\qmax|)=\log\left(\frac{2\Qi}{\pi}\right)-2\log(\rho)$. 
The fractional instantons having $Q\sim \frac{1}{\Nc}$, in the dilute situations, extend over the full 2-torus and are therefore expected to be localized at a single point in the diagram as derived in Sec.~\ref{subseq:classical_solutions}. This point is close to the $\Qi=\frac{1}{\Nc}$ line if the fit is done close to the maximum.

On the lattice, instantons deviate from the continuum BPST solution. At small $\rho$, the lattice spacing modifies the profiles and instantons are not resolved below a certain minimal size $\rho$ in units of the lattice spacing. Large instantons, on the other hand, are affected by finite volume effects in particular related to $l_s$. The deviations between the continuum and the lattice versions of semiclassical contributions can be estimated using numerically generated samples of single objects. 

The parameter space of the $\Qi>\frac{1}{\Nc}$ solutions can be partially explored using the properties of the gradient flow. The Wilson action leads to a decreasing size $\rho$ and increasing height $\qmax$, whereas the overimproved flow leads  to the opposite effect. Both types of gradient flow move the instanton solution along the line of approximately minimal action, i.~e.\ a lattice counterpart of the continuum BPST line. Increasing the width of the instanton with the overimproved flow, one finally approaches a region where the excentricity starts to increase more rapidly and eventually two peaks of fractional instantons emerge.

\begin{figure}
    \begin{subfigure}[t]{0.3\textwidth}
    \includegraphics[width=1.1\textwidth]{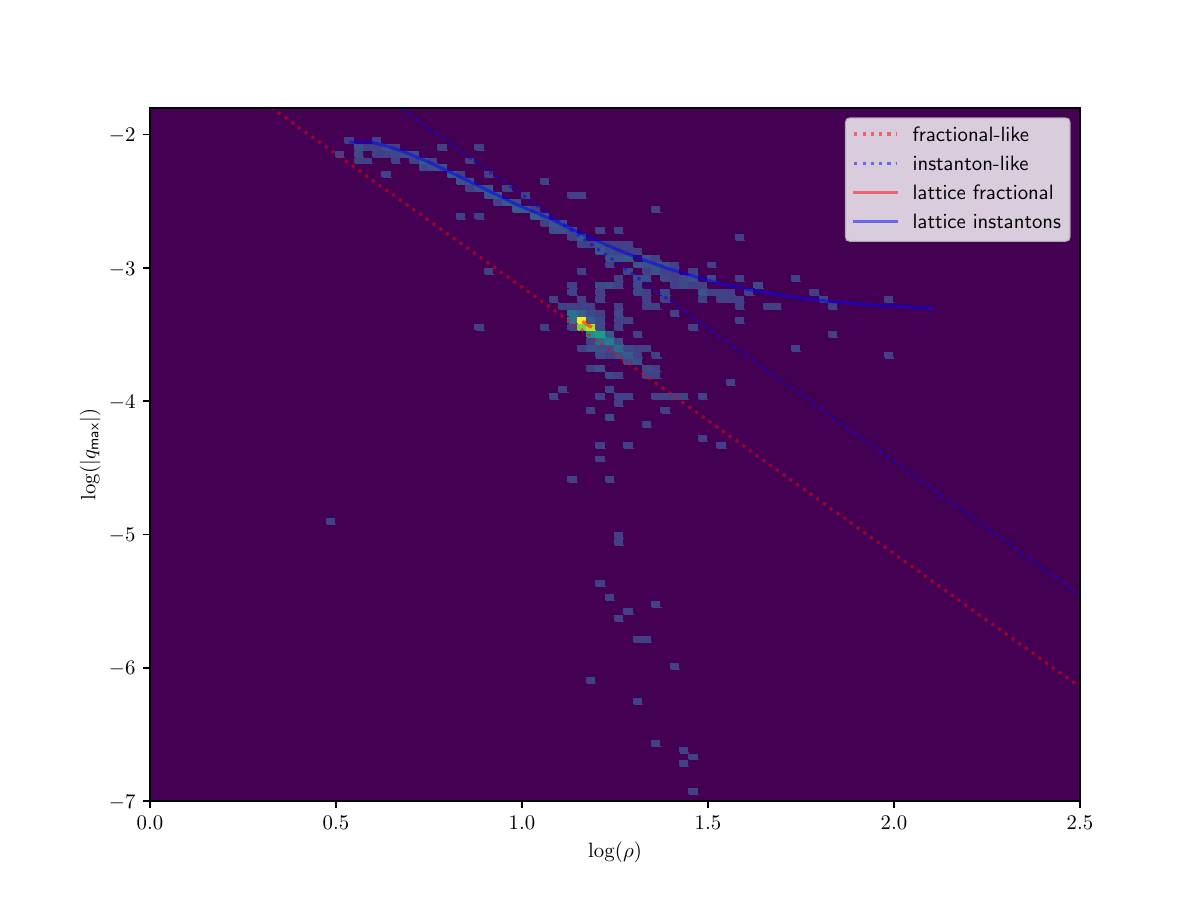
    }
    \caption{\label{fig:ns6largesep}large separation}
    \end{subfigure}
        \begin{subfigure}[t]{0.3\textwidth}
    \includegraphics[width=1.1\textwidth]{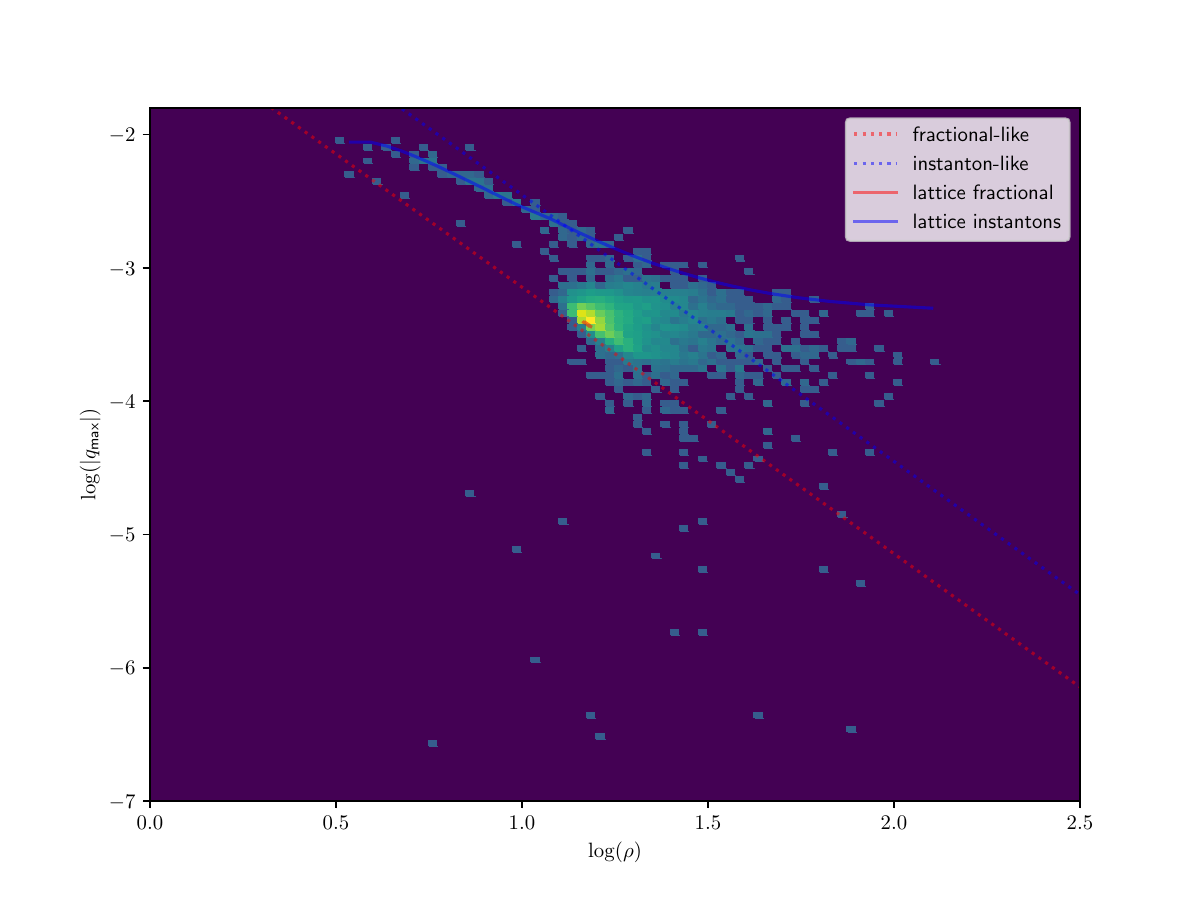}
    \caption{small separation, same sign}
    \end{subfigure}
        \begin{subfigure}[t]{0.3\textwidth}
    \includegraphics[width=1.1\textwidth]{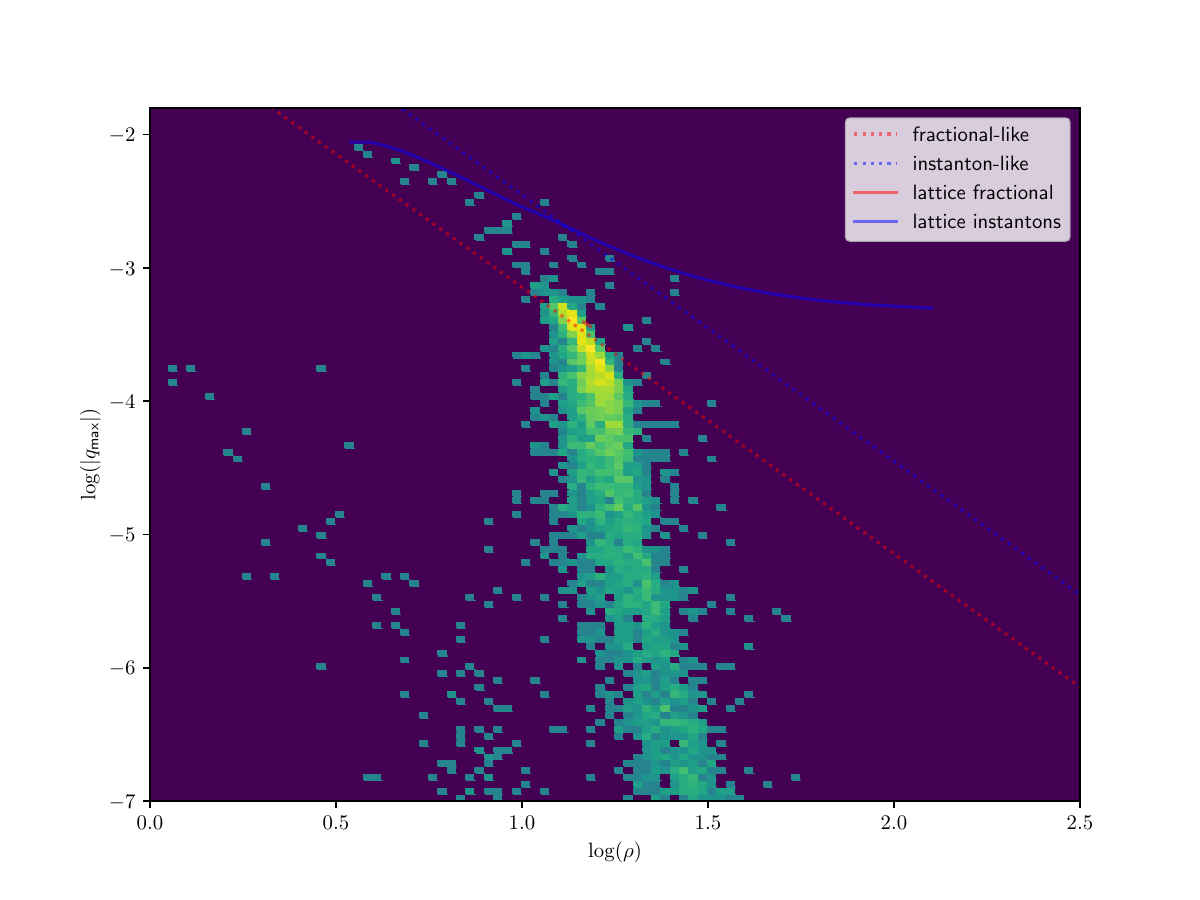}
    \caption{small separation, opposite sign}
    \end{subfigure}
    \caption{Density plot of the fit parameters obtained with same parameters as Fig.~\ref{fig:fit_param_ns6tau15}. This time separated according to the distances and relative sign compared to the next peak. From left to right: peaks with a separation larger than a standard deviation above the average; peaks with more than one standard deviation below the average separation and the same sign; peaks with small separation and opposite sign.}
    \label{fig:fit_param_ns6tau15_sep}
\end{figure}
\begin{figure}
    \centering
        \includegraphics[width=0.5\textwidth]{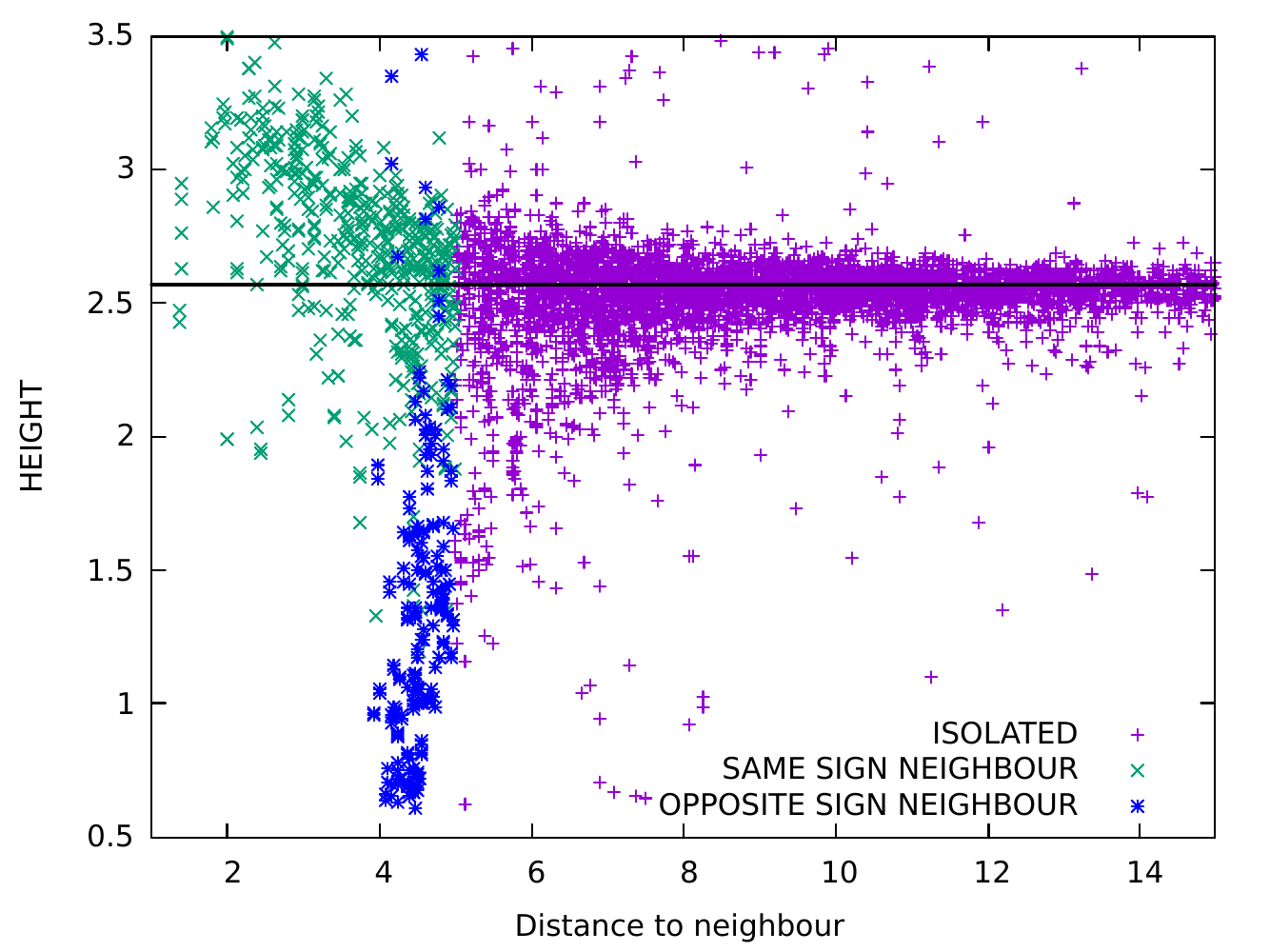}
	    \caption{Distribution showing the correlation between the height of the peak and distance to a neighbouring peak. 
        Same parameters as Fig.~\ref{fig:fit_param_ns6tau15}, but only for a subset of the configurations. Peaks that are at distance larger than 5 from a neighbour are labelled "ISOLATED". Peaks with nearest neighbours of the topological charge sign (in green) show an increase in height. Those with  neighbours of opposite sign (in blue) show a sharp decrease in height. }
	    \label{fig:population}
\end{figure}

\begin{figure}
    \begin{subfigure}[t]{0.5\textwidth}
    \includegraphics[width=\textwidth]{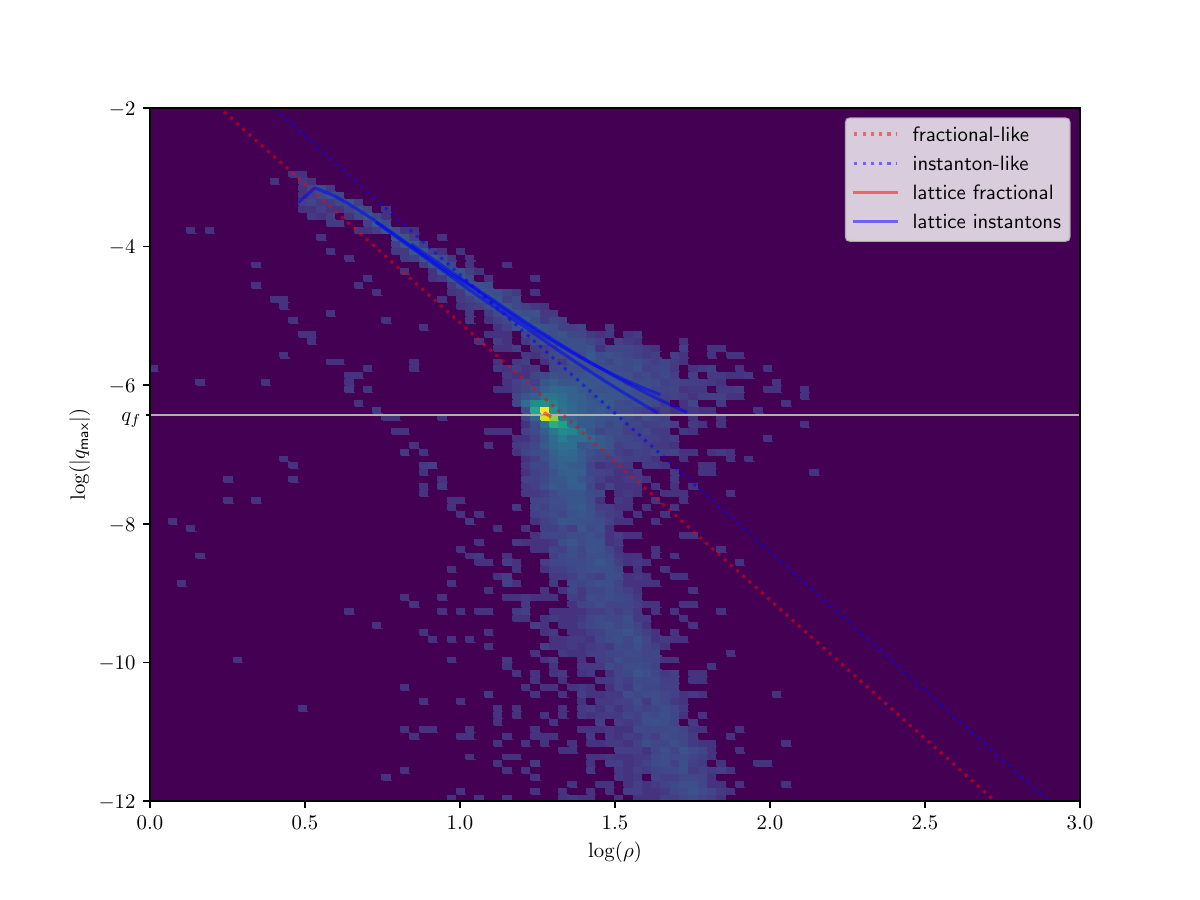}
        \caption{$\Ls=6$}
    \end{subfigure}
        \begin{subfigure}[t]{0.5\textwidth}
    \includegraphics[width=\textwidth]{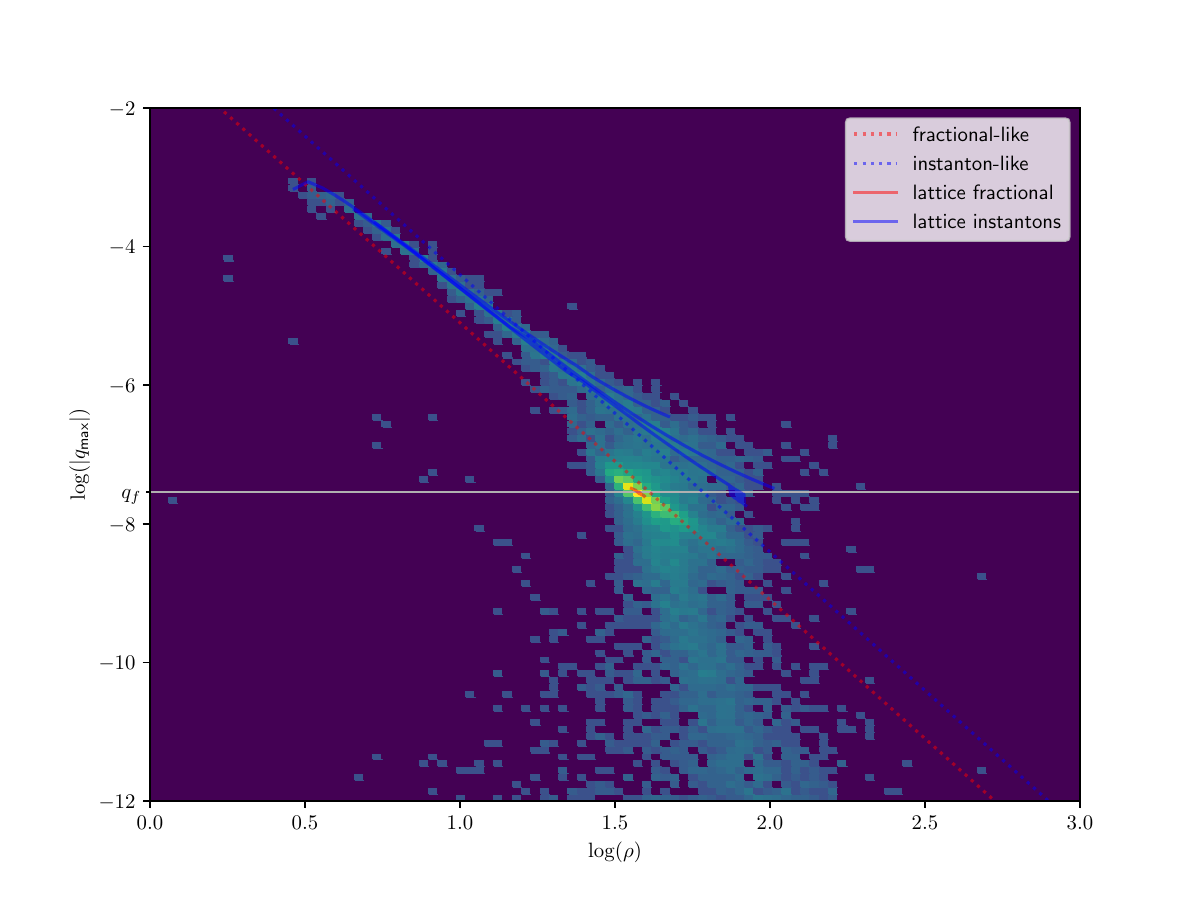}
    \caption{$\Ls=8$}
    \end{subfigure}
        \begin{subfigure}[t]{0.5\textwidth}
    \includegraphics[width=\textwidth]{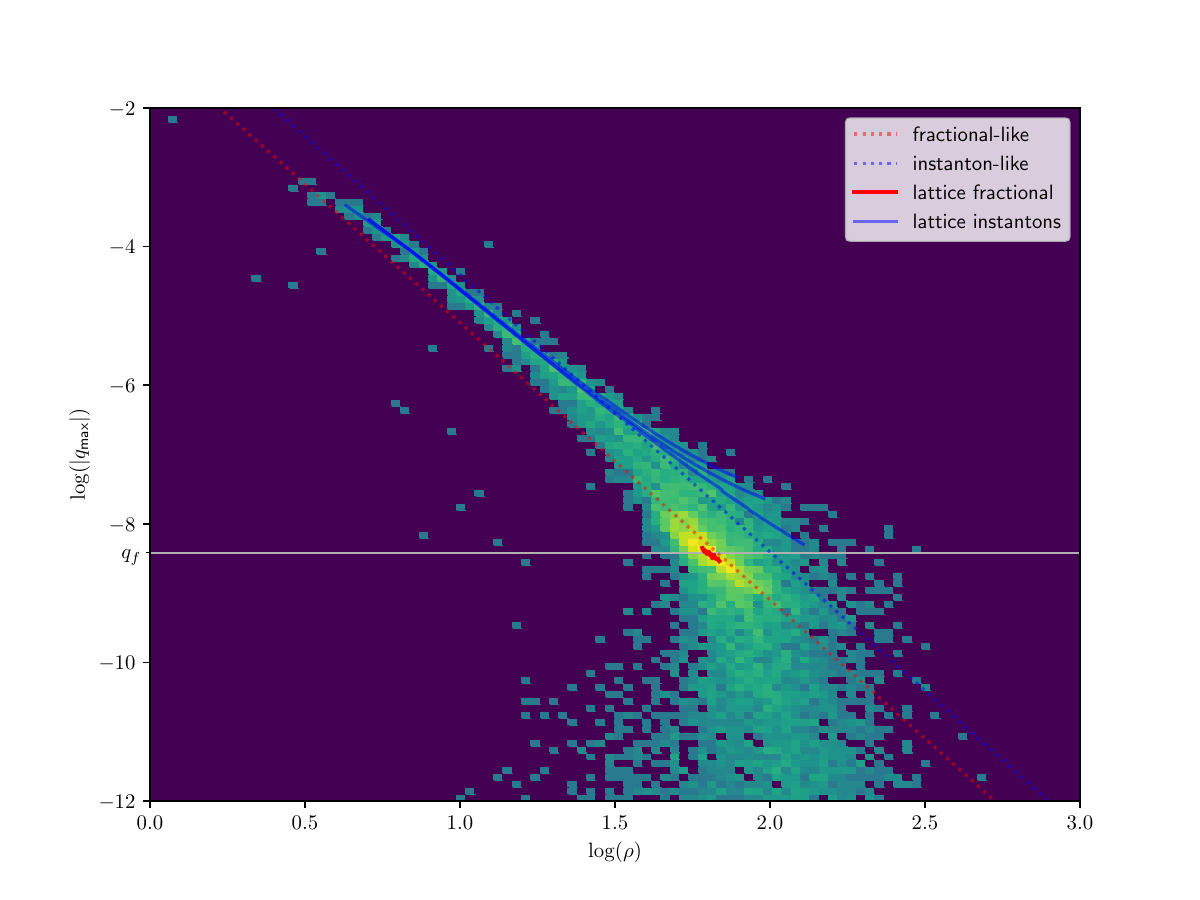}
        \caption{$\Ls=10$}
    \end{subfigure}
        \begin{subfigure}[t]{0.5\textwidth}
    \includegraphics[width=\textwidth]{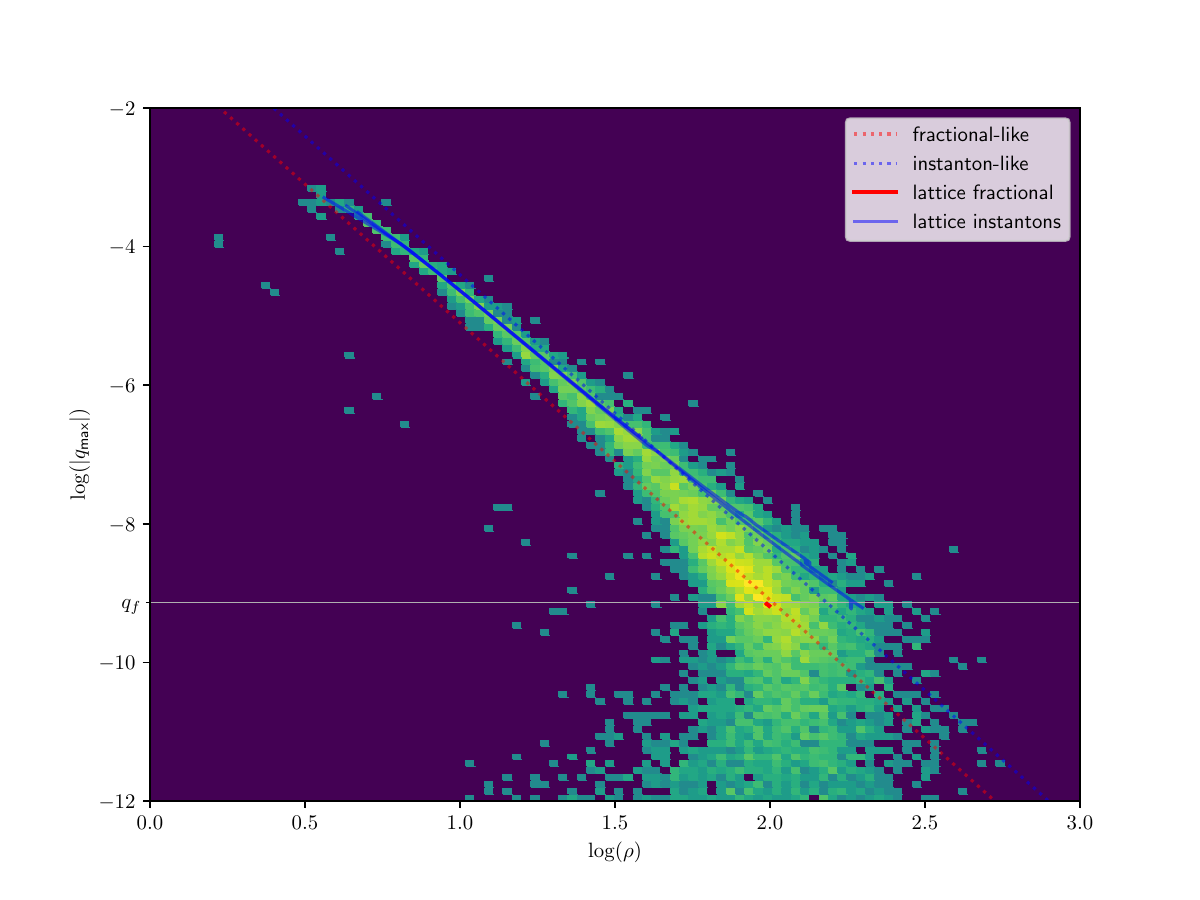}
    \caption{$\Ls=12$}
    \end{subfigure}
        \begin{subfigure}[t]{0.5\textwidth}
    \includegraphics[width=\textwidth]{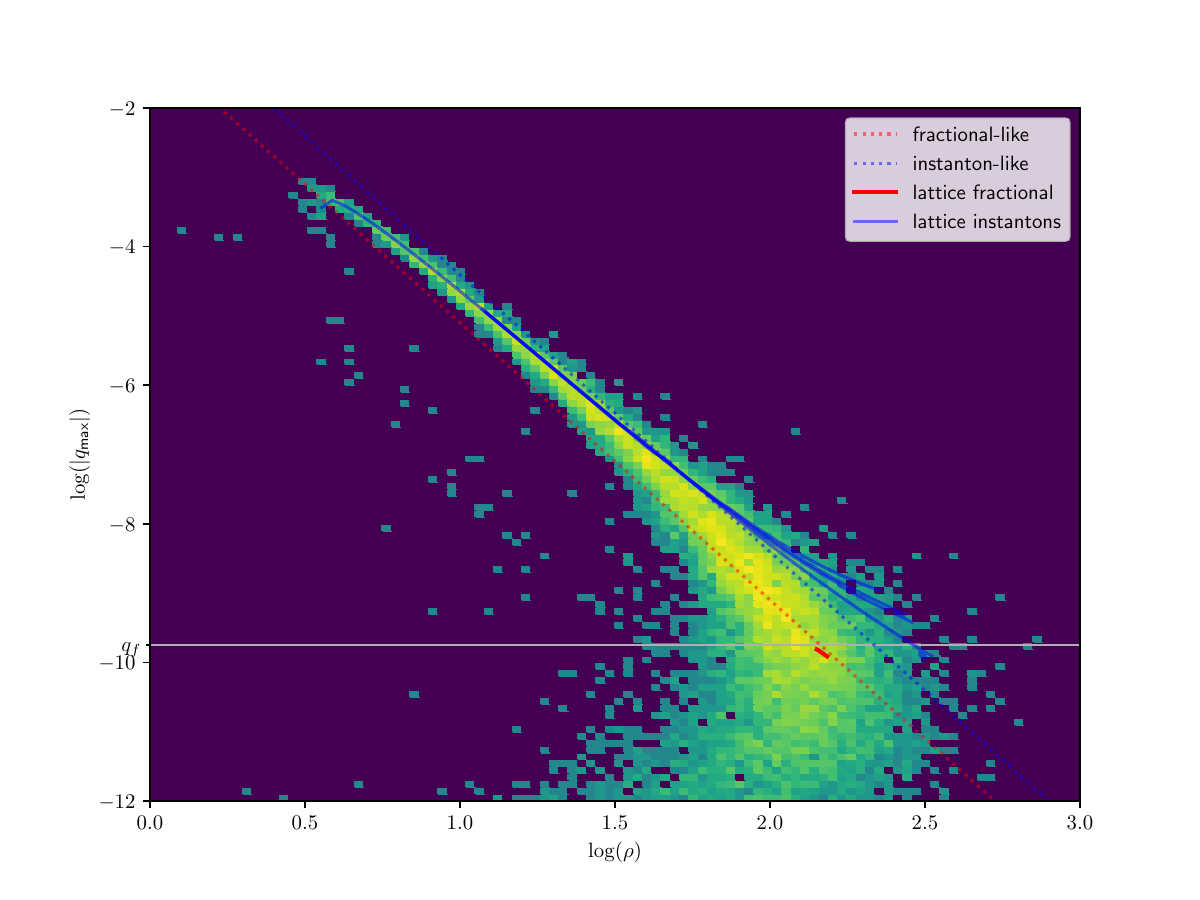}
        \caption{$\Ls=14$}
    \end{subfigure}
        \begin{subfigure}[t]{0.5\textwidth}
    \includegraphics[width=\textwidth]{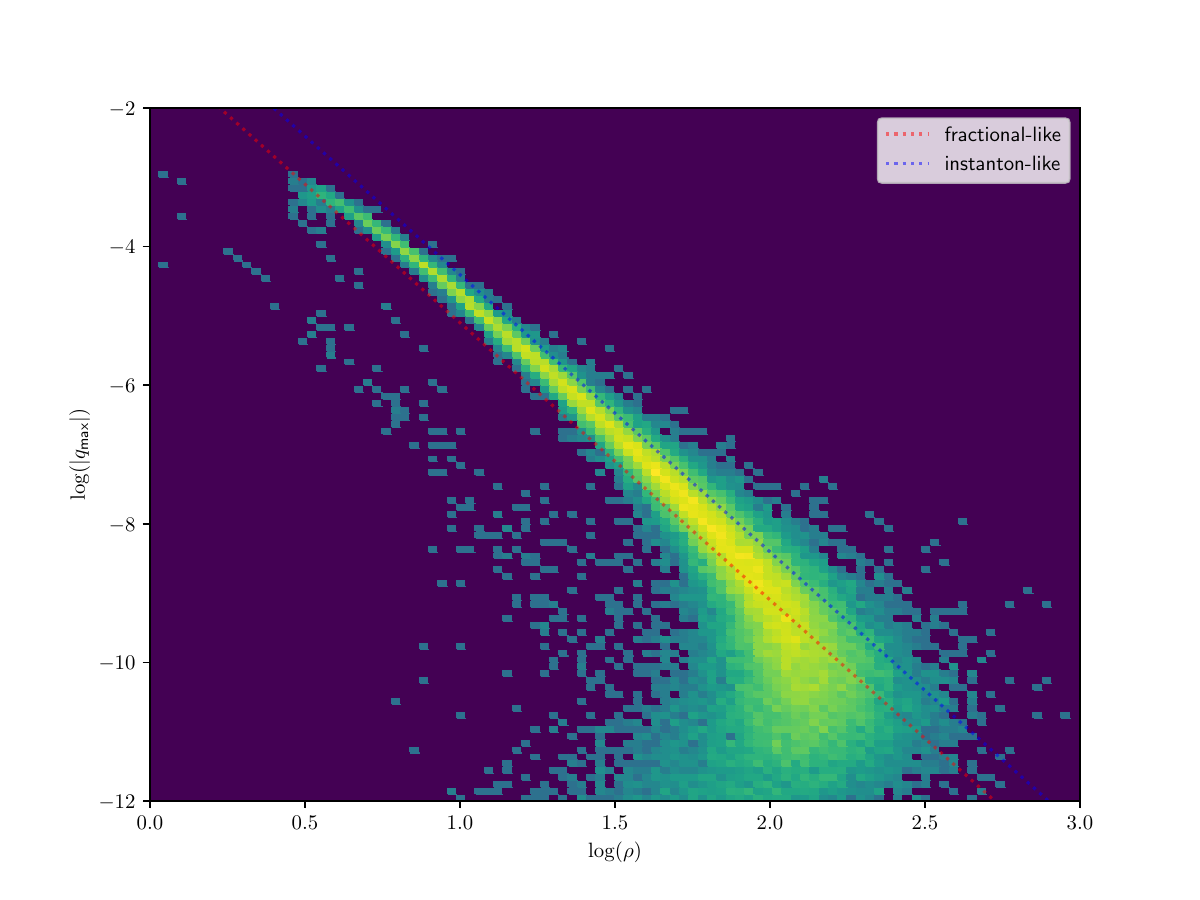}
    \caption{$\Ls=30$}
    \end{subfigure}
    \caption{Density plot of the four dimensional fit parameters for different $\Ls$ ($\beta=2.6$,$t_{gf}=15$). $\Lr=104$ for $\Ls<10$ and $\Lr=64$ otherwise. The $q_f$ value corresponds to the prediction of $Q(0)$ \eqref{eq:q0fit} in lattice units. In order to estimate the uncertainties of the lattice instanton lines, we have taken three different approximate instanton configurations and deformed them with the overimproved and Wilson flow resulting in three different lines.}
    \label{fig:fit_param_nsdep}
\end{figure}
At first we discuss the observed distributions of fit parameters in the case of $\Ls=6$ and $\beta=2.6$ before we consider the $l_s$ dependence. The distributions is shown in Fig.~\ref{fig:fit_param_ns6tau15}. Clearly the majority of the peaks correspond to fractional instantons. A smaller fraction of the distribution can be associated with instantons. 

The fractional instanton contributions seems to be extended into two directions: at constant $|\qmax|$ towards larger sizes and at almost constant sizes towards smaller $|\qmax|$. This can be explained considering neighbouring peaks. If the next neighbour is a fractional with the same sign, the width of the peak distribution and the height should increase. On the other hand, if the next peak is of opposite sign, i.~e.\ a fractional anti-fractional pair, the height is expected to decrease. The latter effect is enhanced by the gradient flow, which tends to annihilate pairs of opposite sign.
In order to verify this expectation, we separate out the contributions with large distances to the next neighbour (more than one standard deviation larger than the average distance). In addition, we separate contributions with the smallest distances to the next neighbour, which are further separated in the ones with the same sign as the neighbouring peak and the ones with opposite sign of the neighbouring peak. The resulting three distributions of fit parameters are shown in Fig.~\ref{fig:fit_param_ns6tau15_sep}. From these density plots it is clear that well separated peaks follow the expected semiclassical predictions, which are either the BPST instanton line or the fractional instanton profile. If peaks are close to each other there is a significant distortion depending on the relative sign. As expected, the densities at small $\qmax$ can be attributed to pair annihilations. Another illustration of this fact is shown in Fig.~\ref{fig:population}: at large separations, the distribution is close to the expected peak height of a fractional instanton, but there are strong deviations as the separation becomes smaller. The shift in height, positive for equal sign neighbours and negative for opposite sign, is consistent with the effect of overlapping profiles. 

The last point in this discussion of the raw fit parameters is a consideration of peak distributions towards the large $l_s$ limit. This provides only a qualitative picture about the infinite volume limit and more precise quantitative statements are presented in Sec.~\ref{seq:results}.

At larger $l_s$, the integration of the smaller directions in our two-dimensional fit approach \eqref{eq:fit2d} leads to a distortion and misidentification of densities since there is a non-trivial dependence on $T_2$. In order to combine a well-defined large $l_s$ and small $l_s$ limit, we apply the following strategy: based on the full four dimensional topological density, we perform a four dimensional fit around each peak according to \eqref{eq:fit4d}. At smaller $l_s$, this leads to a considerable anisotropy of $\rho_\mu$ comparing the $T_2$ and $R_2$ directions. Therefore, we consider $\rho$ instead of $\rhop$ which is the geometric mean of only the two $R^2$ directions.  

The density plots for the parameters are shown in Fig.~\ref{fig:fit_param_nsdep}. Increasing $l_s$ the size of the fractional instantons increases. At the same time, the distribution gets broader. Finally, it becomes almost independent of $l_s$ resembling at $\Ls=14$ already closely the large volume limit ($\Ls=30$). This is consistent with a picture in which $\rho$ at small $l_s$ is determined by the $T_2$ size, while at large $l_s$ the distance between the objects determines $\rho$. 

In Appendix~\ref{sec:ncdependence} we add some comments about the distribution for SU(3) and SU(4).

\subsection{The identification algorithm}
\label{subseq:identifiation_algorithm}
Our further discussion of the fractional instanton two-dimensional gas  requires an explicit determination of the number of fractional instantons for each configuration.
As described in Sec.~\ref{subseq:Distributions}, individual peaks can be identified in the distribution of topological charge and characterized by simple fit parameters. There is a natural correspondence between these parameters and semiclassical objects at least for small $l_s$ and well separated peaks. In this case, fractional instantons can be straightforwardly distinguished from intanton-like contribution and noise, see Fig.~\ref{fig:ns6largesep}.
Deformations have to be taken into account due to lattice and finite volume effects. 
The obvious separation according to \eqref{eq:QiBPST} at the $|\Qi|=3/4$ line is hence not a reasonable choice.
At smaller distances, the overlap of neighbouring objects introduces additional deformations.

We have tested several different approaches, but the simple separation according to the $\qmax$ already provides a reasonable method to distinguish instantons, fractional instantons, and background noise. In fact, the majority of the instantons have twice the peak height of an isolated fractional instanton and at sufficient gradient flow time, there is also a decent separation from the noise level. It is expected that this simple method is insufficient at larger $l_s$. However, the primary focus of the current work is an extensive data scan utilizing straightforward methodology. In a further publication, we plan to consider in more detail improvements at larger $l_s$.  Note that there is in general a difficulty with large size instantons since, given a limited set of data points, the distribution becomes indistinguishable from a superposition of two factional instantons. 

Hence we apply the following simple algorithm for the identification of fractional instantons: The peaks with a height $\qmax$ larger than two times the expected fractional instanton size are identified as instantons. The expected fractional instanton size ($q_\text{frac}$) is determined by a peak in the histogram of $\qmax$, which is close to the point expected from the considerations of isolated fractionals \eqref{ex:maxqsemiclass}. This takes into account possible distortions of the histogram due to the higher density of the objects.
We have chosen a sufficient gradient flow time for a descent noise separation. The threshold value $|\qmax|<20\%\,q_\text{frac}$ determines the neglected remaining background noise peaks.

Errors are estimated by varying upper threshold for the separation between instantons and fractional instanons between $1.75\, q_\text{frac}$ and $2.25\, q_\text{frac}$  and threshold for the noise level between $10\%\,q_\text{frac}$ and $30\%\,q_\text{frac}$. For illustration, we show the result of the final identification in comparison to the raw topological charge density in Fig.~\ref{fig:exampledensitiesns6ns8}. The identification algorithm is able to distinguish the different contributions and provides a reasonable separation of the objects.

The methods explained in this Section provide the basic input for the discussions of the results presented in the following. Note that in some cases we consider the action density instead of the topological charge density, which due to approximate self-duality, provides a proportional signal. In Sec.~\ref{subseq:Properties_gas}, we discussed the properties of an  idealized gas of fractional instantons has been introduced. Most of the calculations were done in the thin abelian vortex approximation (TAVA). This should work quite well in the most dilute situations. 
In addition  now include together with  the numbers  $n_+^f$ ($n_-^f$) of  fractional (anti-) instantons,  also the ones $n_+^I$ ($n_-^I$) for the ($Q=1$ BPST) (anti-)instantons. In some of the observables   the $Q=1$ instantons can be made equivalent to  two fractionals. Hence the correspondence is $n_+=n_+^f+2n_+^I$, $n_-=n_-^f+2n_-^I$.

\section{Analysis of Results}
\label{seq:results}
In this Section, we will present our results  focusing among the small
torus size region. Because of asymptotic freedom, the coupling constant
at these scales is small and the simple dilute gas formulas are
expected to work better. As the torus size grows, the density of objects also increases 
until the dilute gas approximation breaks down.

In Section~\ref{seq:semiclassical} we gave the expected behaviour on the
basis of the semiclassical approximation which we are now going to
confront with our Monte Carlo results. 

\subsection{Poisson distributed 2D gas of vortex-like fractional
instantons}
The nature of the 2D gas is particularly neat for the smaller values of $\Ls=6$ or $8$ at $\beta=2.6$ or higher. 
We use the $S_2$ observable which gives the action-density integrated over the  directions of the small 2-torus. 
This gives a distribution of peaks located at different position in the big torus plane. A snapshot of the topological charge distribution of a typical
 configuration is displayed in \Figref{fig:exampledensitiesns6ns8}. A similar picture is obtained for the action density. Instead of the identification algorithm based on the topological charge, we consider first an investigation using a simple quadratic interpolation of the action density, which can be compared to \eqref{eq:S2fit}. 

\begin{figure}[h!]
    \centering
        \includegraphics[width=0.5\textwidth]{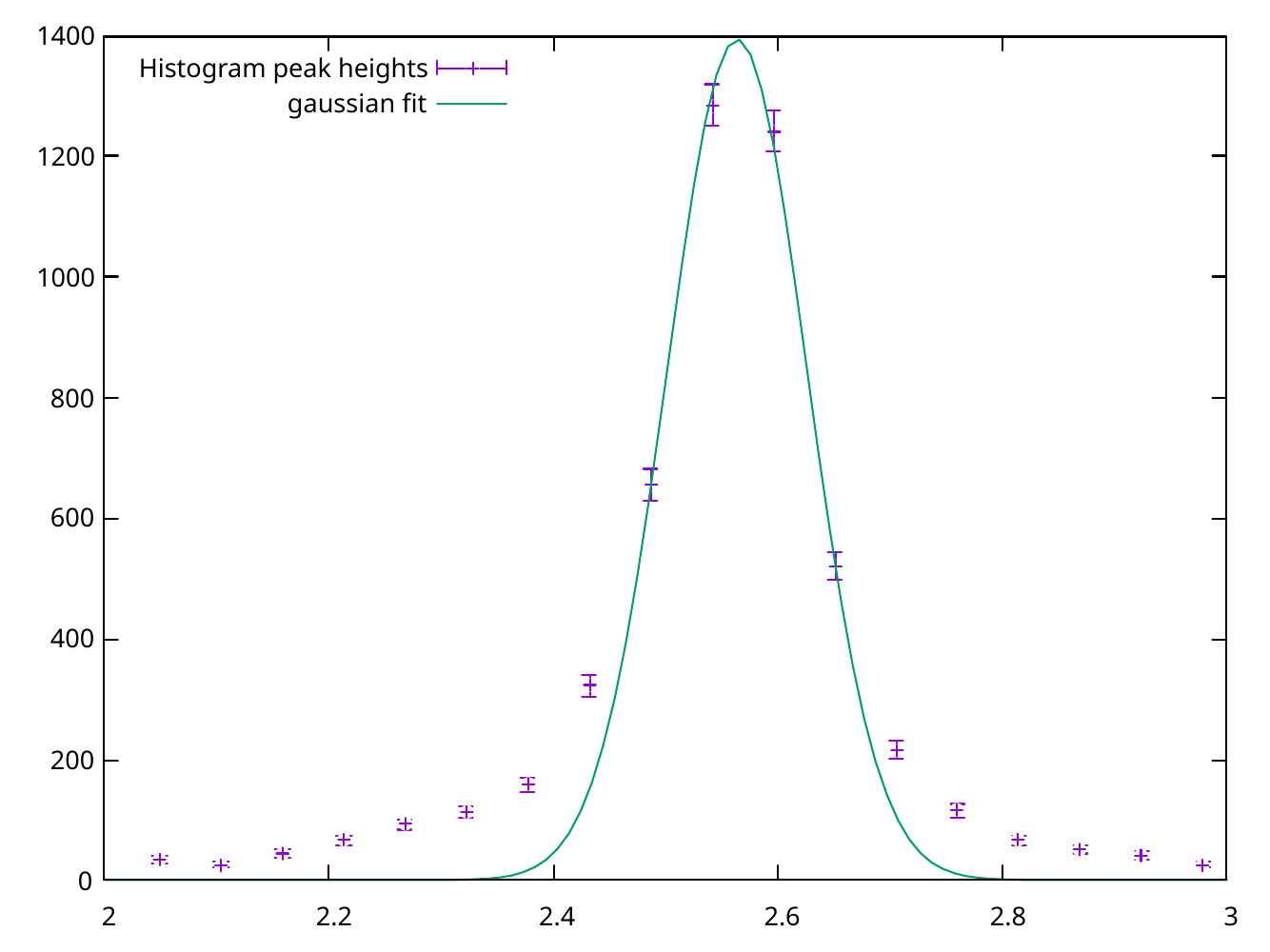}
	    \caption{Histogram of the action density peak heights $H$ o at $\Ls=6$, $\beta=2.6$, $6^2\times 104^2$ lattice. This corresponds  to wider range  histogram  of the topological charge density (log scale) shown in Fig.~\ref{fig:fit_param_ns6tau15}.}
	        \label{fig:histo}
\end{figure}

Analysing a sample of configurations   at
$L_s=6$ and $\beta=2.6$ shows a distinctive peak close to the expected location 
$H(6)=2.567=(100.27-283/6^2)/6^2$ according to \eqref{eq:S2fit}\footnote{Notice that on the lattice one measures the value of the $S_2$
integrated over a plaquette which is $S_2 /L_s^2$.}. The histogram of peak heights is shown in
\Figref{fig:histo}
which in the vicinity of the peak is fitted to a Gaussian
distribution. The fitted peak position is 2.564(2) and the standard
deviation of the Gaussian is 0.088(2). Clearly the distribution is
more spread than what the Gaussian predicts, but the precision of the
peak location leaves little doubt about the nature of the peak. 

The histogram of peak heights has also tails that extend towards higher or lower values. It is important to understand the origin of these deviations in this simple situation since it can help to extend this interpretation to situations occurring at much higher densities. Indeed, this was already discussed in the previous section and illustrated in  Fig.~\ref{fig:population}. There we see that peaks at a distance larger than 5 lattice spacing from a neighbour(in purple) are giving the narrow Gaussian distribution around the expected value. Those closer to a neighbour are split into two sets: if the neighbour has the same topological charge sign (in green) the height tends to higher values. This is easily explained by the overlap of the contributions. The remaining set corresponds to peaks having a closest neighbour of opposite sign (in blue).  In this case we observe that as the distance becomes smaller there is a strong neutralization of the signal leading to a much smaller peak height. We think that this explains very neatly the origin of the extended tails in Fig.~\ref{fig:histo}.

Now we proceed to test the remaining aspects of the distribution
itself: its Poisson nature and the 2D spatial distribution. The latter
is crucial to show that the FI are indeed independent and do not
cluster into molecules which, as mentioned earlier, would kill its
effect on the string tension. 

The characteristics of the trinomial/Poisson distribution was described in
Sec.~\ref{seq:semiclassical}. Although we gave analytic formulas for most distributions we have also used a numerical method to determine the predictions of a  2D gas of 
fractional instantons.  Giving as input only the density one can generate  hundreds of thousands of configurations  in very short times following the probability distribution explained in Sec.~\ref{seq:semiclassical}
and setting the minimum area $A_0$ to a plaquette and the total area to the corresponding one in the lattice simulation. 
Once a FI or AFI is created in a given plaquette, its location is generated uniformly with this plaquette.  One can easily add the constraint of having only an even total number of objects (integer $Q$). In this way all distributions can be determined in terms of a single parameter, the mean number of objects, 
which can be adjusted to match the set of configurations one is trying to compare with. 

Let us now proceed to compare the distributions obtained by analysing the Monte Carlo generated configurations with the analytic and numerical ones of a free 2D gas. We take as an example the case of $\Ls=6$ and $\beta=2.6$ for which 
the density is relatively small.

\begin{figure}[h!]
    \centering
    \includegraphics[width=0.9\textwidth]{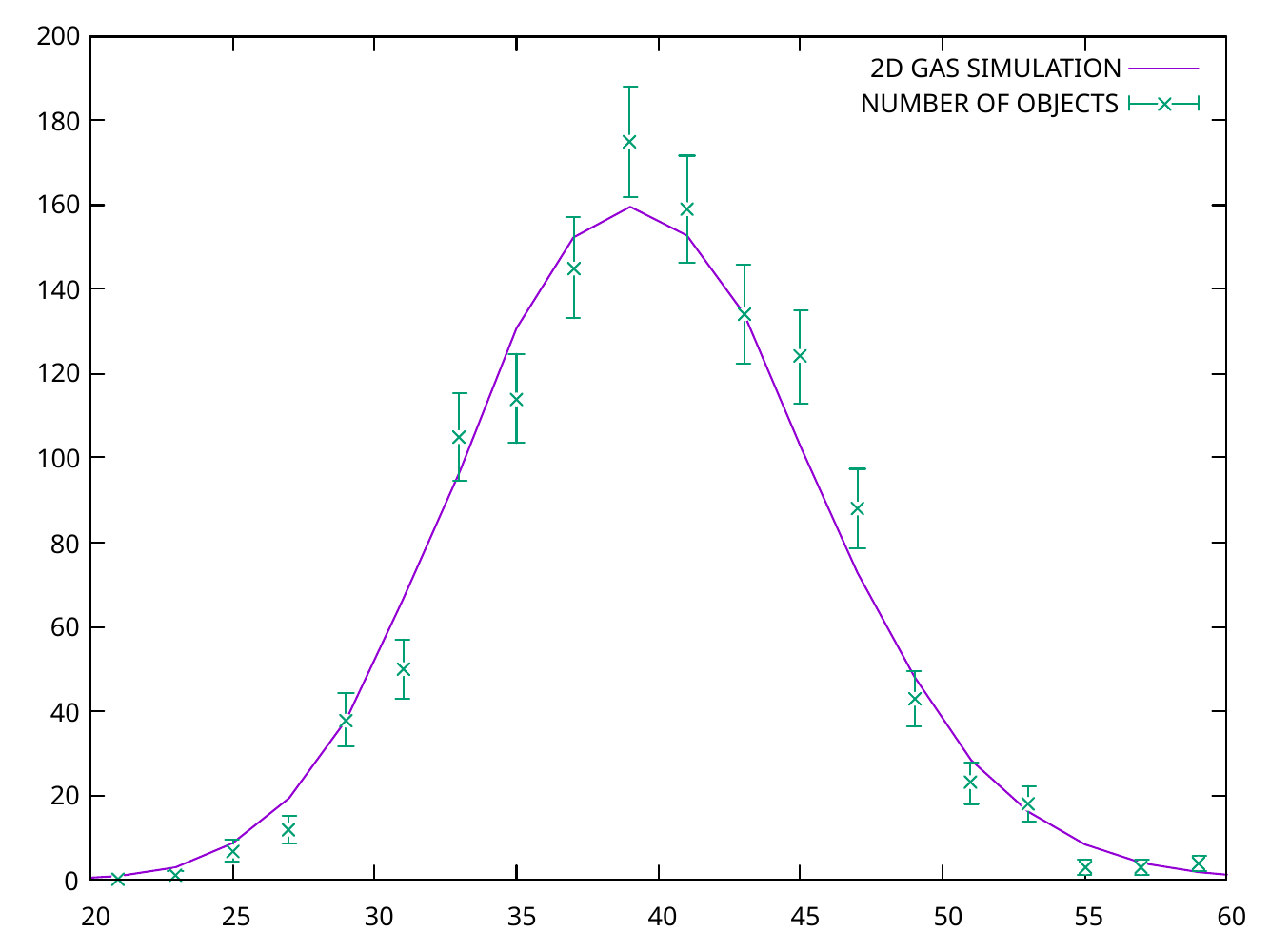 }
    \caption{The plot compares the histogram of the number of objects with the prediction of the 2D Poisson gas (see text) for $\Ls=6$, $\beta=2.6$.}
    \label{fig:poisson}
    \end{figure}

Let us first study the histogram that gives the distribution of the number of objects (FI and AFI). Since the lattice configurations have been obtained with no twist in the big plane we expect only an even number of objects (FI+AFI) per configuration.   
Unfortunately our peak-identification method 
has an uncertainty and odd numbers are also found, although in a smaller
proportion. Nonetheless, this  uncertainty  in the number of peaks is
much smaller than the dispersion of the distribution. This can be seen 
in \Figref{fig:poisson} where we take a bin size of 2 which adds up the even and odd values. 

The actual data points are compared with the results of the 2D gas, drawn as a continuum line. Since by construction, the 2D gas result has no odd number of objects, we use also the same bin size of 2 for this case. The matching is very good and the only input for the 2D gas is that the mean number of objects matches the Monte Carlo configurations.

Another interesting piece of information  is the distribution of minimum distance to a neighbouring peak. As will be argued later this distribution can be used to measure the effect of the flow as well as the possible interactions (attractive or repulsive) among the objects. For that purpose it is interesting to distinguish the cases in which the closest neighbour has the same or opposite sign of the topological charge. For the Poisson gas the two distributions are essentially equal.

\begin{figure}[h!]
    \centering    
\includegraphics[width=0.45\textwidth]{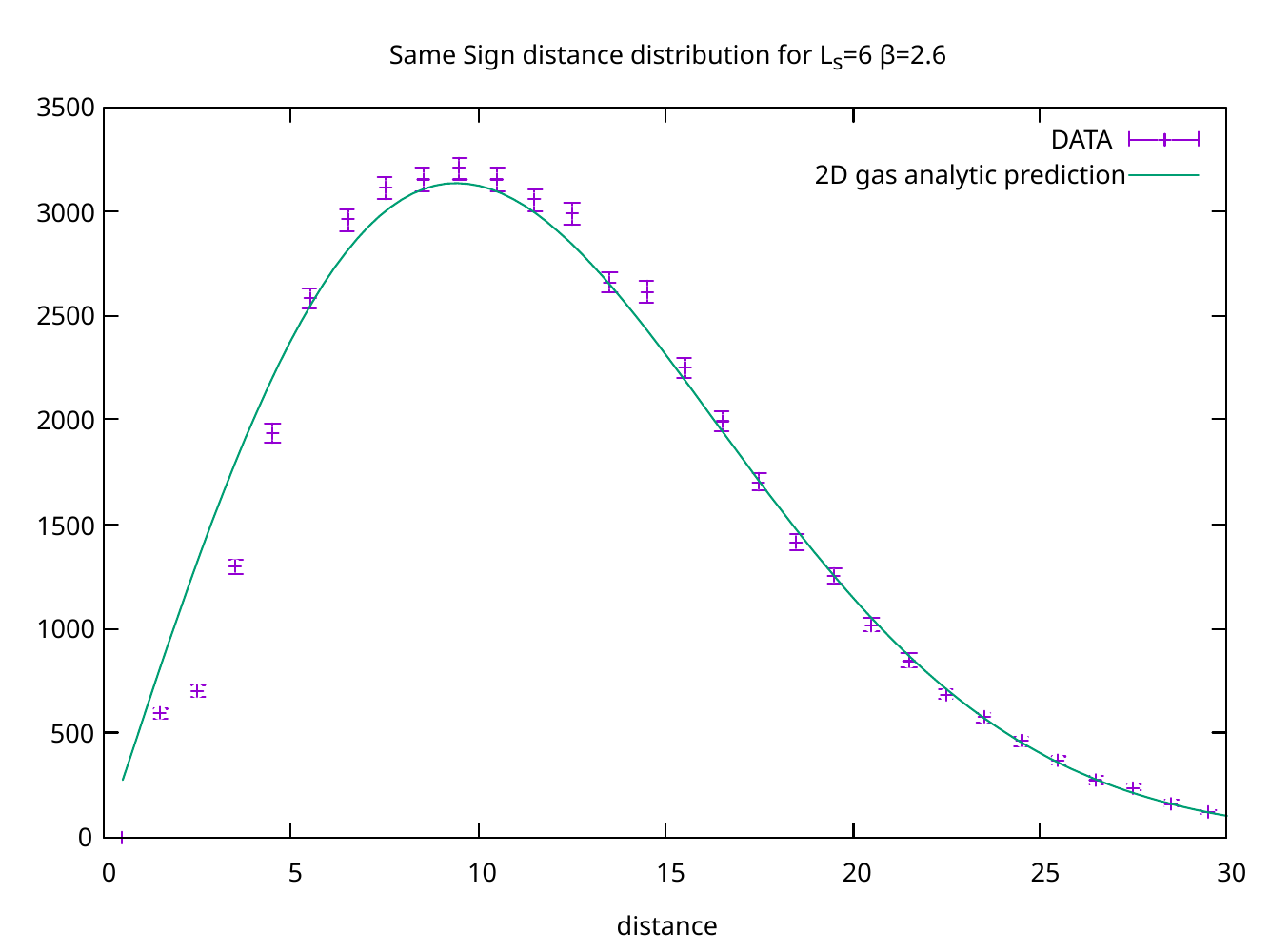}
\includegraphics[width=0.45\textwidth]{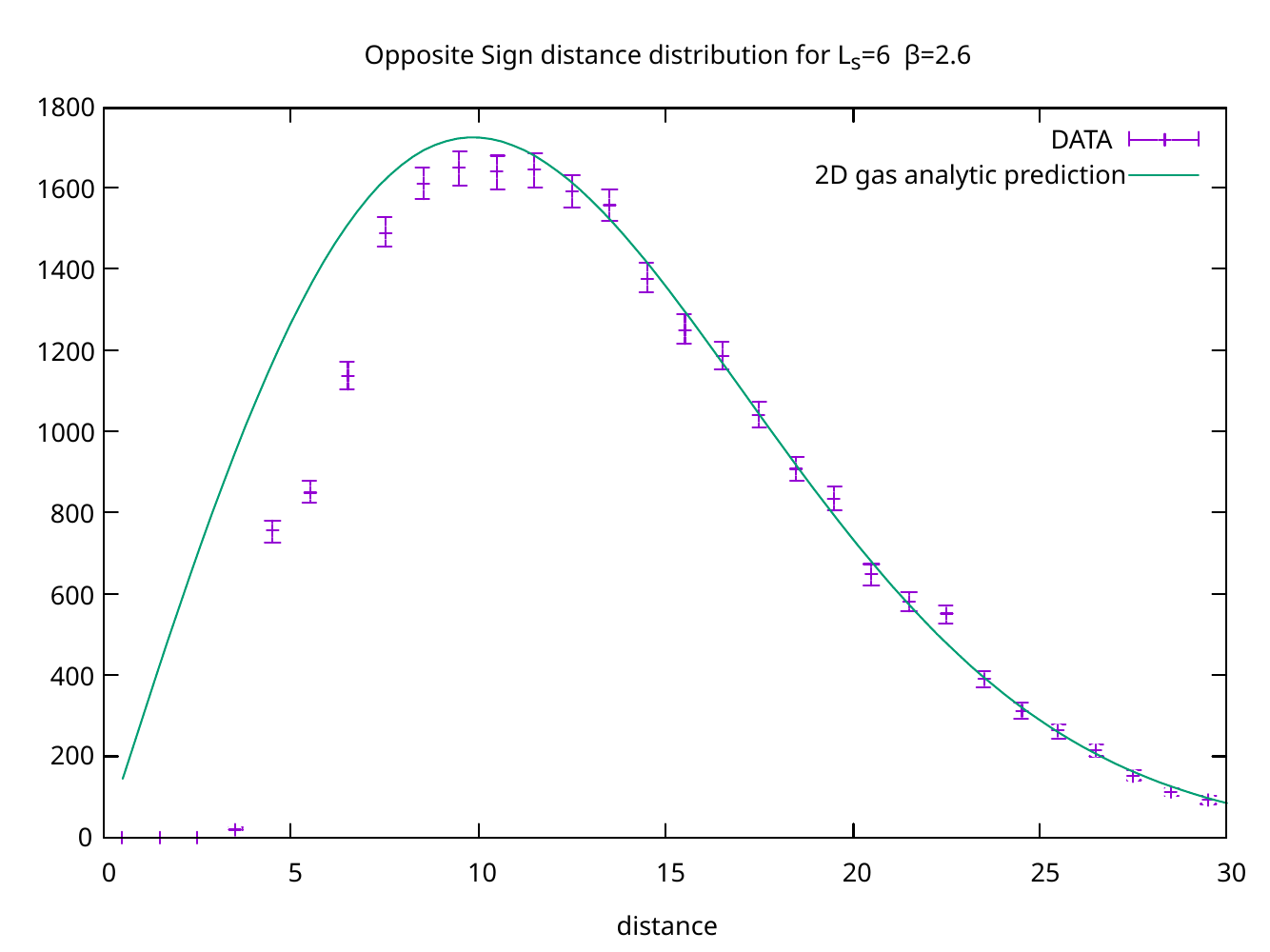}
\caption{Distribution of minimum distances from one object to the closest one depending on whether the closest neighbour has the same or opposite sign of topological charge (for $\Ls=6$, $\beta=2.6$). This is compared with the predictions of the Poissonian 2D gas (see text).}
\label{fig:distdistrib}
\end{figure}

As an example we consider again the case 
for $\Ls=6$, $L_t=104$ and $\beta=2.6$.
The histogram of distances is shown in \Figref{fig:distdistrib}.
The actual data are split into distance to an object of equal sign (left plot)
and to opposite sign (right plot). The results are compared to our analytic formula for a 2D Poissonian gas distribution
\begin{align}
2Ax\ e^{-B x^2}
\end{align}
The value of $A$ and $B$ are extracted from a fit  for distances  $30>x>15$. 
The results are quite informative. For the equal sign case, the value of $B$ corresponds to a density of $\rho_{2D}=0.00359(4)$, while the actual measured density is  $\rho_{2D}=0.00362(3)$. The total normalization is also consistent within errors. The curve matches  very well with the data at large distances. For $d<15$ there seems to be a small depletion at short distances compensated by a small enhancement at intermediate distances. It is hard to know if that is the result of the overlapping of profiles, a consequence of the peak selection or a genuine hard-core repulsion among structures. Notice that the prediction is based on the thin vortex approximation, while our objects have a size.

The opposite sign distance distribution is very different. The fit is also good at large distances but the extracted value of $B$ corresponds to a density of $\rho_2D=0.00328(6)$ which is somewhat smaller than the measured one. Furthermore, the overall integral of the curve is 17\% higher than the actual number of pairs. This can be seen as a decrease of the expected neighbouring opposite sign pairs at short distances.  
There are two related effects that can contribute to this phenomenon. On one side,  there might be a reduction of pairs   due to FI-AFI annihilation, a phenomenon to be expected as a result of the flow. Another possible effect is the strong decrease in peak height observed for close opposite-sign pairs, as shown in Fig.~\ref{fig:population}, making it harder to distinguish from low amplitude noise. 

Notice that even in this rather dilute case the peak of the  distance distribution is at $~10$ in
lattice units which is  less than  twice the size of the small torus (The prediction of our formulas is $1/\sqrt{\pi\rho_{2D}}=9.38$). Furthermore, 
in $\sim$18\% of all cases the distance is equal or smaller than $6$, which
would certainly alter the 2D shape of the object.  

The situation gets worse  as the density increases. For example for 
$\Ls=10$, $L_t=64$ and $\beta=2.6$, our formulas give a  peak of the minimum distance distribution  at $7.39$, which is less than the  actual torus size of $10$. It is clear that we expect a large number of cases with modified shapes due to the distortion induced by the neighbours. As we saw before, this would contribute to the spreading of the height distribution.

\begin{figure}[h!]
    \centering    
\includegraphics[width=0.95\textwidth]{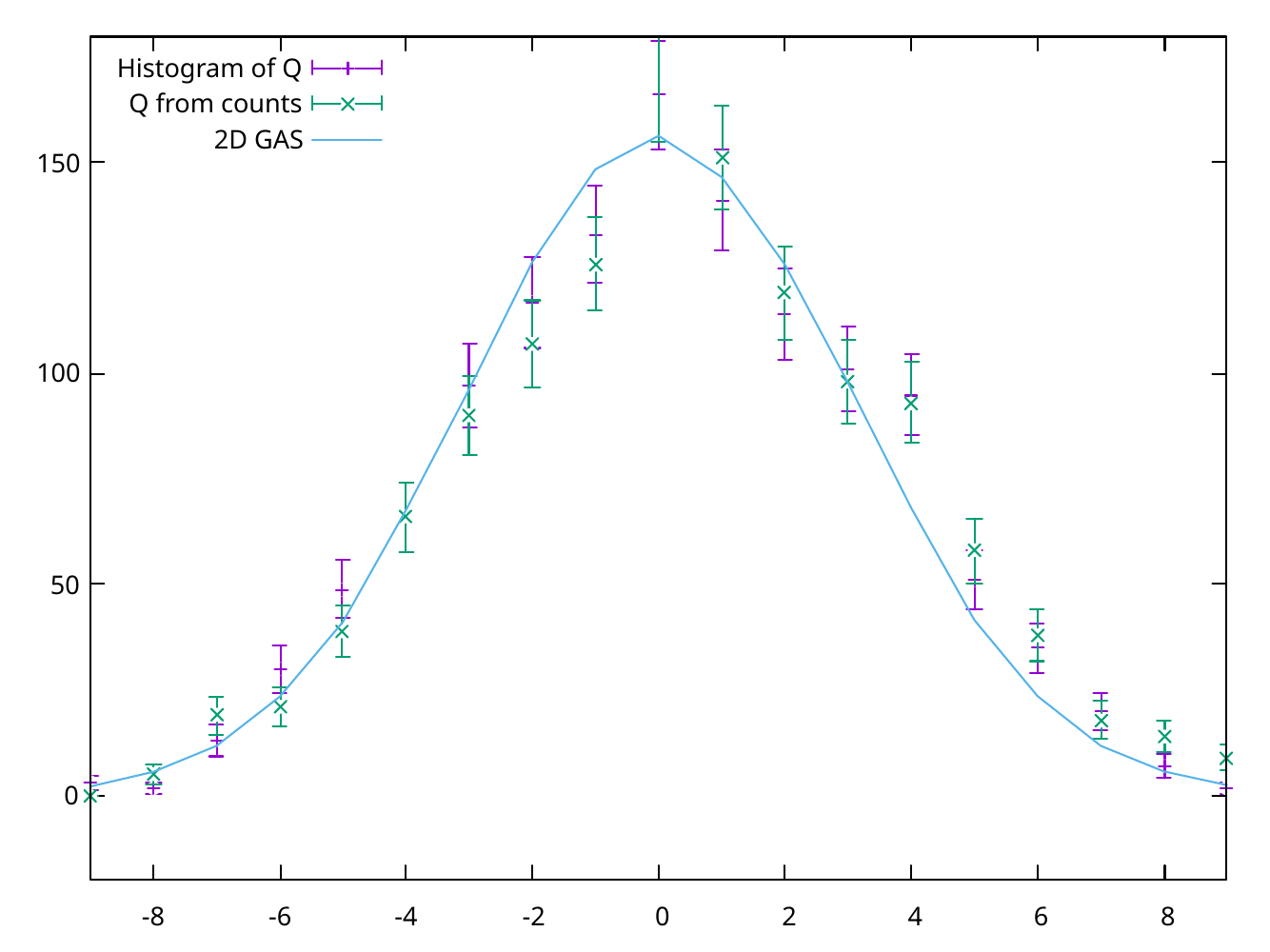}
\caption{Histogram of topological charges measured in our configurations at  $L_s=6$ $\beta=2.6$. The two sets of data points come from a direct measurement of the topological charge or by summing up the charges of the objects identified by our methods. The continuous line is the result of our 2D gas simulation. 
}
\label{fig:topdistrib}
\end{figure}

Another interesting distribution is that of the topological charge. The predictions of the 2D free gas of fractional instantons were given in Sec.~\ref{seq:semiclassical}. This quantity can be measured directly as the integral over the topological charge density and  compared, configuration by configuration, with the result of adding the integer or half integer charges  of  both instantons and fractional instantons  that result from our identification. This provides a test that the identification is correct. For our standard $L_s=6$ $\beta=2.6$ we display in Fig.~\ref{fig:topdistrib} the observed distribution using both methods compared with the distribution generated by  our 2D gas simulation.

\subsection{Density evolution as a function of $\beta$}
\begin{figure}[h!]
    \centering
        \includegraphics[width=0.49\textwidth]{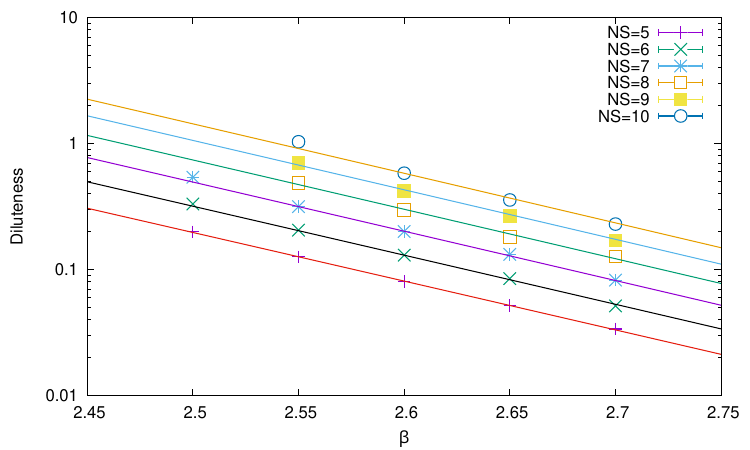}
	    \includegraphics[width=0.49\textwidth]{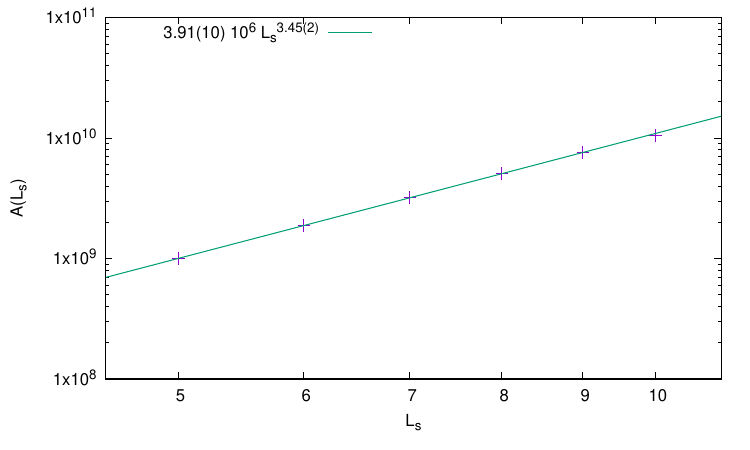}
	        \caption{Comparison of the measured value of the diluteness $D(Ls, \beta)$ with the semiclassical predictions.  LEFT: Comparison of the  data with the beta dependence in Eq.\eqref{semiclas}. RIGHT: Dependence of the prefactor with $\Ls$ and fit to Eq.~\eqref{prefact}.
            }
		    \label{fig:semiclassical_lattice}
		    \end{figure}

We will now confront our results concerning the evolution of the
density of FI and AFI as a function of the lattice coupling at fixed 
value of the small torus lattice size. The semiclassical prediction 
on the lattice implies that the density is proportional to the
probability of creating a fractional instanton, which is proportional
to the Boltzmann weight times a prefactor related to the number of
zero-modes of the adjoint Dirac operator (4 in our case), as follows
\be
\label{semiclas}
 A(L_s)\, \beta^2\, \exp\{-\beta S_L(L_s)/4\}\, ,
\ee
where the FI lattice action $S_L(L_s)$ approximates the continuum
result  of $4 \pi^2$ and its value can be obtained from the studies of a single 
FI given  in Section~\ref{seq:semiclassical}. Notice that Eq.~\ref{semiclas} 
is the same one that was used long time ago to
analyse the density of kink-like fractional instantons that appear in
a $ T_3\times R$ study~\cite{GarciaPerez:1993ab}. Instead of comparing the density
itself we will plot the 2D diluteness observable $D(L_s, \beta)$ (defined in Eq.~\eqref{dilutenessdef})
which is dimensionless but obviously has the same $\beta$ dependence
up to a change in the numerical prefactor $A(L_s)$. We will focus on the region of
diluteness less than 1, where one expects the semiclassical
approximation to work better. Our  results are given in
Fig.~\ref{fig:semiclassical_lattice} and  cover a large range of values of the
diluteness and of $\beta$ and $L_s$ values. For the calculation $Q=1$ instantons are
counted as 2 fractional instantons.

The lines are the results of a one parameter fit to Eq.~\eqref{semiclas}. 
Only the values of $\beta>2.5$ have been fitted and the
resulting $\chi^2$ values are of order 1. For smaller $\beta$ there is
a small excess. Notice that the data points follow very closely the fit 
lines which look like parallel straight lines, because they are dominated 
by the exponential dependence. The agreement is very remarkable and extends
also up to rather large values of $L_s$. 

The values of $A(L_s)$ resulting from the fit should approach  the 
predictions of the leading order renormalization group as explained in
the theoretical discussions of Sec.~\ref{subseq:Continuum_limit}. In the left part of Fig.~\ref{fig:semiclassical_lattice}  
we plot the resulting values in a log-log plot. We fit the values of $A(L_s)$ to  a power of 
$L_s$ which shows as straight lines in this plot:
\be
\label{prefact}
A(L_s)=A\, L_s^\alpha
\ee
The fit is very good  having $\sqrt{\chi^2}= 0.9$ and giving an exponent 
$\alpha=3.45(2)$ which is remarkably close to the prediction of 
leading order renormalization group $11/3=3.67$.  The fit gives the value of the coefficient $A=3.91(10)\, 10^6$.

In summary,  our data of low diluteness and high $\beta$ values 
follows the semiclassical predictions on the lattice and in the continuum. 

\begin{figure}[h!]
\centering
\includegraphics[width=0.8\linewidth]{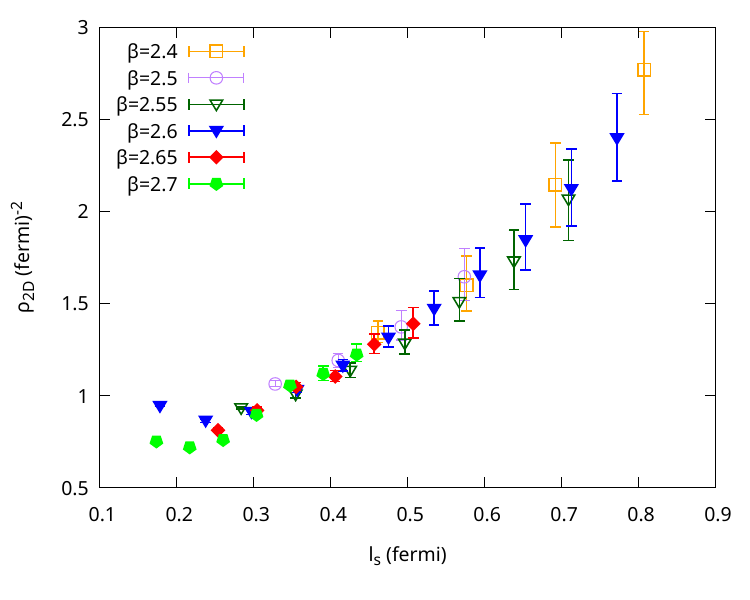}
\caption{Evolution of the density of objects $\rho_{2D}$ with the physical torus size $l_s$. Instantons are counted as two fractional instantons. The error bars contain the systematic part coming from the identification algorithm as explained at the end of Sec.~\ref{subseq:identifiation_algorithm}}
\label{fig:scaling_density}
\end{figure}

Another way to show that our results are actually continuum results is to
express the density $\rho_{2D}(l_s)$ as a function of the torus size $l_s$, 
both expressed in physical units. This is displayed in Fig~\ref{fig:scaling_density}. Notice that the results scale very well at low values of the density 
and for $\beta>2.5$. Deviations appear at larger values while deviations at small $l_s$ are due to lattice artifacts introduced once the small size of the $T_2$ box gets closer to the scale of the lattice spacing.

\subsection{Relation with the string tension}
\begin{figure}[h!]
\centering
\includegraphics[width=0.9\textwidth]{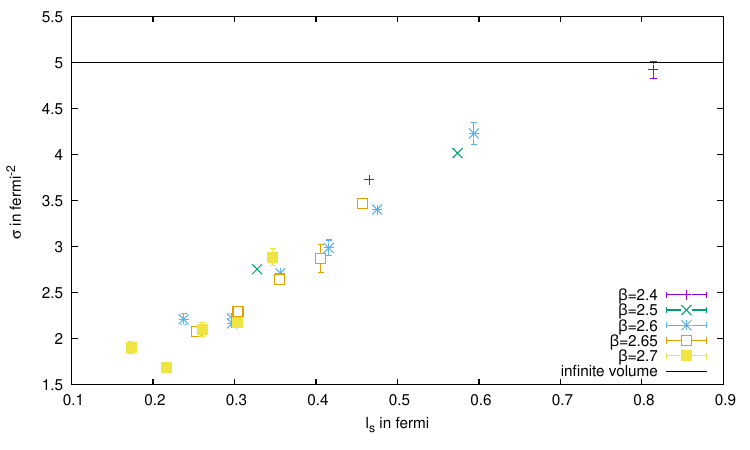}
\caption{Effective string tension defined through Creutz ratios (see text) as a function of the 2-torus size $l_s$. This has been obtained at the flow times of Tab.~\ref{tab:confs}.}
\label{fig:creutzflowed}
\end{figure}

In this subsection we show our results concerning Creutz ratios and
string tension. Our goal is to verify the predictions of the
semiclassical formulas which establish a relation between the
two-dimensional density of fractional instantons and the string tension. 
The string tension  in lattice units is given by the infinite size
limit of square Creutz ratios. It is clear that this limit cannot be
attained at finite volume. There are indeed distortions to the area law 
when the size of the loop approaches the size of our large torus. Thus
we are limited to analysing square Creutz ratios $\chi(R)$ up to sizes 
at most one fourth or one fifth of the size of the large torus.
On the opposite hand, the value of $R$ should not be smaller than $L_s$
since that is the size of the fractional instanton at low densities. 
In Fig.~\ref{fig:creutzflowed} we display the effective string tension in
physical units, defined as the average of Creutz ratios in
the interval $R \in [ 1.1\, l_s , 2.0\, l_s ] $, as a function of $l_s$ in
fermi. Although the results show a certain degree of dispersion the
growth with $l_s$ is obvious. For comparison we include a horizontal
line which marks the value of $5$ fermi$^{-2}$ corresponding to the
infinite volume string tension used as a unit for the lattice spacing.

\begin{figure}[ht!]
\centering
\includegraphics[width=0.9\textwidth]{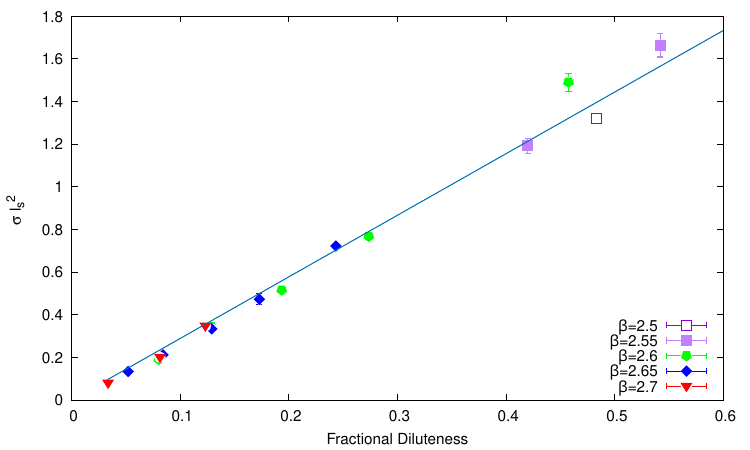}
\caption{Plot of the effective string tension versus the 2D density of fractional instantons. Both quantities are multiplied by $l_s^2$. The continuous line is 2.89 x. (Results at flow times of Tab.~\ref{tab:confs})}
\label{fig:correlation}
\end{figure}

The simple Thin Abelian Vortex Approximation (TAVA) predicts that the string tension is given by two times the density 
of fractional instantons. To test this, we compare the effective string tension measured from Creutz ratios as explained above with the density of fractional instantons and anti-instantons. A nicer plot is obtained if we multiply both the density and the string tension by $l_s^2$.
Our results are plotted in Fig.~\ref{fig:correlation}. The horizontal axis, labelled fractional diluteness, is given by the 2D density times $l_s^2$. 
We emphasize that $Q=1$ instantons are not included in the count. The linear correlation of the two quantities is obvious from the figure. 
However, the best fit to the proportionality constant is 2.89 rather than 2. This can be due in part to the fact that the range of Creutz ratios is rather small and the effect of the finite size of the fractional instantons shifts these to a value between 2 and 2.4, as verified in our measurements on smooth configurations. The higher value might indicate that the  density has been reduced due to annihilation of pairs during the flow (we will see the effects on the flow in the following section).

The results that we have presented show two remarkable facts. First of all, 
the fact that the effective   string tension appearing
in Fig.~\ref{fig:creutzflowed} at $l_s\sim 0.8$ fermi is indeed very close 
to the value of the string tension at infinite volume. On the other hand we have  clearly shown that the very same effective string tension is proportional to the density of fractional instantons. 
It is the combination of these two facts that was taken as an indication that the Confinement property of the infinite volume theory has its origin in fractional instantons.

Nonetheless, we realize  that our estimates of both the string tension  and densities are subject to corrections of different nature which should be clarified. We already emphasized that our estimate of the string tension is based on Creutz ratios which are not asymptotically large. This is expected to give estimates that are larger by 10 to 20 \% (as we learned from the smooth configuration Fig.~\ref{fig:CreutzCompare}).  A much more important issue is the effect of the flow. This is expected to reduce the number of objects altering the densities. 
  For the overimproved flow the
reduction comes mainly from annihilation of nearby
 fractional instanton-antiinstanton pairs. For Wilson flow one also expects the evaporation of small instantons, although this is not
expected to affect the string tension. In the next subsection we will
analyse these effects.

\subsection{Flow dependence of the results}
 
\begin{figure}[ht!]
\centering
\includegraphics[width=0.9\textwidth]{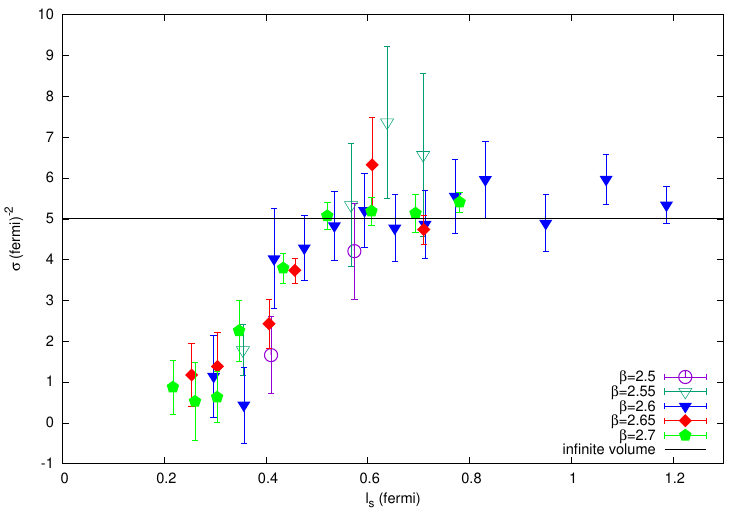}
\caption{Evolution of the string tension with $l_s$ as obtained at zero flow time from Wilson loops and Creutz ratios APE-smeared over the directions transverse to the plane of the loop ($t_{gf}=0$).}
\label{fig:unflowed_Creutz}
\end{figure}

 Indeed, it is possible to compute the string tension for the
 configurations directly without the flow. There are various ways in which this can be done. A rather standard way led to the results presented before in Fig.~\ref{fig:fit_stringtension}. From this Figure it is clear that the string tension grows for small torus sizes and seems to stabilize to its infinite volume value at $l_s\sim$ 0.8 fermi, in agreement with our measurements done on the flowed configurations. One can also  compute the string tension from square Creutz ratios $\chi(R)$. The procedure is to extrapolate the results to infinite size. In the standard infinite volume case the Creutz ratios show a linear dependence in $1/R^2$ for large enough  sizes with a coefficient which is not far from $\pi/6$, which is the expected Lüscher term~\cite{Luscher:1980fr} arising universally from the
 transverse fluctuations of a string with contour on the Wilson loop.
 One expects the behaviour to be modified in our setting for large 
 values of $R$ and small torus sizes $l_s$, since the fluctuations are
 bounded. In the absence of a more precise formula, we have simply
 performed a linear and a quadratic fit to the square Creutz ratios
 $\chi(R)$  as a function of $1/R^2$. The results are collected in
 Fig.~\ref{fig:unflowed_Creutz}. 
Errors are big  because large  Creutz ratios are very noisy
quantities, but it is clear that the results are consistent with the previous estimates. 
Using alternative methods, which assume a certain form of the static quark anti-quark potential, one obtains similar results presented in Fig.~\ref{fig:fit_stringtension}.
Once more  at around $l_s~\sim 0.7\ \mathrm{fermi}$ the string tension
reaches a value which is very close to the asymptotic one.   

\begin{figure}[ht!]
\centering
\includegraphics[width=0.9\textwidth]{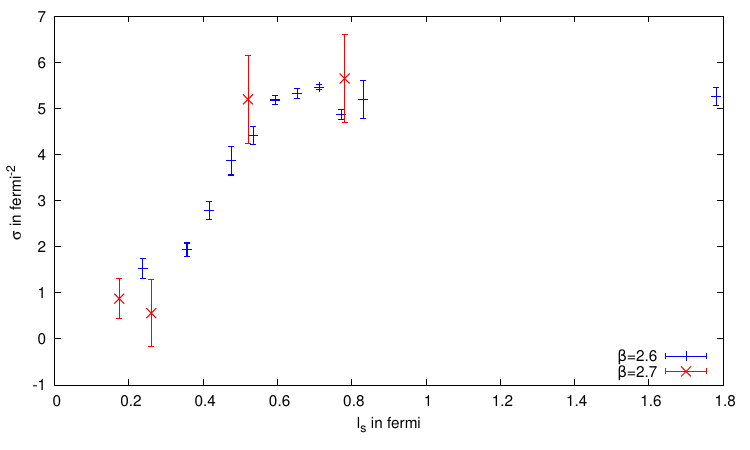}
\caption{String tension as a function of $l_s$ as obtained by extrapolation of the flowed Creutz ratios to zero flow time.}
\label{fig:st_extrap}
\end{figure}

Another alternative calculation makes use of the technique employed in 
Ref.~\cite{Gonzalez-Arroyo:2012euf} which gave very good results. The idea is
based on the fact that Creutz ratios are lattice approximations to 
the double derivative of the logarithm of the Wilson loop. These are 
ultraviolet finite observables. Hence,  one expects that if one
applies gradient flow for very small times $\tau$, it is possible to
extrapolate back to zero flow time without the occurrence of
singularities. Furthermore, even at very small flow times the
fluctuations of the Creutz ratios are strongly suppressed. The
question then is what is the correct functional dependence to be used 
for the extrapolation. In the continuum within perturbation theory 
the flow time $\tau$ introduces  a damping of momenta in propagators 
of the form $\exp\{-\tau p^2/2\}$. Creutz ratios are indeed computable
in perturbation theory and one can use the flowed perturbative formulas 
to predict the $\tau$ dependence of the ratios. The resulting formula
has the form
\be
\chi(R,\tau)=\chi(R,0) (1-e^{-b/\tau}) 
\ee
This shows a plateau and then a decrease. We fit our ratios to this
formula within a region in which the ratios  have only decreased by a
few percent. Typically this occurs when the smearing radius $\sqrt{8
\tau}$ is 0.1-0.5$R$.  Still, the decrease in statistical errors is dramatic.
The resulting $\chi(R,0)$ values are treated in the same way as the 
ones obtained directly. 
Unfortunately the method requires storing the evolution of all
Creutz ratios and configurations for the corresponding interval of
flow times. Thus, we used the method only as a test for a reduced
sample of configurations and parameters. The results are given in
Fig.~\ref{fig:st_extrap}. The displayed errors include both statistical and systematic added in quadrature. The latter are estimated by the spread of the results as applied to both smeared and smeared Wilson loops. 
The values seem to be slightly higher 
that the ones obtained previously with other methods, but the pattern is the same: there is a rise of the string tension  with
$l_s$ ending at around $0.7\ \mathrm{fermi}$. The values then flatten and seem to tend smoothly to infinite volume. Notice that we have added the result obtained at $\Ls=30$ corresponding to $l_s\sim 1.8$. The value there is slightly higher than the nominal value of 
5 $\mathrm{fermi^{-2}}$, but this value is taken by results of other authors which have errors of similar size.

As we have seen, within the errors of our estimates, the flow has a rather small effect on the string tension shape and values. A much more delicate point is how does the flow affect the density of fractional instantons. It is certainly expected that the density decreases with the flow time. This follows from the mechanisms described earlier in subsection~\ref{subseq:MC_Filtering}: Pair annihilation and fusion and evaporation of $Q=1$ instantons. The latter aspect does not affect the density of fractional instantons, but fusion does. For Wilson flow one expects that nearby pairs of equal sign fractional instantons fuse to produce a single $Q=1$ which does not contribute to the density. 

\begin{figure}[H]
\centering
\includegraphics[width=0.8\textwidth]{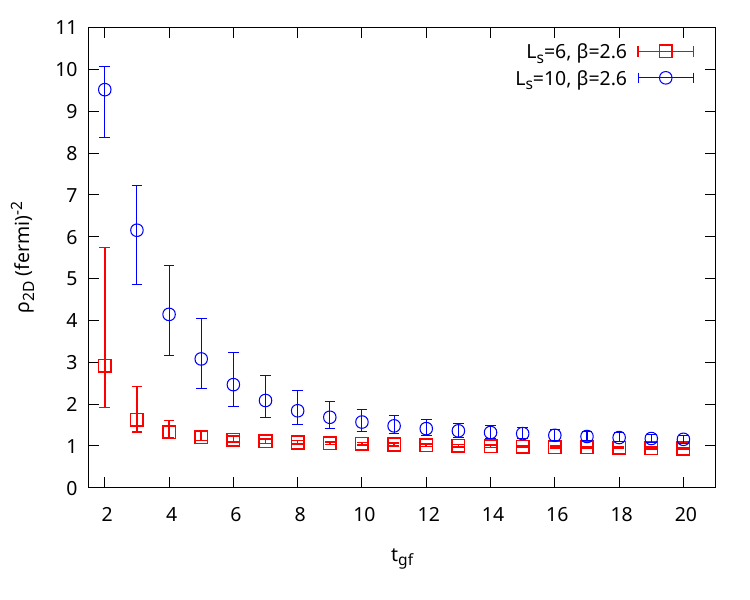}
\caption{Dependence of the fractional instanton density $\rho_{2D}$ on the flow time $t_{gf}$ for two different ensembles at $l_s=0.24$ for the small lattice and $l_s=0.59$ for the large one.}
\label{Fig:scaling_dens_flow}
\end{figure}

It is impossible to perform a measurement of the density at zero flow time, since the signal is completely covered by the low frequency noise. However, what we can do is to monitor how the density depends on the flow time to pin down its effect on the correlation between this density and the measured string tension. For this purpose, we made a detailed study on the gradient flow dependence for a particular set of ensembles. The results are collected in  Fig~\ref{Fig:scaling_dens_flow}. As expected, the fractional instanton density decreases monotonically with the flow time $t_{gf}$. At small flow times, the errors are quite large, and tend to give much higher values because the configurations are polluted by noise which is misidentified as fractional instantons. For large flow time, the noise has decreased enough such that the identification algorithm performs well. The mild dependence left at large flow time is due to the effects mentioned above. From Fig.~\ref{Fig:scaling_dens_flow} we can see that our choice of $t_{gf}=15$ for $\beta=2.6$ in Tab.~\ref{tab:confs} is such that we end in the region where the $t_{gf}$ dependence is mild even for still quite large torus size $l_s\sim0.6$.

\begin{figure}[H]
\centering
\includegraphics[width=0.80\textwidth]{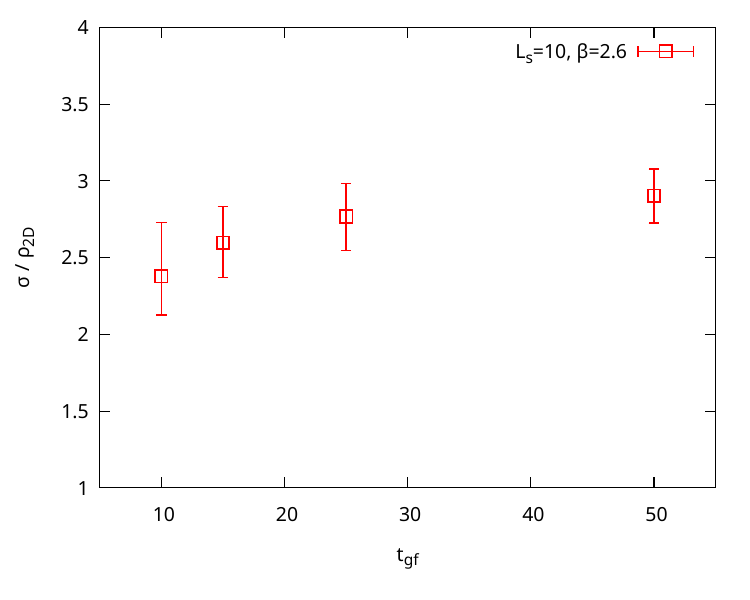}
\caption{Flow time dependence of $\sigma/\rho_{2D}$, the ratio between the effective string tension and the fractional instanton density. The plot shows that the ratio grows rather mildly  during the flow, thus exhibiting  a strong correlation between the two quantities.}
\label{Fig:flow_string_dens}
\end{figure}

For these ensembles we have also measured 
the string tension by means of the average of Creutz ratios of size $R\in [1.1,2] l_s$ for various values of the flow time. The ratio of this result with the density is depicted in 
Fig.~\ref{Fig:flow_string_dens} for one of the ensembles at the larger value of $l_s$. Despite the big change  in density, the ratio shows a rather mild growth. Indeed, this growth is consistent with the interpretation given earlier to 
the rather large slope obtained in Fig.~\ref{fig:correlation} at $t_{gf}=15$.
The ratio at smaller flow times comes closer to the expected value slightly above 2.

\section{Going to larger torus sizes}
\label{seq:larger_torus}
In the previous sections we have studied  the evolution of the 
Yang-Mills system in $T_2\times R^2$ as a function of the size of the torus. Our results have concentrated upon the low density regime where one expects the semiclassical predictions to hold. We have verified that this is indeed the case  by checking the various distributions and observables.  As the size grows, the density increases and our results predict when the gas ceases to be dilute. This occurs as the size becomes roughly larger than the 0.6 fermi. This matches with the estimate made many years ago~\cite{RTN:1993ilw,Gonzalez-Arroyo:1995ynx} 
when studying the evolution within a 
$T_3\times R$ geometry. Extending the study of  the system 
beyond that point is very important in order to relate the understanding developed in the semiclassical analysis with the behaviour of the system at infinite volume. In particular, our previous study has traced the origin of a non-zero string tension to the presence of fractional instantons that behave as center vortices and induce Confinement.  Interestingly enough at the edge of the semiclassical region the value of the string tension is not far from the value 
measured at infinite volume by other research groups. It is this similarity and the subsequent mild evolution of the string tension towards its infinite volume value that led one of the present authors and his students to propose a possible explanation~\cite{Gonzalez-Arroyo:1995ynx} that can be referred to as  Fractional Instanton Liquid Model (FILM). It is important to review the main elements of the proposal to determine what specific predictions arise from this scenario in studying the evolution at larger torus sizes. 

The quantum fluctuations imply that larger fractional instantons are more probable than small ones. This is not surprising since the same happens for instantons, as established by 't Hooft in his famous semiclassical analysis~\cite{tHooft:1976snw}. However, the maximum size is limited by the torus. Thus, within the semiclassical regime the fractional instanton fill in the torus completely and they simply become more dense as a 2D gas. However, 
as the gas ceases to be dilute the distance among objects becomes competitive with the torus size and the anisotropy of the distribution starts to disappear. It is to be expected that eventually the distance among objects is a physical scale determined in terms of the characteristic scales of the theory, like the Lambda parameter, and decoupled from the physical size of the geometrical container. 
It is in this way that one can understand that the result tends to a limit at infinite volume and isotropy is recovered. It makes a lot of sense that this occurs rather quickly once the gas ceases to be dilute. 

Notice that fractional instanton do not exist isolated (i.e. surrounded by the classical vacuum) so that their size is directly connected to the distance to its neighbours. In the dilute setting the neighbours are the replicas induced by the torus geometry, but in the non-dilute case the closest neighbours might come from all directions.    The picture that emerges as the size becomes larger is that of a $4D$ liquid. 

\begin{figure}[H]
    \centering
    \includegraphics[width=0.49\linewidth]{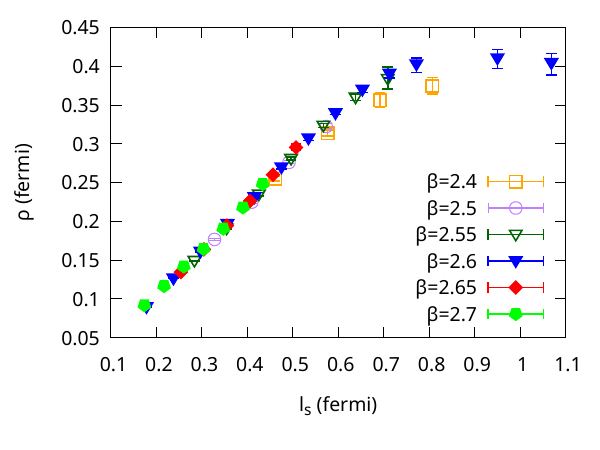}
    \includegraphics[width=0.49\linewidth]{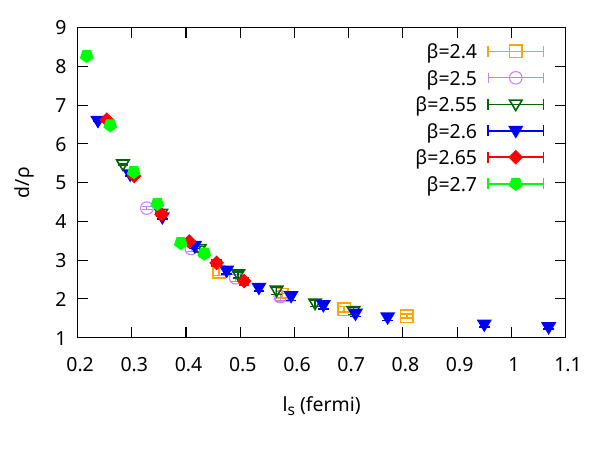}
    \caption{LEFT: Evolution of the size of the fractional instantons $\rho$ as a function of the size of the $T_2$ box $l_s$. In this case we discarded completely the contribution coming from the instantons as they have a different parameter space. RIGHT: Evolution of the mean distance between fractional instantons and their size. The data hints that at large $l_s$ the size of the fractional is given by their mean distance $\rho\sim d$.}
    \label{fig:scaling_size}
\end{figure}

In the present work we have  monitored both the size and the distance between the fractional instantons as a function of the torus size in physical units. Our results are collected in Fig.~\ref{fig:scaling_size}. On the left plot  we can see how at the edge of the semiclassical regime $l_s\sim 0.7$ fermi, the size of the fractional instantons  departs from the naive linear dependence $\rho\sim l_s$. In fact, it seems that this size $\rho$ couples to the mean distance between topological objects, displayed in the right plot. Furthermore, this distance seems to scale and flatten out in physical units in agreement with the expectations of the model.

Unfortunately, our methods of identification of fractional instantons rely heavily on the height distribution of the 2D integrated profile $S_2$. This is highly inadequate to study the 4D liquid distribution expected to hold at larger torus sizes. Four-dimensional  topological charge/action density heights, sizes and distance to neighbours have to be used for a proper identification in this 
new context. 

There is an additional difficulty 
because at  high densities the proportion of ordinary $Q=1$ instantons  grows. However, one can still think of  these instantons as overlapping pairs of fractional instantons. Instantons of much smaller size can still appear, but they are suppressed by having a smaller quantum weight.  In any case
distinguishing |Q|=1 instantons from fractional instantons remains a challenge. One might still use the ratio of action per peak as an indicator of the percentage of fractional instantons versus instantons as was done in the  $T_3\times R$ case~\cite{GonzalezArroyo:1995ex}.

Another technical complication which becomes very relevant at high densities is the method used to suppress high frequency quantum fluctuations. We have commented about the effects of Wilson flow and using alternatives is particularly relevant for the high density case.

Concerning the main observables  there are also important changes. It
is clear that the density of fractionals have to be computed in 4D
$\rho_{4D}$ instead of 2D. Thus, at fixed lattice spacing and large
torus size, the number of fractionals should grow like $\Ls^2$. One has also to revisit the expected relation between the density of fractional instantons and the value of the string tension.  Given a large Wilson
loop one can fix the minimal 2D surface having the loop as boundary, having area $A$. 
It is really the number of fractional instantons cut by this surface
what really matters for the calculation of the expectation value and 
assuming a fixed fractional instanton size $l_I$, this number should
be proportional to $l_I^2\times \rho_{4D}  \times A$. Still the surface 
will sometimes cut the fractional instantons at the center and
sometimes at the edge. It is also much more doubtful that one can use a
purely abelian methodology to determine the effect of the fractional
instantons cut by the plane on the expectation value of the Wilson
loop. After all they are neither thin nor abelian. Our naive estimate 
done in Ref.~\cite{Gonzalez-Arroyo:1995ynx} incorporated these effects by applying  a certain 
reduction factor $0<x<1$ to the size of the fractionals making the
relation $\sigma= 2 x^2 l_I^2\times \rho_{4D}$. This implies a loss of
predictive power although $x$ is expected not to depend on the lattice
spacing. One can envisage a possible independent determination of $x$
in the fashion presented in Ref.~\cite{Gonzalez-Arroyo:1996eos}, namely creating a
configuration as a gas of  fractional instantons and measuring the
string tension in it.

\section{Conclusions and future prospects}
\label{sec:conclusions}
In this paper, we have explored the evolution of the SU(2) Yang-Mills system
living in a $T_2\times R^2$ space-time as a function of the size of
the 2-torus $l_s$. Due to the twisted boundary conditions, the torus is permeated by one unit of `t Hooft flux
which prevents the system from breaking the $Z_2\times Z_2$ center
symmetry at small torus sizes. In this regime  one expects that many
long-range properties of the system are well-explained by
semiclassical methods. In particular, this is given by a gas made of 
certain local solutions of the classical equations of motion discovered many
years ago in Ref.~\cite{GonzalezArroyo:1998ez} and named vortex-like fractional instantons.
These objects extend over the $T_2$ torus and are localized in the
$R^2$ plane. Thus, the gas is actually a two-dimensional one. We have 
explained the semiclassical expectations on the  properties of this gas and its 
implications on global observables of the system such as the string tension or 
the topological charge density. We have then made a very extensive
investigation of the system using lattice gauge theory methods. In
particular, we have Monte Carlo generated  many configurations, covering
a large range of lattice sizes and lattice
couplings, in order to test both the semiclassical behaviour of the
lattice system as well as its continuum limit.  The results presented
in Sec.~\ref{seq:results} confirm that for torus linear sizes of  0.2-0.6 fermi 
($l_s\sqrt{\sigma}\in[0.09,0.27]$) the system behaves as expected. 
The density of objects matches closely the predictions of the theory 
and allows the determination of a basic parameter measuring the
strength of quantum fluctuations around the solution. The parameter
scales according to the predictions of the beta function of the system
confirming that the behaviour is actually a continuum result.
Particularly interesting is the evolution of the string tension as a
function of the torus size. The Wilson loop around a single vortex-like
fractional instanton gives -1, and hence it behaves as a center-vortex.
Thus in the dilute situation the system amounts to  a two-dimensional
$Z_2$ gauge theory, which is a confining  theory with a string tension
given by twice the density of the gas. We call this limit the
thin-abelian-vortex approximation (TAVA). In our case the vortices have a
size, within which the solution reveals its non-abelian character. This
produces alterations for small Wilson loops and higher densities that
our results exhibit. For low densities the gas displays the expected
properties of a Poissonian gas not only in the number-count but also
in the spatial distribution. 

The previous results have been obtained after applying gradient flow 
to the Monte Carlo configurations. This process erases short distance
fluctuations up to  a distance scale called smearing radius. We have
chosen this radius of the order of the size of the fractional
instanton solution, allowing its easy identification but without
alter the distribution of these objects at  larger scales. The
methodology allows the obtention of smooth images of the configurations
giving rise to spectacular videos, which we make available in our
supplementary material.  

We have also described in the previous section some results obtained
for larger torus sizes ($l_s$>0.6 fermi). At these sizes the gas stops
to be  dilute and the semiclassical analysis is far more complicated. 
Understanding the transition to infinite volume is crucial to relate
the physical insight and understanding achieved in our study on the
origin of properties such as Confinement  to those that hold at T=0 in
the infinite system. An important hint is that the value of the string
tension attained at the edge of the semiclassical regime is very close
to the one known to hold at infinite volume. This is the same type of
finding that was obtained many years ago~\cite{GarciaPerez:1993ab, GarciaPerez:1993jw}
when following a similar strategy within a $T_3\times R$ geometry
rather than our  $T_2\times R^2$ path. In that case the semiclassical 
description is given in terms of a one-dimensional gas of kink-like
fractional instantons which play a crucial role in restoring the
center-symmetry and generating a non-zero string tension. The search
for an explanation led to the proposal of a model of the Yang-Mills
vacuum basically describable as a fractional instanton liquid
model~\cite{GonzalezArroyo:1995zy,GonzalezArroyo:1995ex}. The 2D gas develops into a 4D liquid whose
characteristic scales are given by physical values not involving the
size of space. In this way the full 4D rotational invariance is
recovered and the boundary conditions become irrelevant. This becomes
an updated version of the scenario proposed many years ago in Ref.~\cite{Callan:1977qs}
but with the singular merons being replaced by bona-fide smooth
solutions of the classical equations of motion: fractional instantons. 
Exploring this possibility with our methods is very appealing. However, there
are considerable improvements that have to be implemented to achieve
this goal. In the present work the identification of objects is greatly simplified by the
fact that the size is dictated by the geometry and the relevant
location is two-dimensional. Furthermore, the gradient flow method could
alter the shapes and distribution given that at higher densities the size of
the objects and the distances among them have similar values. There are
known ways to  achieve ultraviolet noise reduction which are less 
distortive that the Wilson flow method, mostly employed in our present
paper. Unfortunately, they involve a considerable upgrade in the
resources needed. In our opinion an investigation of the larger sizes 
should incorporate all these  new methods to be really  effective. Our
plan is to start working along this direction, but its completion will
probably take some time.

It is important to emphasize that our results do not necessarily imply that vortices underlie the Confinement property in the infinite volume theory. The vortex-like structure is enforced by the particular choice of geometry used. In the same way the $T_3\times R$ geometry led to similar conclusions, but the structure of the semiclassical objects was very different (kink-like FI).
Similarly an $S_1\times R$ geometry leads naturally to calorons and monopoles. Several recent works have studied how to interpolate among different geometries~\cite{Poppitz:2022rxv,Wandler:2024hsq,Hayashi:2024yjc}. In this respect geometry plays a similar (biasing) role to gauge fixing in more traditional lattice studies of confinement~\cite{Ambjorn:1999ym}. In our opinion, the detailed structure of the fractional instanton liquid is largely unknown and there are at present no strong reasons to believe that they are predominantly arranged into 0-1-2-3-4 dimensional substructures. 
What is crucial, with respect to the abelian-like  gauge fixing  arguments, is that these objects carry topological charge and 
are locally fully-fledged solutions of the non-abelian Yang-Mills equations of motion. 

There are other possible extensions of our work which can be addressed
and that we plan to consider. A very basic one is going over to larger
number of colours $N$. One of the
nice features of the fractional instanton scenario is that their
semiclassical weight  survives the large $N$ limit.  A few tests have been done and included in the
present paper. This has many attractive features. The generalization of 
the vortex-like fractional instantons to higher $N$ are
known~\cite{Montero:1999by}. The objects are wider and easier to distinguish
from ordinary instantons. Furthermore, the proportion of instantons to
fractional instantons is expected to drop considerably. Other authors have also conducted studies of different nature centered on the SU(3) case~\cite{Itou:2018wkm,Anber:2025yub}.

\section*{Acknowledgments}
G. B.\ and I.S.\ are  funded by the Deutsche Forschungsgemeinschaft (DFG) under Grant No.~432299911 and 431842497. 
A.G-A acknowledges
support by the Spanish Research Agency (Agencia Estatal de Investigaci\'on) through the grant IFT
Centro de Excelencia Severo Ochoa CEX2020-001007-S, funded by MCIN/AEI/10.13039
/501100011033, and by grant PID2021-127526NB-I00, funded by MCIN/AEI/10.13039/
501100011033 and by “ERDF A way of making Europe”.
Part of the computing time for this project has been provided by the compute cluster ARA of the University of Jena.
The authors gratefully acknowledge the Gauss Centre for Supercomputing e.~V.\ (www.gauss-centre.eu) for funding this project by providing computing time on the GCS Supercomputer SuperMUC-NG at Leibniz Supercomputing Centre (www.lrz.de).

\bibliographystyle{JHEP}
\bibliography{main}
\appendix
\section{Dependence of the distributions on gradient flow parameters}
\label{sec:gfdependence}
\begin{figure}[H]
    \begin{subfigure}[t]{0.3\textwidth}
    \includegraphics[width=1.1\textwidth]{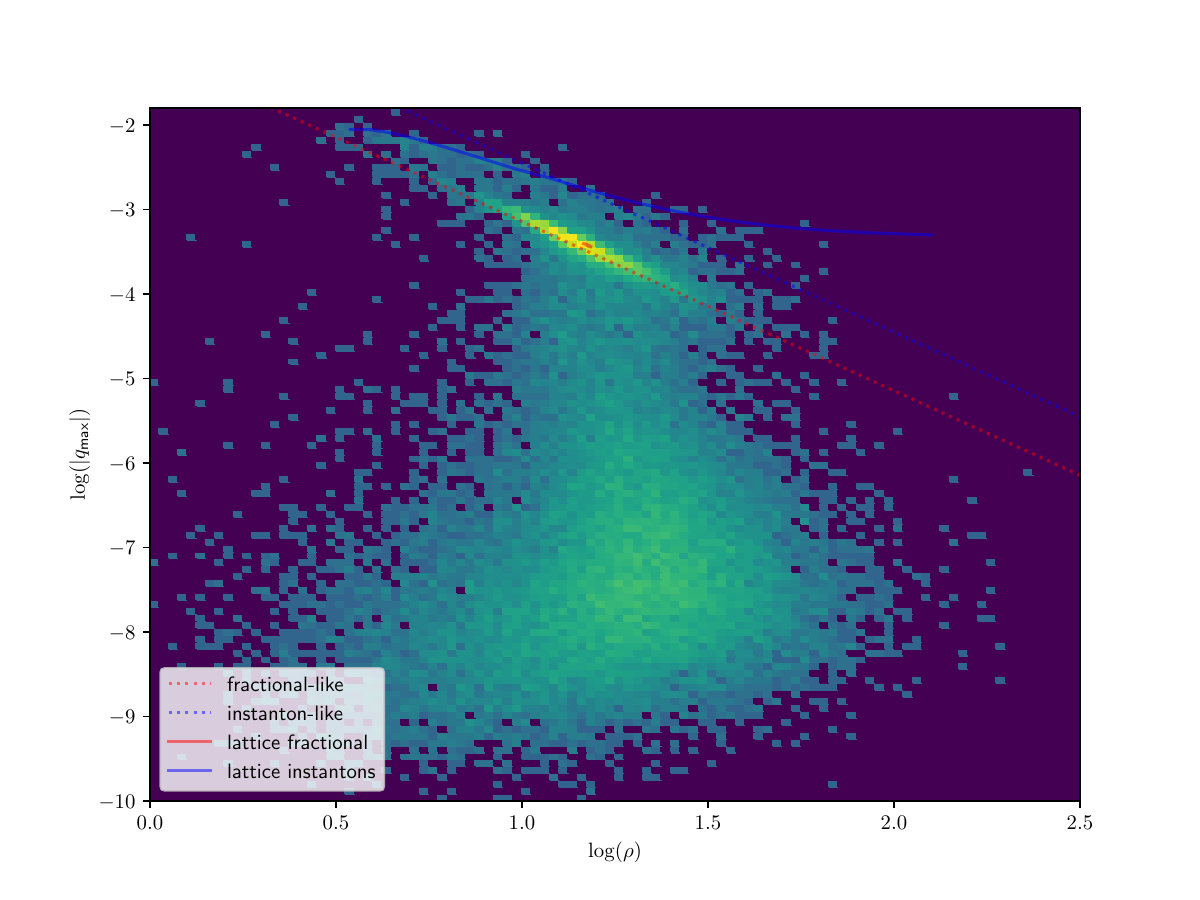}
    \caption{$t_{gf}=4$}
    \end{subfigure}
        \begin{subfigure}[t]{0.3\textwidth}
    \includegraphics[width=1.1\textwidth]{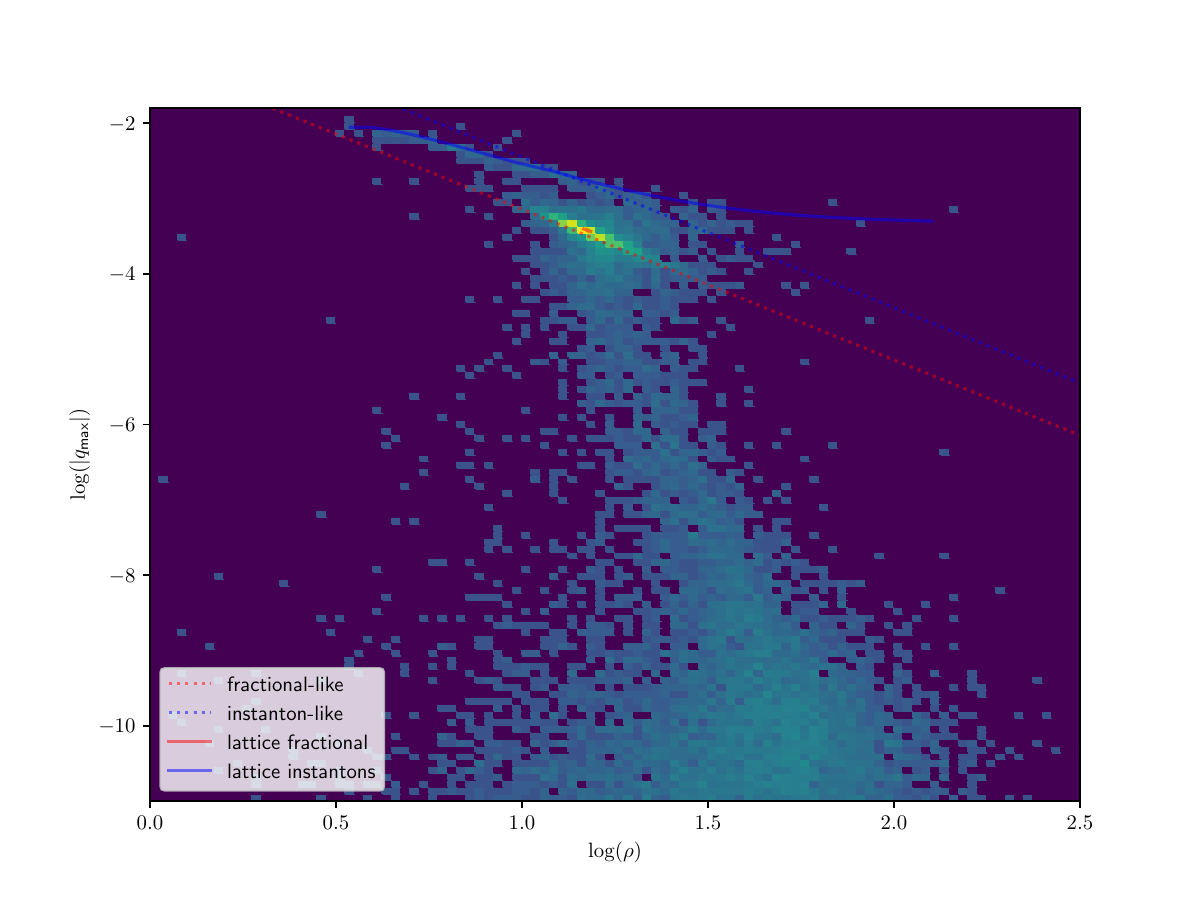}
    \caption{$t_{gf}=8$}
    \end{subfigure}
        \begin{subfigure}[t]{0.3\textwidth}
    \includegraphics[width=1.1\textwidth]{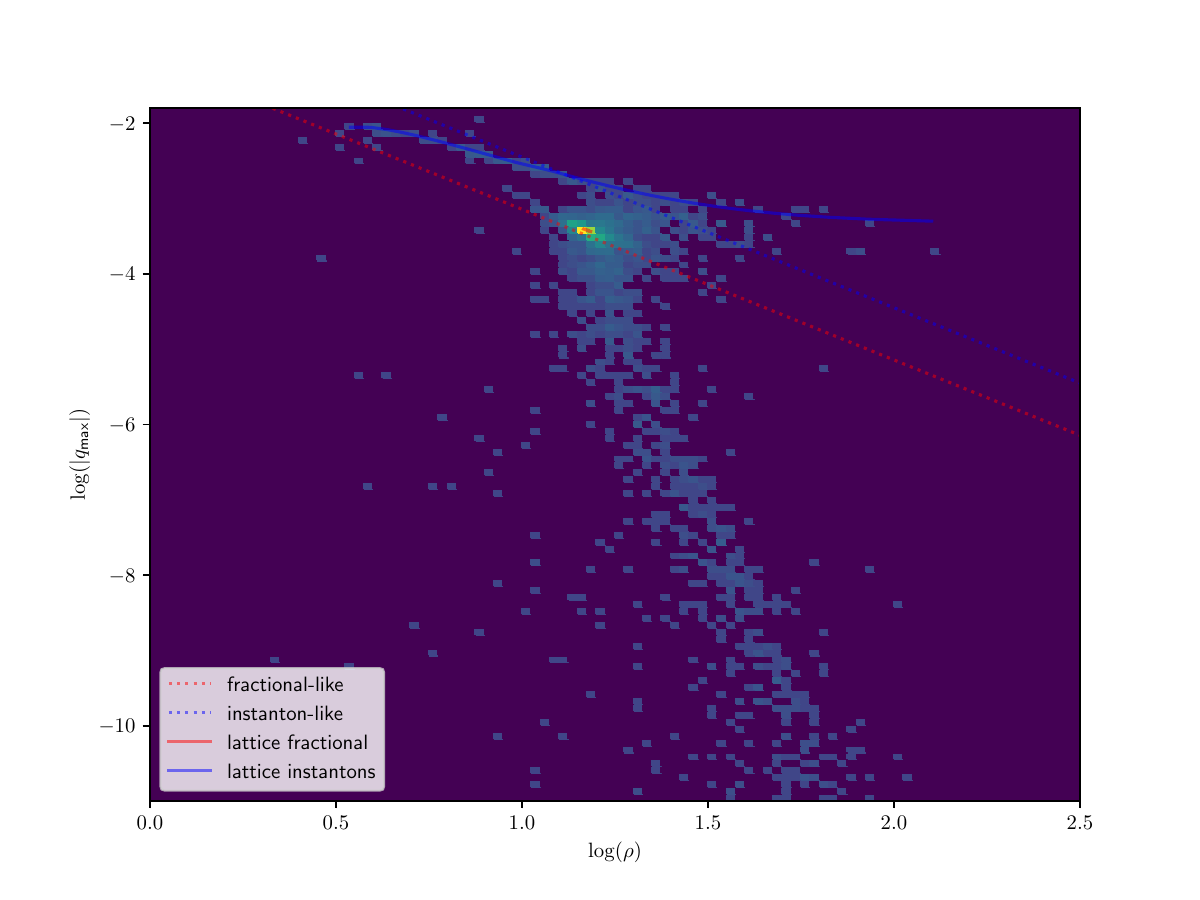}
    \caption{$t_{gf}=15$}
    \end{subfigure}
    \caption{Density plot of the fit parameters obtained with same parameters as Fig.~\ref{fig:fit_param_ns6tau15} but for different gradient flow times $t_{gf}$. Only a subset of the configuration has been considered.}
    \label{fig:fit_param_ns6taudep}
\end{figure}

The distribution of fractional instantons depends on the gradient flow time and on the choice of the lattice action in the gradient flow. For a subset of the total configurations, we have repeated the analysis at different flow times $t_{gf}$. The result is shown in Fig.~\ref{fig:fit_param_ns6taudep} with an increased scale of the y-axis to show the noise level at small $\qmax$.

At small $t_{gf}$, the level of the background noise indicated by the contribution at very small $\qmax$ is much higher. In addition the single peak of the fractional instanon gets blurred into a wider stripe by the noise. In these examples, the noise is even at $t_{gf}=4$ well separated from the fractional instanton peak, but at larger $l_s$ and smaller  $t_{gf}$ the identification becomes more ambiguous. Increasing the flow time, the noise gets reduced and the semiclassical features become clearer. At the same time, however, fractional instanton anit-instanton pairs are annihilated and the tail of the distribution is reduced.

We have also tested the overimproved flow as an alternative to the standard gradient flow with the Wilson action. In the regime at small $l_s$ and the specific flow times considered in our current analysis, there is no major difference between the two flow types. However, we expect a relevant difference towards larger $l_s$. One very small effect is that the instanton contributions are slightly shifted towards larger radius with the overimproved flow.

\section{Distribution of fit parameters at larger $N_c$ (SU(3) and SU(4))}
\label{sec:ncdependence}
\begin{figure}
    \begin{subfigure}[t]{0.5\textwidth}
    \includegraphics[width=\textwidth]{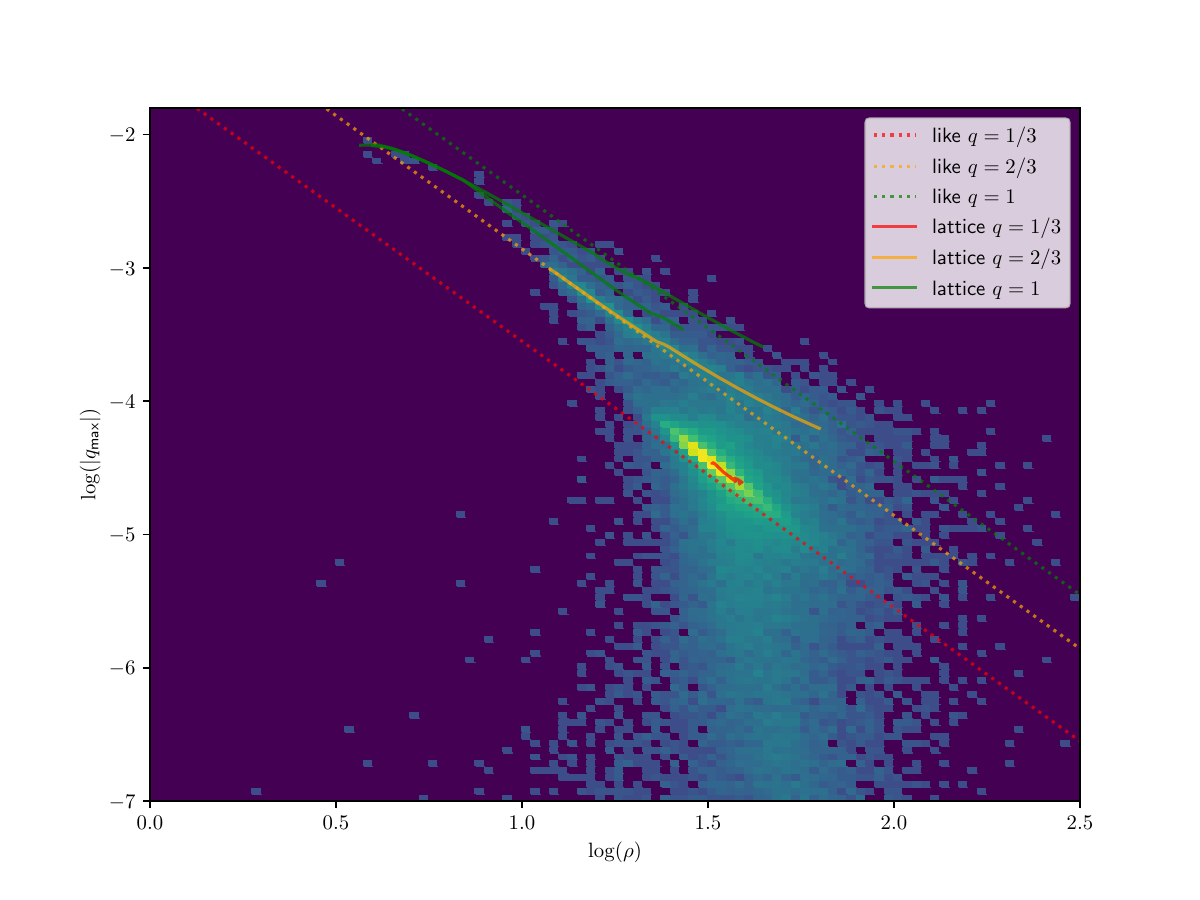}
    \caption{SU(3)}
    \end{subfigure}
        \begin{subfigure}[t]{0.5\textwidth}
    \includegraphics[width=\textwidth]{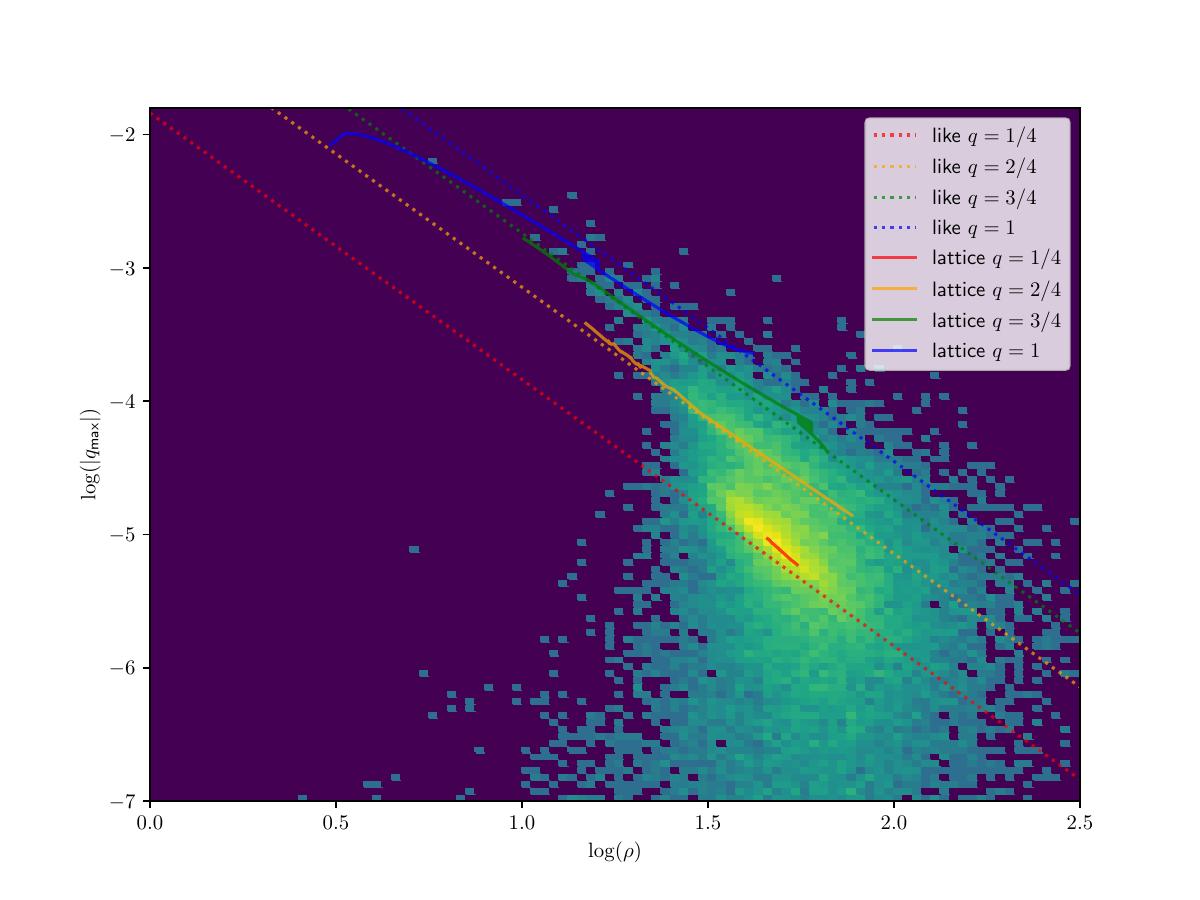}
    \caption{SU(4)}
    \end{subfigure}
    \caption{Density plot of the fit parameters obtained in case of SU(3) and SU(4) Yang-Mills theory. In both cases a $N_s=6$, $N_r=64$ lattice configuration. $\beta=6.3$ for SU(3) and $\beta=11.5$ for SU(4). In the case of SU(3) we have added two possible deformations of instantons as ``lattice $q=1$'' and their evolution with the gradient flow.}
    \label{fig:fit_param_ns6ncdep}
\end{figure}
We have investigated the topological densities on $T_2\times R^2$ also at larger $N_c$ for SU($N_c$) gauge theories. In the current work, we don't aim for a complete analysis of these cases, but we want to provide some of the results in order to discuss possible differences. The larger $N_c$ cases allow for several choices of the twist. We focus here on the same twist as in the SU(2) case.

Semiclassical configurations are generated at large $\beta$ and sufficient gradient flow. Using appropriate boundary conditions one can generate examples for the different objects with fractional $\Qi$. The semiclassical configurations can be deformed using different types of gradient flow.

A significant difference compared to SU(2) are the additional possible $\Qi<1$ configurations apart from the minimal $\frac{1}{N_c}$ topological charge. In the case of SU(3), for example, there are additional semiclassical objects with topological charge $2/3$. The instantons have, consequently, several possibilities to be decomposed into these objects. In a distribution at small $l_s$, which consists mostly out of $Q=1/N_c$ fractional instantons, the $Q=1$ contributions are less likely to be generated.

\section{Further characteristics of fractional instanton distribution}
\label{sec:sig_frac_inst}
We have investigated further observables in order to characterize the fractional instanton distributions. A more detailed discussion of fractional instanton solutions on $T_2\times R^2$ will be presented in a further publication and we therefore only briefly mention some properties, which can be investigated on the lattice.
\begin{figure}
    \centering
    \includegraphics[width=0.7\textwidth]{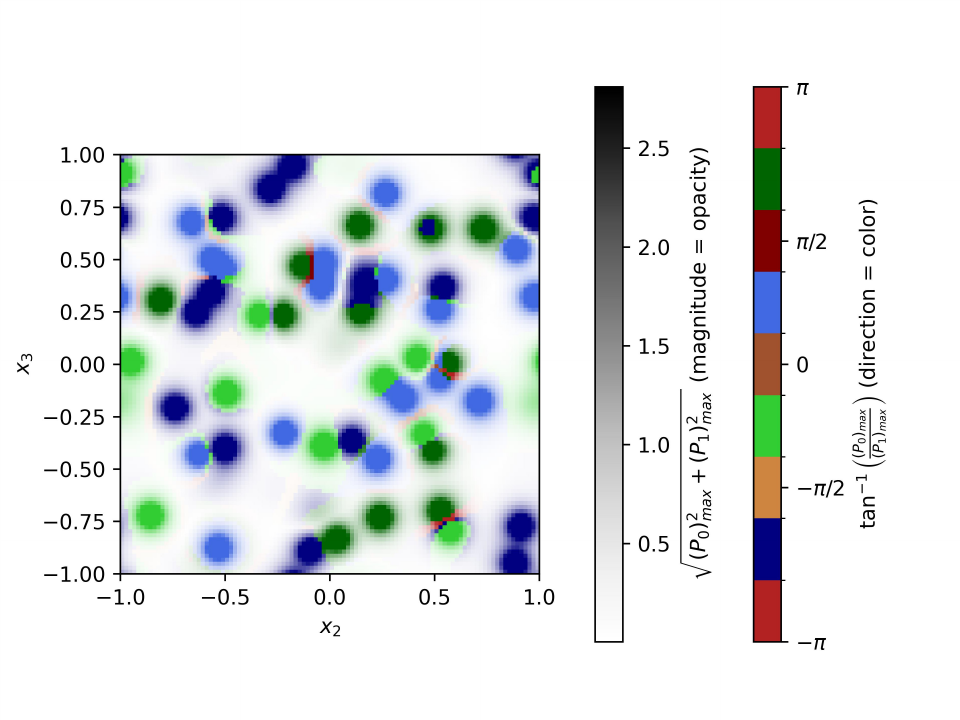}
    \caption{Polyakov loop at $Ns=6$, $\beta=2.6$. Same configuration as Fig~\ref{fig:examplens6} }
    \label{fig:pl_frac_inst}
\end{figure}

One example are the Polyakov loops in the two short directions $P_0(x_1,x_2.x_3)$ and $P_1(x_0,x_2,x_3)$. At each point in $R^2$ $(x_2,x_3)$ we take the value at the maximum of its modulus. Choosing $x_1$ or $x_2$ such that $|P_0|$ or $|P_1|$ are maximized gives a two-dimensional distribution of two numbers, a vector field. We represent it by the modulus and the angle $\phi$ of the corresponding vector. Semiclassics predicts that the vector is maximized at the center of each fractional instanton, which means that $|P_0|=|P_1|=2$. These provide four independent orientations  at the center points depending on the relative sign of $P_1$ and $P_2$: $(+,+)$ corresponding to $\phi=\pi/4$; $(+,-)$ corresponding to $\phi=-\pi/4$; $(-,-)$ corresponding to $\phi=-3\pi/4$; $(-,+)$ corresponding to $\phi=3\pi/4$. Fig.~\ref{fig:pl_frac_inst} shows that this is indeed in agreement with the results.

\section{Improved fit of fractional instanton distributions}
\begin{figure}
    \begin{subfigure}[t]{0.5\textwidth}
    \includegraphics[width=\textwidth]{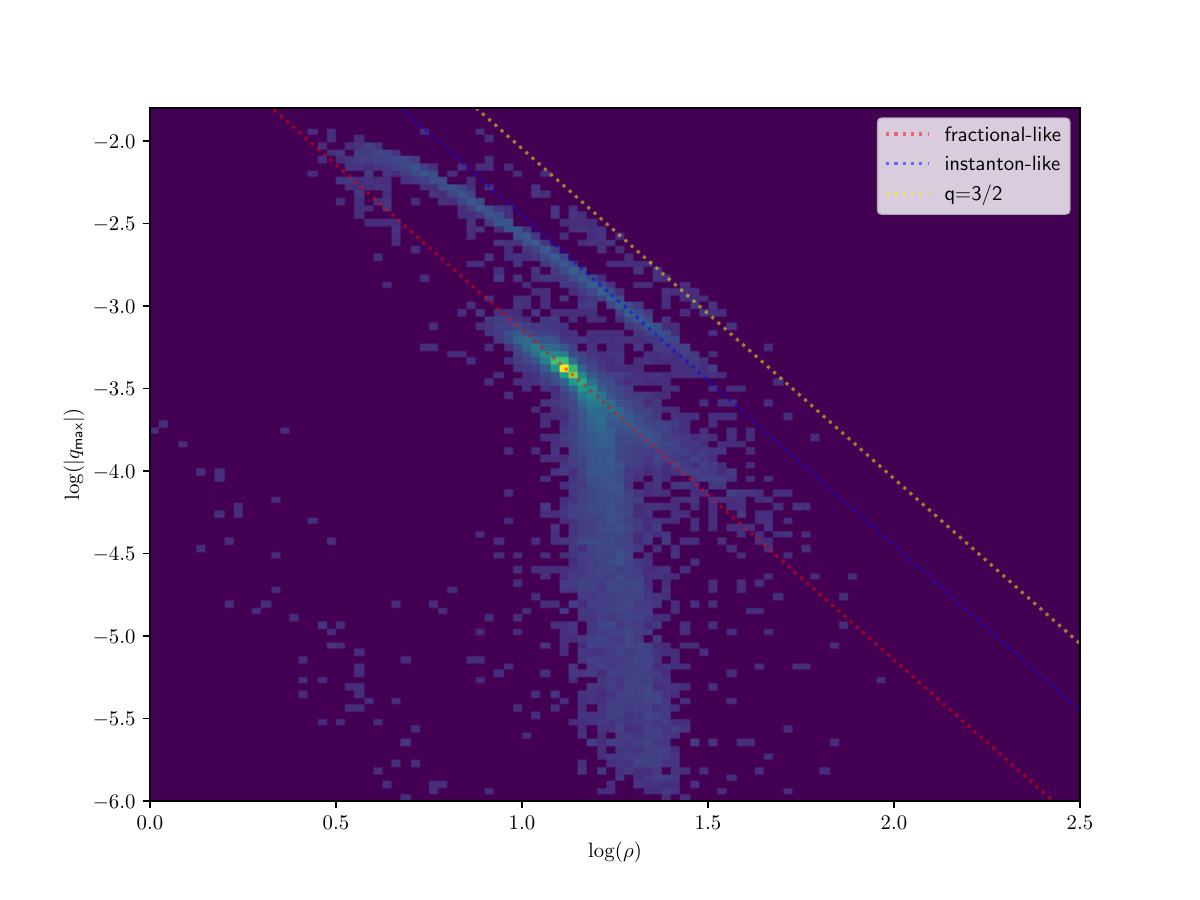}
    \caption{$\Ls=6$}
    \end{subfigure}
        \begin{subfigure}[t]{0.5\textwidth}
    \includegraphics[width=\textwidth]{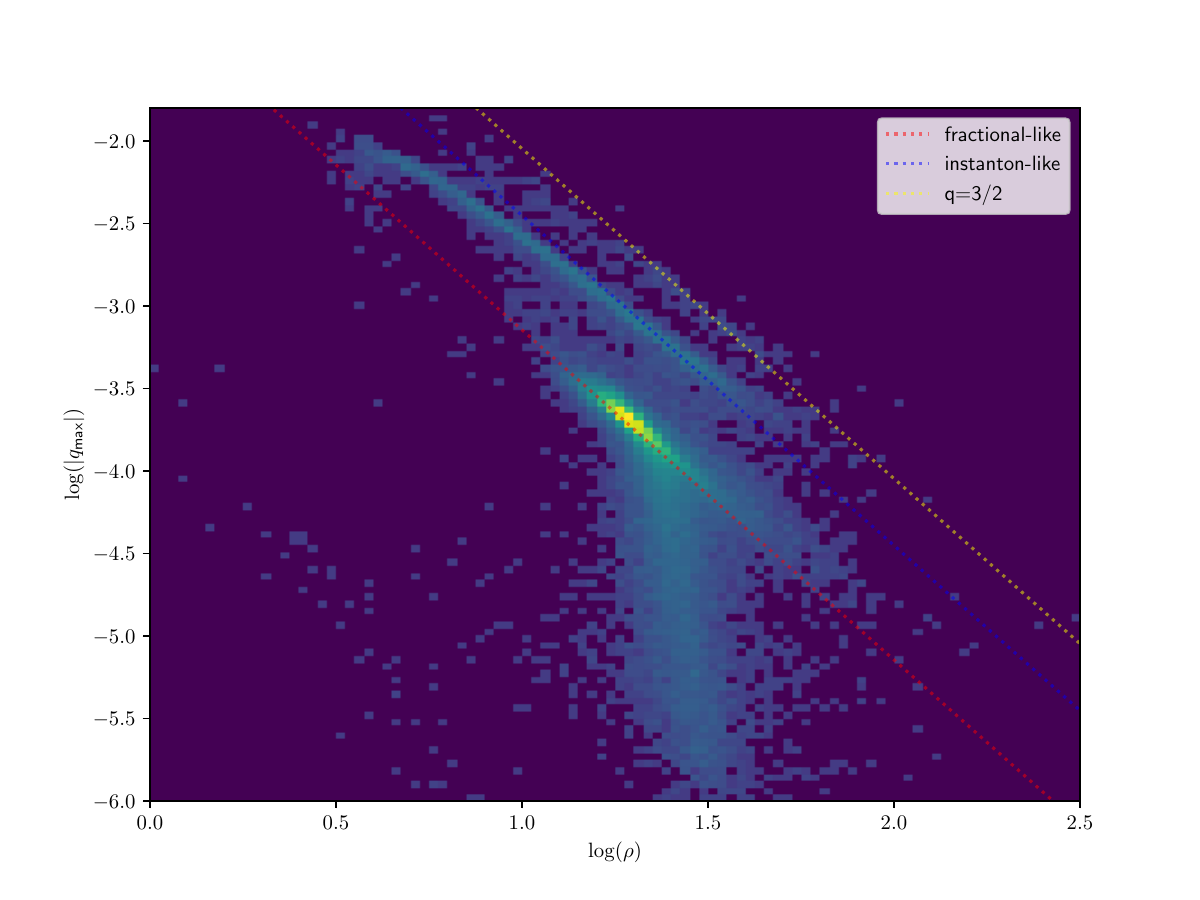}
    \caption{$\Ls=7$}
    \end{subfigure}
    \caption{Fit of the peak contributions at $\Ls=6,7$; $\Lr=104$; $\beta=2.6$; and $t_{gf}=15$. In contrast to the previous presentation in Fig.~\ref{fig:fit_param_ns6tau15}, all fit peaks are included in a single fit.}
    \label{fig:improved_fit}
\end{figure}
We have already tested methods to improve the fit of the fractional instanton contributions. It is possible to include the linear superposition of several objects in a single fit approach. In this way, one considers all of the peaks in a single fit. We have done first tests on such kind of approaches and the results are illustrated by Fig.~\ref{fig:improved_fit}. Note that we have considered only a single radius parameter $\rho=\rho_2=\rho_3$ and considered only the part of the distribution above the noise level.
The results indicate a clearer separation of the different topological contributions. However, it seems that more advanced methods are required in order to clarify the picture at large $l_s$.

\section{Extra tables}

\begin{table}[h!]
    \centering
    \begin{tabular}{ |c|c|c| } 
        \hline
        \Ls & \Lr & $\beta$ \\
        \hline
        3 & 104 & 2.6\\
        4 & 104 & 2.4\\
        4 & 104 & 2.5\\
        4 & 104 & 2.55\\
        4 & 104 & 2.6\\
        4 & 104 & 2.65\\
        4 & 104 & 2.7\\
        5 & 104 & 2.4\\
        5 & 104 & 2.5\\
        5 & 104 & 2.55\\
        5 & 104 & 2.6\\
        5 & 104 & 2.65\\
        5 & 104 & 2.7\\
        6 & 104 & 2.4\\
        6 & 104 & 2.5\\
        6 & 104 & 2.55\\
                \hline
\end{tabular}
\hspace{0.5cm}
\begin{tabular}{ |c|c|c| } 
        \hline
        \Ls & \Lr & $\beta$ \\
        \hline
        6 & 104 & 2.6\\
        6 & 104 & 2.65\\
        6 & 104 & 2.7\\
        7 & 104 & 2.4\\
        7 & 104 & 2.5\\
        7 & 104 & 2.55\\
        7 & 104 & 2.6\\
        7 & 104 & 2.65\\
        7 & 104 & 2.7\\
        8 & 104 & 2.6\\
        8 & 104 & 2.65\\
        8 & 104 & 2.7\\
        8 & 64 & 2.55\\
        9 & 104 & 2.6\\
        9 & 104 & 2.65\\
        9 & 104 & 2.7\\
        \hline
    \end{tabular} 
\hspace{0.5cm}
\begin{tabular}{ |c|c|c| } 
        \hline
        \Ls & \Lr & $\beta$ \\
        \hline
        9 & 104 & 2.8\\
        9 & 64 & 2.55\\
        10 & 104 & 2.6\\
        10 & 104 & 2.65\\
        10 & 104 & 2.7\\
        10 & 64 & 2.55\\
        10 & 64 & 2.6\\
        11 & 64 & 2.6\\
        12 & 64 & 2.6\\
        13 & 64 & 2.6\\
        14 & 104 & 2.6\\
        18 & 104 & 2.7\\
        20 & 64 & 2.6\\
        30 & 64 & 2.6\\
        \hline
        \multicolumn{2}{c}{}\\
        \multicolumn{2}{c}{}\\
\end{tabular} 
    \caption{Complete collection of all runs relevant for our Monte-Carlo simulations summarized in Tab.~\ref{tab:confs}.}
    \label{tab:confsfull}
\end{table}

\end{document}